%% file: pyratbayI.tex
\documentclass[fleqn, usenatbib]{mnras}

\usepackage{newtxtext}
\usepackage[T1]{fontenc}
\usepackage{ae,aecompl}
\usepackage{comment}
\usepackage{graphicx}	
\usepackage{amsmath}	
\usepackage{amssymb}	
\usepackage{verbatim}
\usepackage{natbib}
\usepackage{etoolbox}
\usepackage{longtable}
\usepackage{ragged2e}

\usepackage[frozencache=true,cachedir=.]{minted}

\let\oldumu=\umu
\renewcommand\umu{\ifmmode\oldumu\else\math{\oldumu}\fi}
\newcommand\micro{\umu}
\renewcommand\micron{\micro m}
\newcommand\microns{\micron}

\let\oldsim=\sim
\renewcommand\sim{\ifmmode\oldsim\else\math{\oldsim}\fi}
\let\oldpm=\pm
\renewcommand\pm{\ifmmode\oldpm\else\math{\oldpm}\fi}
\newcommand\by{\ifmmode\times\else\math{\times}\fi}
\newcommand\ttt[1]{10\sp{#1}}
\newcommand\tttt[1]{\by\ttt{#1}}

\newbox{\wdbox}
\newcommand\added[1]{{\if\trackchanges1\bf\fi{#1}}}

\catcode`@=11

\renewcommand\math[1]{$#1$}

\let\atab=&

\let\oldmsp=\sp
\let\oldmsb=\sb
\def\sp#1{\ifmmode
           \oldmsp{#1}%
         \else\strut\raise.85ex\hbox{\scriptsize #1}\fi}
\def\sb#1{\ifmmode
           \oldmsb{#1}%
         \else\strut\raise-.54ex\hbox{\scriptsize #1}\fi}
\newbox\@sp
\newbox\@sb
\def\sbp#1#2{\ifmmode%
           \oldmsb{#1}\oldmsp{#2}%
         \else
           \setbox\@sb=\hbox{\sb{#1}}%
           \setbox\@sp=\hbox{\sp{#2}}%
           \rlap{\copy\@sb}\copy\@sp
           \ifdim \wd\@sb >\wd\@sp
             \hskip -\wd\@sp \hskip \wd\@sb
           \fi
        \fi}

\newcommand\SST{{\em Spitzer Space Telescope}}
\newcommand\Spitzer{{\em Spitzer}}

\newcommand\HST{{\em HST}}
\newcommand\Hubble{{\em Hubble}}
\newcommand\Webb{{\em James Webb Space Telescope}}
\newcommand\JWST{{\em JWST}}
\newcommand\ariel{{\em Ariel}}
\newcommand\chisq{\ifmmode{\chi\sp{2}}\else\math{\chi\sp{2}}\fi}
\newcommand\redchisq{\ifmmode{ \chi\sp{2}\sb{\rm red}}
                    \else\math{\chi\sp{2}\sb{\rm red}}\fi}
\newcommand\Teq{\ifmmode{T\sb{\rm eq}}\else$T$\sb{eq}\fi}
\newcommand\Tb{\ifmmode{T\sb{\rm b}}\else$T$\sb{b}\fi}
\newcommand\mjup{\ifmmode{M\sb{\rm Jup}}\else$M$\sb{Jup}\fi}
\newcommand\rjup{\ifmmode{R\sb{\rm Jup}}\else$R$\sb{Jup}\fi}
\newcommand\msun{\ifmmode{M\sb{\odot}}\else$M\sb{\odot}$\fi}
\newcommand\rsun{\ifmmode{R\sb{\odot}}\else$R\sb{\odot}$\fi}
\newcommand\Rs{\ifmmode{R\sb{\rm s}}\else$R\sb{\rm s}$\fi}
\newcommand\mearth{\ifmmode{M\sb{\oplus}}\else$M\sb{\oplus}$\fi}
\newcommand\rearth{\ifmmode{R\sb{\oplus}}\else$R\sb{\oplus}$\fi}
\newcommand\Rp{\ifmmode{R\sb{\rm p}}\else$R\sb{\rm p}$\fi}

\newcommand\kb{\ifmmode{k\sb{\rm B}}\else$k\sb{\rm B}$\fi}
\newcommand\molhyd{H$\sb{2}$}
\newcommand\methane{CH$\sb{4}$}
\newcommand\water{H$\sb{2}$O}
\newcommand\carbdiox{CO$\sb{2}$}

\newcommand\ammonia{NH$\sb{3}$}

\newcommand\kayser{cm$\sp{-1}$}
\newcommand\repack{\textsc{repack}}
\newcommand\pyratbay{\textsc{Pyrat Bay}}

\newcommand\TEA{\textsc{TEA}}
\newcommand\rate{\textsc{rate}}
\newcommand\MCC{\textsc{mc3}}
\newcommand\petit{\textsc{petitRADTRANS}}
\newcommand\taurex{\textsc{TauREx}}
\newcommand\aura{\textsc{Aura}}
\newcommand\poseidon{\textsc{poseidon}}
\newcommand\twoohnineb{HD~209458\,b}
\newcommand\tli{\textsc{tli}}

\newcommand\atmosphere{\textsc{atmosphere}}
\newcommand\spectrum{\textsc{spectrum}}
\newcommand\opacity{\textsc{opacity}}
\newcommand\mcmc{\textsc{mcmc}}
\newcommand\pcloud{\ifmmode{p\sb{\rm top}}
                 \else\math{p\sb{\rm top}}\fi}

\newcommand\vs{\emph{vs.}}
\newcommand\der{\ifmmode{\rm d}\else\math{\rm d}\fi}

\usepackage{minted}
\usepackage{threeparttable}

\defcitealias{MacDonaldMadhusudhan2017mnrasRetrievalHD209b}{MM2017}
\defcitealias{PinhasEtal2019mnrasHotJupiterSampleRetrieval}{P2019}

\title[The Pyrat Bay Framework for Exoplanet Modeling]
{The Pyrat Bay Framework for Exoplanet Atmospheric Modeling:\\
A Population Study of \textit{\textbf{{ Hubble}}}/WFC3 Transmission Spectra}

\author[Cubillos \& Blecic]{
Patricio~E.~Cubillos$^{1}$\thanks{E-mail: patricio.cubillos@oeaw.ac.at}
and 
Jasmina~Blecic$^{2,3}$
\\
$^{1}$Space Research Institute, Austrian Academy of Sciences,
Schmiedlstra{\ss}e 6, Graz 8042, Austria\\
$^{2}$Department of Physics, New York University Abu Dhabi,
PO Box 129188 Abu Dhabi, UAE\\
$^{3}$Center for Astro, Particle, and Planetary Physics (CAP$^3$),
New York University Abu Dhabi, PO Box 129188 Abu Dhabi, UAE
}

\pubyear{2021}

\begin{document}
\label{firstpage}
\pagerange{\pageref{firstpage}--\pageref{lastpage}}
\maketitle

\begin{abstract}
We present the open-source {\pyratbay} framework for exoplanet
atmospheric modeling, spectral synthesis, and Bayesian retrieval.  The
modular design of the code allows the users to generate atmospheric 1D
parametric models of the temperature, abundances (in thermochemical
equilibrium or constant-with-altitude), and altitude profiles (in
hydrostatic equilibrium); sample ExoMol and HITRAN line-by-line
cross sections with custom resolving power and line-wing cutoff values;
compute emission or transmission spectra considering cross sections from
molecular line transitions, collision-induced absorption, Rayleigh
scattering, gray clouds, and alkali resonance lines; and perform
Markov chain Monte Carlo atmospheric retrievals for a given transit or
eclipse dataset.  We benchmarked the {\pyratbay} framework by
reproducing line-by-line sampling of ExoMol cross sections,
producing transmission and emission spectra consistent with {\petit}
models, accurately retrieving the atmospheric properties of
simulated transmission and emission observations generated with
{\taurex} models, and closely reproducing {\aura} retrieval
analyses of the space-based transmission spectrum of {\twoohnineb}.
Finally, we present a retrieval analysis of a population of transiting
exoplanets, focusing on those observed in transmission with the {\HST}
WFC3/G141 grism.  We found that this instrument alone can confidently
identify when a dataset shows {\water}-absorption features; however,
it cannot distinguish whether a muted {\water} feature is caused by
clouds, high atmospheric metallicity, or low {\water} abundance.  Our
results are consistent with previous retrieval analyses.  The
{\pyratbay} code is available at PyPI (\mintinline{shell}{pip install
pyratbay}) and conda.  The code is heavily documented
(\href{https://pyratbay.readthedocs.io}
{https://pyratbay.readthedocs.io}) and tested to provide maximum
accessibility to the community and long-term development stability.
\end{abstract}

\begin{keywords}
planets and satellites: atmosphere --
radiative transfer --
methods: statistical --
software: public release
\end{keywords}

\section{Introduction}
\label{sec:intro}

Since the discovery of the first planet orbiting around a solar-type
star other than our sun \citep{MayorQueloz1995natJupiterExoplanet},
exoplanets have shaken our conception of the universe.  The more than
4000 exoplanets known to date have shown a diversity far larger than
what we expected from our own solar system.  The unprecedented
conditions found in hot-Jupiter or super-Earth planets, for example,
represent an invaluable opportunity to widen our understanding of
atmospheric and planetary physics.

In terms of characterization, spectro-photometric time-series
observations of transiting exoplanets have taken a central stage.
When planets transit in front of their host stars, they block a
fraction of the observed stellar flux.  The amplitude of this signal
is proportional to the square of the planet-to-star radius ratio.
When planets pass behind their host stars, the planetary emission is
blocked by the star.  The amplitude of this signal is proportional to
the planet-to-star flux ratio.  The observational and theoretical
techniques used to characterize transiting exoplanets have shown a
sustained and fast-paced development over time.  Characterization
studies from space-based observatories (the focus of this article)
have quickly evolved from single-target, single-broadband observations
\citep{CharbonneauEtal2000apjHD209458bTransit}, to
population studies of multi-wavelength 
spectroscopic observations
\citep[e.g.,][]{SingEtal2016natHotJupiterTransmission,
  IyerEtal2019apjWFC3transmissionSample,
  BarstowEtal2017apjHotJupiterRetrieval,
  FuEtal2017apjWFC3statistics,
  TsiarasEtal2018ajPopulationStudy,
  FisherHeng2018mnrasWFC3SampleRetrieval,
  PinhasEtal2019mnrasHotJupiterSampleRetrieval,
  WelbanksEtal2019apjMasssMetallicityTrends,
  GarhartEtal2020ajSpitzerEclipsesStatistics}, covering from the
  ultraviolet to the infrared region of the spectrum.

By modeling how light interacts with a planetary atmosphere (via the
radiative-transfer equation), transit and eclipse observations help us
determine the physical properties of the atmosphere, as different
species produce unique spectral features as a function of wavelength,
temperature, and pressure.  Furthermore, the inferred physical state
of a planetary atmosphere gives us an insight into the planet's
formation, evolution, chemistry, and dynamics.

To estimate the atmospheric properties in a statistically robust
manner, many studies adopt the Bayesian retrieval approach.  This
approach evaluates the posterior distribution of a parametric
atmospheric model given a dataset of an observed spectrum.  A Bayesian
sampler---e.g., a Markov chain Monte
Carlo \citep[MCMC,][]{MetropolisEtal1953jchphMCMC} or nested
sampling \citep{Skilling2004aipcNestedSampling}--- explores the
parameter space by evaluating a large number of model samples, guided
by the prior and likelihood functions.  The sampler generates a
posterior distribution from which one can obtain the parameters'
best-fitting values and credible intervals (the uncertainties).

{\citet{MadhusudhanSeager2009apjRetrieval,
MadhusudhanSeager2010apjRetrieval}} introduced the retrieval approach
to the exoplanet field for atmospheric characterization.  Soon after,
other independent retrieval schemes emerged, e.g.,
NEMESIS \citep{IrwinEtal2008jqsrtNEMESIS},
SCARLET \citep{Benneke2015arxivSCARLETretrieval},
POSEIDON \citep{MacDonaldMadhusudhan2017mnrasRetrievalHD209b},
ATMO \citep{EvansEtal2017natWASP121bEmissionHST,
WakefordEtal2017sciHATP26bTransmission},
\textsc{HyDRA} \citep{GandhiMadhusudhan2018mnrasHyDRA},
and \textsc{ARCiS} \citep{OrmelMin2019aaARCiS}.  Recently, a number of
open-source retrieval packages have facilitated the accessibility and
cross-comparison of retrieval frameworks, e.g.,
CHIMERA \citep{LineEtal2013apjRetrievalI},
PLATON \citep{ZhangEtal2019paspPLATON},
{\petit} \citep{MolliereEtal2019aaPetitRADTRANS},
\textsc{Helios-r2} \citep{KitzmannEtal2020apjHelios-r2}, and
{\taurex} \citep{AlRefaieEtal2019arxivTaurex3}.

Developing a comprehensive atmospheric modeling framework is a
challenging task, as it requires the knowledge of a vast number of
physical processes (e.g., radiative transfer, chemistry, cloud
formation, dynamics, etc.) and advanced statistical analyses.
Furthermore, both the physical and statistical analyses are
computationally intensive tasks, demanding an efficient implementation
of tools.  To overcome these limitations, researchers resort to a wide
range of often strong assumptions and simplifications, for example:
adopt 1D atmospheric parameterizations for 3D objects, neglect stellar
contamination due to starspots or
faculae \citep[e.g.,][]{RackhamEtal2017apjStellarHeterogeneityI}, or
assume isobaric properties throughout the
atmosphere \citep{HengKitzmann2017mnrasSemiAnalyticalRT}.  The
inadequacy of these assumptions has already been
debunked \citep[e.g.,][]{RocchettoEtal2016apjJWSTbiases,
BlecicEtal2017apj3Dretrieval, CaldasEtal2019aaTransmission3Deffects,
WelbanksMadhusudhan2019ajRetrievalDegeneracies,
WelbanksEtal2019apjMasssMetallicityTrends,
TaylorEtal2020mnrasEmissionBiases,
MacDonaldEtal2020apjColdTemperatureRetrievals}.  However, the limited
signal-to-noise ratio, spectral coverage, and resolving power of
current instruments do not statistically justify the use of more
sophisticated models.  This will not be the case for the next
generation of observatories, like the {\Webb} or the {\it Atmospheric
Remote-sensing Infrared Exoplanet Large-survey} ({\ariel}) mission, as
they will enable a more precise characterization of planetary
atmospheres.

The flexibility of the retrieval approach allows for a wide variety of
modeling assumptions, ranging from freely-independent
parameterizations (where atmospheric properties vary independently of
each other) to self-consistent calculations (where the different
properties are related through physical principles).  In general, more
self-consistent models require fewer free parameters, which simplifies
the posterior exploration.  However, such an approach makes strong
assumptions that may not be applicable to a given case.  The adopted
assumptions can even dominate the outcome of a retrieval study.  Thus,
it is highly valuable to compare and contrast the results from
multiple retrieval frameworks that apply different theoretical and
statistical approaches.  Only recently the community has started to
carry out large-scale cross-validation studies of these retrieval
frameworks \citep[][]{BaudinoEtal2017apjBenchmarking,
BarstowEtal2020mnrasRetrievalComparison} and multiple-independent
analyses of a same
observation \citep[][]{KilpatrickEtal2018apjWASP63bWFC3,
VenotEtal2020apjWASP43bWebbPredictions}.  In this context, where
retrieval results are strongly model-dependent and there is a need for
analysis intercomparison, we present the Python Radiative-transfer in
a Bayesian framework ({\pyratbay}), a modular open-source package for
planetary atmospheric modeling, emission and transmission spectral
synthesis, and Bayesian atmospheric retrieval.

In Section \ref{sec:code}, we describe the {\pyratbay} atmospheric
modeling and retrieval package.  In Section \ref{sec:benchmark} we
benchmark the {\pyratbay} package by comparing its outputs to other
datasets and open-source packages.  In Section \ref{sec:analysis} we
present the {\pyratbay} atmospheric retrieval analysis of transmission
spectra for 26 exoplanet datasets obtained from space-based
observations.  Finally, in Section \ref{sec:summary} we summarize our
conclusions and discuss future developments.

\begin{figure}
\centering
\includegraphics[width=\linewidth, clip, trim=0 10 0 0]
{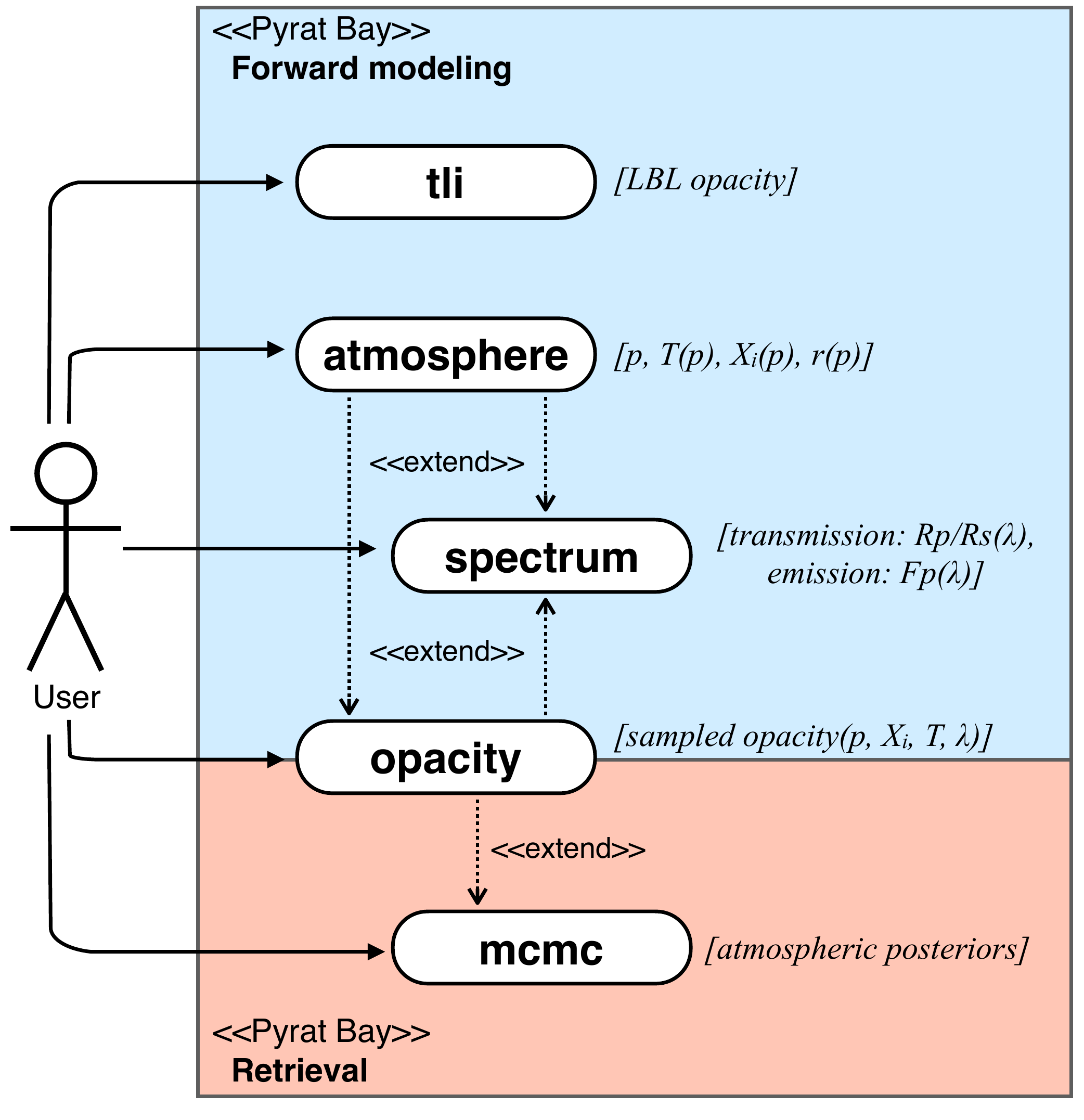}
\caption{
    {\pyratbay} Unified Modeling Language (UML) use case diagram, and
    the main outputs from each case (italics).  Users can execute the
    {\pyratbay} run modes (white bubbles) either individually or
    chained (dashed lines) in a single command (e.g., compute a
    temperature profile, abundances, and spectrum all at once), or
    execute a case with custom inputs provided (e.g., compute a
    spectrum from a custom provided atmospheric model).}
\label{fig:usecase}
\end{figure}

\section{The Pyrat-bay Modeling Package}
\label{sec:code}

The Python Radiative-transfer in a Bayesian framework ({\pyratbay})
implements state-of-the-art forward and retrieval-modeling tools to
sample line-by-line cross sections, computes one-dimensional
atmospheric models, generates transmission and emission spectra, and
performs Bayesian atmospheric retrievals.  Figure \ref{fig:usecase}
shows the current {\pyratbay} running modes available to the user.
Briefly, the {\tli} mode reformats cross-section line lists that are
available online into transition line information (TLI) files, which
is the format used by {\pyratbay} to compute spectra. The
{\atmosphere} mode computes atmospheric pressure ($p$), temperature
($T$), volume-mixing-ratio abundance ($X\sb{\rm i}$), and altitude
($r$) profiles.  The {\spectrum} mode computes transmission spectra of
stellar light as it passes through the limb of a planet during a
transit, or emission spectra of a planet integrated over the observed
hemisphere.  The {\opacity} mode samples the TLI line-by-line cross
sections into a pressure, temperature, and wavelength grid.  Lastly,
the {\mcmc} mode performs Bayesian atmospheric retrievals, constrained
by a transmission or emission set of observations.  The following
sections describe in more detail the design of the {\pyratbay} package
(Sec.~\ref{sec:design}), the underlying physics
(Secs.~\ref{sec:atmosphere}--\ref{sec:spectrum}), and the statistical
approach (Sec.~\ref{sec:mcmc}).

\subsection{The Design}
\label{sec:design}

Given the large number of interconnected routines required to
implement a comprehensive retrieval framework, we closely followed the
best-practices for software development when designing the {\pyratbay}
package.  In particular, we adopted the Python Enhancement Proposals
(PEP) for style (PEP 8), philosophy (PEP 20), documentation (PEP 257),
and versioning (PEP 440).  We designed the {\pyratbay} framework to
provide maximum accessibility to the community by utilizing mostly the
Python Standard Library packages.  In this way, we minimized the usage
barriers imposed by external dependencies that may have
hardware-specific dependencies (e.g., CUDA on NVIDIA GPU) or that have
a long chain of dependencies (e.g., \textsc{PyMultiNest}).  We
dedicated a significant effort to document and test the {\pyratbay}
package ($\sim$40\% of the written code corresponds to tests and
documentation).  Simultaneously, along with feature development, we
implemented an exhaustive unit and integrated testing suit using
the \textsc{pytest} package, automated through the Travis Continuous
Integration framework.  This not only prevents the developers to
introduce bugs, but also enforces a better code design (hard-to-test
code is indicative of overtly complicated code).  In summary, we
designed {\pyratbay} as a sustainable code, to endure the rapid-pace
development seen in the exoplanet field.

The code is executable either from the command line or interactively.
Most of the code is implemented in Python object-oriented programming,
including C-code extensions for computationally-intensive low-level
routines to retain high performance.  Running the code interactively
enables the user to access most of the input, intermediate, and output
variables involved in a given calculation.  Beyond the core
functionality shown in Figure \ref{fig:usecase}, we designed the
application programming interface (API) of the package such that most
intermediate routines are self-sufficient and fully documented.  This
provides the user access to a broad range of routines to compute
specific processes that are of common interest for planetary
atmospheric research (e.g., hydrostatic equilibrium, equilibrium
temperatures, spectral band-integration, etc.).

The {\pyratbay} code is compatible with Python 3.6+, and has been
tested to work on Linux and OS X machines.  The software is available
for direct installation from the Python Package Index (PyPI) as:
\begin{minted}{shell}
    pip install pyratbay
\end{minted}
or from conda:
\begin{minted}{shell}
    conda install -c conda-forge pyratbay
\end{minted}
The source code is hosted at the Github version-control
repository \href{https://github.com/pcubillos/pyratbay}
{https://github.com/pcubillos/pyratbay}, and the documentation is
hosted at \href{https://pyratbay.readthedocs.io}
{https://pyratbay.readthedocs.io}.  A significant fraction of the
{\pyratbay} code is based on the GNU GPLv2 radiative-transfer
code \textsc{Transit}
\citep{Rojo2006phdThesis}, thus the {\pyratbay} code is available
under the open-source GNU GPLv2 license.

\subsection{The Atmosphere}
\label{sec:atmosphere}

The {\pyratbay} code provides routines to either construct planetary
atmospheric profiles or read custom profiles given by the user.  The
subsections below describe the pressure, temperature, volume-mixing
ratio, and altitude profile models that are provided by the code.

\subsubsection{Pressure Profile}
\label{sec:pressure}

The pressure ($p$) is the independent variable of the atmosphere in
this framework.  {\pyratbay} provides a routine to compute a pressure
profile as an equi-spaced array in log-pressure, determined by the
pressures at the top and bottom of the atmosphere and the number of
layers.

\subsubsection{Temperature Profile}
\label{sec:temperature}

{\pyratbay} provides three parametric temperature-profile models.  The
first model is a simple isothermal profile, which has a single
parameter ($T_0$) that sets a constant temperature at all layers in
the atmosphere:
\begin{equation}
T(p) = T_0.
\end{equation}

The second model is the three-channel Eddington approximation model
of \citet{LineEtal2013apjRetrievalI}, which is based on the analytic
formulation by \citet{Guillot2010aaRadiativeEquilibrium}.  This model
has six parameters: $\log_{10}(\kappa')$, $\log_{10}(\gamma_1)$,
$\log_{10}(\gamma_2)$, $\alpha$, $T_{\rm irr}$, and $T_{\rm int}$.
The temperature profile is given as:
\begin{equation}
T\sp{4}(p) = \frac{3 T\sb{\rm int}\sp{4}}{4} \left(\frac{2}{3} + \tau\right) 
    + (1-\alpha) \frac{3 T\sb{\rm irr}\sp{4}}{4} \xi_1(\tau)
    +    \alpha  \frac{3 T\sb{\rm irr}\sp{4}}{4} \xi_2(\tau),
\end{equation}
with
\begin{equation}
\xi_i(\tau) = \frac{2}{3} + \frac{2}{3\gamma_i}
  \left[1 + \left(\frac{\gamma_i\tau}{2}-1\right)e^{-\gamma_i\tau}\right] +
  \frac{2\gamma_i}{3} \left(1-\frac{\tau\sp{2}}{2}\right)E\sb{2}(\gamma_i\tau),
\end{equation}
where $E\sb{2}(\gamma\sb{i}\tau)$ is the second-order exponential
integral; $T\sb{\rm int}$ is the internal heat temperature; and
$\tau(p) = \kappa' p = \kappa p /g$ is the thermal optical depth for a
given atmospheric gravity, $g$.  Note that this implementation uses
the variable $\kappa'\equiv\kappa/g$ instead of $\kappa$ as a free
parameter, since the gravity is redundant from a parametric point of
view, as it acts only as a scaling factor between $p$ and $\tau$.  The
irradiation temperature parameter $T_{\rm irr}$ determines the stellar
flux absorbed by the atmosphere.  For a given system it can be
estimated as:
\begin{equation}
T\sb{\rm irr} = \left(\frac{1-A}{f}\right)^{1/4}
                \left( \frac{R\sb{\rm s}}{2a}\right)\sp{1/2} T\sb{\rm s},
\end{equation}
where $T\sb{\rm s}$ and $R\sb{\rm s}$ are the stellar effective
temperature and radius, respectively, $a$ is the orbital semi-major
axis, $A$ is the Bond albedo, and $f$ is a day--nightside energy
redistribution factor ranging from 0.5 (no redistribution to night
side) to 1.0 (redistribution over entire surface).

Lastly, the third model is the parametric profile
of \citet{MadhusudhanSeager2009apjRetrieval}, which has six
parameters: $\log_{10}(p_1)$, $\log_{10}(p_2)$, $\log_{10}(p_1)$,
$a_1$, $a_2$, and $T_0$, that describe a temperature profile separated
into three regions (layers 1 through 3 from top to bottom).  Following
the nomenclature of \citet{MadhusudhanSeager2009apjRetrieval}, the
sub-indices 0, 1, and 3 refer to the top of layers 1, 2, and 3,
respectively.  The temperature profile is given as:
\begin{equation}
T(p) = \left\{
\begin{array}{lll}
T_0 + \left[\frac{1}{a_1}\ln(p/p_0)\right]^2 & \text{if } p < p_1
   & (\rm layer\ 1) \\
T_2 + \left[\frac{1}{a_2}\ln(p/p_2)\right]^2 & \text{if } p_1 \le p < p_3
   & (\rm layer\ 2) \\
T_3   & \text{if } p \ge p_3 & (\rm layer\ 3)
\end{array} \right.
\end{equation}
A thermally inverted profile will result when $p_1<p_2$; whereas a
non-inverted profile will result when $p_2<p_1$.

\subsubsection{Chemical Composition Profiles}
\label{sec:abundances}

{\pyratbay} offers two models to compute 1D volume-mixing-ratio
abundances: uniform- and thermochemical-equilibrium abundance
profiles.  The uniform-abundance model creates constant-with-altitude
volume-mixing-ratio profiles for each species, with values specified
by the user.  {\pyratbay} also allows the user to compute abundances
in thermochemical equilibrium via the {\TEA}
package \citep{BlecicEtal2016apsjTEA}, with custom elemental
metallicities (all metal abundances scaled by a same factor) or custom
individual elemental abundances.  Given the pressure, temperature, and
elemental composition of an atmosphere, {\TEA} finds the equilibrium
solution by minimizing the Gibbs free energy of the system.

\subsubsection{Altitude Profile}
\label{sec:hydroeq}

{\pyratbay} computes the atmospheric altitude (radius) profile of the
pressure layers assuming hydrostatic equilibrium.  We consider the
ideal-gas law and an altitude dependent gravity $g(r)=GM\sb{\rm
p}/r\sp{2}$, where $G$ is the gravitational constant and $M\sb{\rm p}$
is the mass of the planet, to solve the hydrostatic-equilibrium
equation as:
\begin{equation}
\label{eq:hydrostatic}
\frac{\der r}{r\sp{2}} = -\frac{k_{\rm B}T}{\mu GM\sb{\rm p}} \frac{\der p}{p},
\end{equation}
where $k_{\rm B}$ is the Boltzmann constant, and $\mu$ is the mean
molecular mass of the atmosphere.  Note that both $\mu$ and $T$ are
pressure-dependent variables.  Adopting an altitude-dependent gravity
is particularly necessary for low-density planets, where the scale
height may change significantly across the atmosphere.

To obtain the particular solution of this differential equation, the
user needs to supply a pair of radius--pressure reference values to
define the boundary condition $r(p\sb{0}) = R\sb{0}$.  Note that the
selection of the $\{p\sb{0},R\sb{0}\}$ pair is arbitrary.  A
meaningful selection would be the location near the transit radius
(for transmission spectroscopy) or the photosphere (for emission).
Given the degeneracies existing between the temperature and
composition of an atmosphere with the reference $\{p\sb{0},R\sb{0}\}$
pair \citep{Griffith2014rsptaDegenerateSolutions}, atmospheric
retrievals should fit for the reference point when solving the
hydrostatic-equilibrium equation.

\subsubsection{Hill Radius}
\label{sec:toa}

If the user provides the stellar mass and orbital semi-major axis,
{\pyratbay} computes the Hill radius of the planet:
\begin{equation}
R_{\rm H} = a\sqrt[3]{\frac{M\sb{\rm p}}{3M\sb{\rm s}}},
\end{equation}
where $M\sb{\rm s}$ is the mass of the host star.  Above this
altitude, the atmospheric particles are no longer bound to the
planetary gravitational potential, and thus the number density
decreases much faster with altitude than under hydrostatic
equilibrium.  To account for this, if necessary, the code truncates
the top of the planetary atmospheric model at the Hill radius.

\subsection{The Cross Sections}
\label{sec:opacities}

To solve the radiative-transfer equation, we need to compute the
optical depth $\tau$ (unitless), which describes how opaque the
atmosphere is across a given path ${\rm d}s$:
\begin{equation}
\tau(\nu) = \int e(\nu) {\rm d}s,
\end{equation}
where $e(\nu)$ is the extinction coefficient (units of cm$^{-1}$) and
$\nu$ is the wavenumber (i.e., the inverse of the wavelength $\lambda
= 1/\nu$).  Typically, several sources contribute to the extinction
coefficient:
\begin{equation}
e(\nu) = \sum_i n_i \sigma_i + \sum_j n_{j1}\,n_{j2}\,\sigma_j,
\end{equation}
where $n_i$ is the number density of the interacting species (units of
molecule cm$^{-3}$), and $\sigma$ is the absorption cross section
(units of cm$^{2}$ molecule$^{-1}$) of the contributing source (we
have intentionally omitted the spectral dependency in $\sigma$ to
avoid cluttering).  We note that the cross section is closedly related
to the often-used (mass) absorption coefficient $\kappa_\lambda$
(units of cm$^{2}$ g$^{-1}$, also known as opacity) by the relation
$\sigma_\lambda n = \kappa_\lambda \rho$, where $\rho$ is the species
mass density. The following sections describe the most relevant
cross-section sources in planetary atmospheres.

\subsubsection{Line-by-line Cross Section}
\label{sec:lbl_opacity}

Molecular cross sections dominate the radiative-transport properties
in the atmospheres of a wide variety of substellar and low-temperature
stellar objects, since their radiation peak at infrared
wavelengths \citep[see, e.g.,][]{TennysonEtal2016jmsExomol}.
Atmospheric species interact with light by scattering, absorbing, and
emitting photons, which change the molecules' rotational, vibrational,
and electronic quantum states.  Non-ionizing absorption and emission
of photons occur at discrete wavelengths determined by the energy
difference between transition states (line transitions), where each
species has a unique set of energy transitions.  Furthermore, Doppler,
collisional, and natural broadening shape the lines into a Voigt
profile \citep[see, e.g.,][]{GoodyYung1989bookAtmosphericRadiation},
causing the extinction-coefficient spectrum of a molecule to vary
strongly with temperature, pressure, and wavelength.

The integrated intensity of a line transition (in cm$\sp{-1}$) can be
expressed as
\begin{equation}
\label{eq:intensity}
\small
S\sb{j} = n\sb{i}
          \frac{\pi e\sp{2}}{m\sb{e}c\sp{2}} \frac{(gf)\sb{j}}{Z\sb{i}(T)}
          \exp\left(-\frac{h c E\sb{\rm low}\sp{j}}{\kb T}\right)
            \left\{1-\exp\left(-\frac{hc\nu\sb{j}}{\kb T}\right)\right\},
\end{equation}
where $gf\sb{j}$, $\nu\sb{j}$, and $E\sp{j}\sb{\rm low}$ are the
weighted oscillator strength, wavenumber, and lower-state energy level
of the line transition $j$, respectively; $Z\sb{i}$ and $n\sb{i}$ are
the partition function and number density of the isotope $i$,
respectively; $e$ and $m\sb{e}$ are the electron's charge and mass,
respectively; $c$ is the speed of light, $h$ is Planck's constant; and
$\kb$ is the Boltzmann's constant.

\begin{table}
\begin{minipage}{\linewidth}
\centering
\caption{Line-by-line Cross-section Databases}
\label{table:lbl_opacity}
\begin{tabular*}{\linewidth} {@{\extracolsep{\fill}} lll}
\hline
Database             & Format & References \\
\hline
HITEMP (\water, {\carbdiox}) & HITRAN & \citet{RothmanEtal2010jqsrtHITEMP} \\
Li (CO)              & HITRAN & \citet{LiEtal2015apjsCOlineList} \\
Hargreaves (\methane) & HITRAN & \citet{HargreavesEtal2020apjsHitempCH4} \\
POKAZATEL (\water)   & ExoMol & \citet{PolyanskyEtal2018mnrasPOKAZATELexomolH2O} \\
YT10to10 (\methane)  & ExoMol & \citet{YurchenkoTennyson2014mnrasExomolCH4} \\
BYTe (\ammonia) & ExoMol & \citet{YurchenkoEtal2011mnrasNH3opacities}, \\
                &        & \citet{Yurchenko2015jqsrtBYTe15exomolNH3} \\
VOMYT (VO)   & ExoMol  & \citet{McKemmishEtal2016mnrasVOMYTexomolVO} \\
TOTO (TiO)   & ExoMol  & \citet{McKemmishEtal2019mnrasTOTOexomolTiO} \\
Harris (HCN) & ExoMol  & \citet{HarrisEtal2006mnrasHCNlineList, HarrisEtal2008mnrasExomolHCN} \\
Partridge (\water)   & Kurucz    & \citet{PartridgeSchwenke1997jcpH2O} \\
Schwenke (TiO)       & Kurucz    & \citet{Schwenke1998fadiTiO} \\
Plez (VO)            & Plez      & \citet{Plez1998aaTiOLineList} \\
\hline
\end{tabular*}
\end{minipage}
\end{table}

To date, the ExoMol {\citep[e.g.,][]{TennysonEtal2016jmsExomol,
TennysonYurchenko2018atomsExomol}} and
HITRAN \citep[e.g.,][]{RothmanEtal2010jqsrtHITEMP,
GordonEtal2017jqsrtHITRAN2016} groups have led the efforts to generate
the molecular cross-section data required to model substellar objects
(i.e., up to a few thousands of Kelvin degrees).  {\pyratbay} is
compatible with the format from both of these groups, as well as older
formats (Table \ref{table:lbl_opacity}).  To extract the cross-section
data from these line-by-line databases, {\pyratbay} reformats the
online available files into a transition line information (TLI) file,
containing the $gf\sb{j}$, $\nu\sb{j}$, and $E\sp{j}\sb{\rm low}$ data
for each transition, a tabulated $Z\sb{i}$ as a function of
temperature, along with other physical properties of the molecules
(isotopic ratios, masses).

Since the most complex molecules can have up to billions of line
transitions \citep[e.g.,][]{YurchenkoTennyson2014mnrasExomolCH4}, the
memory and computational demand can render a radiative-transfer
calculation unfeasible.  To prevent this issue, {\pyratbay} is
compatible with the open-source {\repack}
package\footnote{\href{https://github.com/pcubillos/repack}
{https://github.com/pcubillos/repack}} \citep{Cubillos2017apjRepack}.
{\repack} identifies and retains the strong line transitions that
dominate the absorption spectrum in a given temperature range,
compressing the total cross-section contribution from the remaining
weaker lines into a continuum cross-section spectrum.  The output list
of strong lines preserves {\sim}99\% of the original cross-section
information, but reduces the number of considered transitions by a
factor of {\sim}\ttt{2}.

The total line-by-line extinction coefficient is then computed as:
\begin{equation}
\label{eq:e_lbl}
e(\nu) = \sum_j S_j V_j(\nu),
\end{equation}
where $V_j(\nu)$ is the Voigt profile of the line $j$. Note that the
index $j$ runs over all lines, and thus, there is an implicit sum over
all contributing species $i$.  To speed up the calculation of
Eq.~(\ref{eq:e_lbl}), {\pyratbay} pre-calculates a grid of Voigt
profiles as a function of the Lorentz and Doppler widths.  Then, for
any given line transition, the code selects the appropriate profile
according to the Dopper and Lorentz width of the line (which depend on
the wavelength, atmospheric temperature and pressure, and molecular
properties).

Furthermore, during the radiative-transfer calculation
(Sec.~\ref{sec:radtransfer}) {\pyratbay} can compute the molecular
cross sections in two ways: either by directly evaluating the
individual line-by-line contribution from the TLI files, or by
interpolating from a pre-computed cross-section grid.  For any given
species, this grid contains the molecular cross sections (in cm$^{2}$
molecule$^{-1}$ units) sampled over a temperature, pressure, and
wavenumber grid.  Interpolating from this grid significantly speeds a
radiative-transfer calculation compared to a direct line-by-line
calculation.

\subsubsection{Collision-induced Absorption}
\label{sec:cia_opacity}

Collision-induced absorption (CIA) arises from transient electric
dipole moments generated from collisions between atmospheric
species \citep{AbelEtal2011jpcaH2H2CIA}.  Even spectroscopically
inactive species show significant CIA, which scales proportionally to
the square of the density (i.e., proportional to the number density of
the colliding species).  Thus, for {\molhyd}-dominated atmospheres,
{\molhyd}--{\molhyd} and {\molhyd}--He are the main source of CIA.
CIA spectra varies quasi-continuously with wavelength, and thus, can
be efficiently tabulated as a function of temperature and wavelength.

Table \ref{table:cia_opacity} lists the CIA sources available to be
used in {\pyratbay}.  For the Borysow data, the code provides the
aggregated CIA cross section for {\molhyd}--{\molhyd} (covering the
$60-7000$~K and $0.6-500$~{\micron} ranges) and for {\molhyd}--He
($50-3500$~K and $0.3-100$~{\micron}).  For the HITRAN data, which
include CIA cross sections for several species pairs, the code
provides routines to format the online HITRAN files into the
{\pyratbay} format.  The CIA files contain the cross-section $\sigma$
in cm$^{-1}$ amagat$^{-2}$ units, tabulated as a function of
temperature and wavenumber.

The total CIA extinction coefficient is computed as:
\begin{equation}
e(\nu) = \sum_i n_{i1}\,n_{i2}\,\sigma_{i},
\end{equation}
where the index $i$ represents a species pair, with $n_{i1}$ and
$n_{i2}$ the number densities of the interacting species.

\begin{table}
\begin{minipage}{\linewidth}
\centering
\caption{Collision Induced Absorption Databases}
\label{table:cia_opacity}
\begin{tabular*}{\linewidth} {@{\extracolsep{\fill}} lll}
\hline
Database  & References \\
\hline
HITRAN                         & \citet{RichardEtal2012jqsrtCIA},
                                 \citet{KarmanEtal2019icarHITRANupdateCIA} \\
Borysow ({\molhyd}--{\molhyd}) & \citet{BorysowEtal2001jqsrtH2H2highT},
                                 \citet{Borysow2002jqsrtH2H2lowT} \\
Borysow ({\molhyd}--He)        & \citet{BorysowEtal1988apjH2HeRT},
                                 \citet{BorysowEtal1989apjH2HeRVRT}, \\
                               & \citet{BorysowFrommhold1989apjH2HeOvertones} \\
\hline
\end{tabular*}
\end{minipage}
\end{table}

\subsubsection{Rayleigh-scattering Haze Cross Section}
\label{sec:rayleigh}

Rayleigh scattering occurs when particles much smaller than the
wavelength of radiation interact with photons.  Rayleigh cross section
scales as $\lambda\sp{-4}$, and thus, is relevant at short
wavelengths.  For gas-giant exoplanets, Rayleigh scattering typically
becomes one of the dominant sources of opacity at wavelengths shorter
than {\sim}1 {\micron}.

{\pyratbay} provides non-parametric Rayleigh models for the most
abundant particles in gas-giant atmospheres, H, {\molhyd}, and
He \citep[][]{DalgarnoWilliams1962apjRayleighH2,
Kurucz1970saorsAtlas}, which cross section ($\sigma$, in
cm$^2$~molecule$^{-1}$ units) are given as:
\begin{equation}
\sigma_{\rm H}(\lambda) =
    \frac{5.799\tttt{-45}}{(\lambda/{\rm cm})^4}
    + \frac{1.422\tttt{-54}}{(\lambda/{\rm cm})^6}
    + \frac{2.784\tttt{-64}}{(\lambda/{\rm cm})^8},
\end{equation}
\begin{eqnarray}
\nonumber
\sigma_{\rm He}(\lambda) =
    \frac{5.484\tttt{-46}}{(\lambda/{\rm cm})^4}
    \bigg(1 + \frac{2.44\tttt{-11}}{(\lambda/{\rm cm})^2}
    \phantom{aaaaaahh} \\
    +\ \frac{5.94\tttt{-10}}{(\lambda/{\rm cm})^2[(\lambda/{\rm cm})^2-2.9\tttt{-21}]}\bigg)^2,
\end{eqnarray}
\begin{equation}
\sigma_{\rm H2}(\lambda) =
    \frac{8.14\tttt{-45}}{(\lambda/{\rm cm})^4}
    + \frac{1.28\tttt{-54}}{(\lambda/{\rm cm})^6}
    + \frac{1.61\tttt{-64}}{(\lambda/{\rm cm})^8}.
\end{equation}

{\pyratbay} also implements the parametric model of
\citet{LecavelierEtal2008aaRayleighHD189733b}, which allows the
user to adjust the strength ($f_{\rm ray}$) and power-law index
($\alpha_{\rm ray}$) of the scattering cross section as:
\begin{equation}
\sigma(\lambda) = f_{\rm ray} \sigma_0
    \left(\frac{\lambda}{\lambda_0}\right)^{\alpha},
\end{equation}
where $\sigma_0=5.31\tttt{-27}$~cm$^2$~molecule$^{-1}$ is the Rayleigh
cross section of {\molhyd} at $\lambda_0=0.35$~{\micron}.
The total Rayleigh extinction coefficient is then computed as:
\begin{equation}
e(\nu) = \sum_i n_{i} \sigma_{i}.
\end{equation}

\subsubsection{Alkali Resonance Lines Cross Section}
\label{sec:alkali}

{\pyratbay} follows the formalism
of \citet{BurrowsEtal2000apjBDspectra} to account for the sodium and
potassium resonant-line cross sections.  This procedure calculates a
line profile considering two distinct regimes at the core and wings of
the profile.  For the core of the line, the model adopts the classical
Voigt line-broadening profile, valid within a range ($\Delta\sigma$)
from the line center, $\nu\sb{0}$, up to a detuning frequency
determined by the atomic--molecular interaction potentials.  Assuming
van der Waals potentials, \citet{BurrowsEtal2000apjBDspectra} found
$\Delta\sigma = 30 (T/500)\sp{0.6}$ cm$\sp{-1}$ for Na and
$\Delta\sigma = 20 (T/500)\sp{0.6}$ cm$\sp{-1}$ for K.  Beyond
$\Delta\sigma$, the model adopts a decreasing power law $|\nu
- \nu\sb{0}|\sp{-3/2}$ for the line shape, according to statistical
theory, with an exponential cutoff far from the line center.

Our implementation uses the resonance D-doublet line parameters
($\nu\sb{0}$, $E\sb{\rm low}$, and $gf$) from the Vienna Atomic Line
Data Base (VALD) \citep{PiskunovEtal1995aapsVALDdatabase}.  The
collisional-broadening halfwidth is calculated from impact
theory \citep{IroEtal2005aaHD209458bRadiativeModel} as: $\gamma=0.071
(T/2000)\sp{-0.7}$ cm$\sp{-1}$ atm$\sp{-1}$ for Na and $\gamma=0.14
(T/2000)\sp{-0.7}$ cm$\sp{-1}$ atm$\sp{-1}$ for K.  We adopt the
cutoff prescription from \citet{BurrowsEtal2000apjBDspectra},
$\exp(-q|\nu - \nu\sb{0}|hc/\kb T)$, where $q$ is an unknown parameter
expected to be on the order of unity for Na.  Our implementation
adopts $q=1$ for both Na and K.

The alkali extinction coefficient for a line-transition $j$ is then
computed as:
\begin{equation}
e_j = \left\{
\begin{array}{ll}
S_j V_j(\nu) & \text{if } |\nu - \nu_{0}| < \Delta\sigma_{0} \\
S'_j |\nu-\nu_{0}|^{-3/2} \exp(-\frac{hc|\nu-\nu_{0}|}{\kb T})
                  & \text{if } |\nu - \nu_{0}| \ge \Delta\sigma_{0}
\end{array} \right.
\end{equation}
where $S_j$ and $V_j$ are the line strength and Voigt profile as in
Eq.~(\ref{eq:e_lbl}), and $S'_j$ is a constant to ensure continuity at
$\nu-\nu_{0} = \Delta\sigma_{0}$.

\subsubsection{Clouds Cross Section}
\label{sec:cloud}

Currently, {\pyratbay} implements a simple gray cloud deck model,
where the atmosphere becomes instantly opaque at all wavelengths at a
given pressure level {\pcloud}.  The extinction coefficient of the
cloud model is computed as:
\begin{equation}
e(\nu) = \left\{
\begin{array}{ll}
0      & \text{if } p <   {\pcloud} \\
\infty & \text{if } p \ge {\pcloud}
\end{array} \right.
\end{equation}
Future works will present more complex cloud schemes using
Mie-scattering theory \citep[see,][]{KilpatrickEtal2018apjWASP63bWFC3,
VenotEtal2020apjWASP43bWebbPredictions}.

\subsection{The Spectrum}
\label{sec:spectrum}

\subsubsection{Spectral Sampling}
\label{sec:sampling}

{\pyratbay} works in wavenumber space.  Internally, the code always
evaluates the molecular Voigt profiles (Section \ref{sec:lbl_opacity})
over a constant sampling-rate grid.  This allows the code to use a
pre-computed grid of Voigt profiles at a sampling rate high enough to
resolve all of the line transitions in the atmosphere
($\Delta \nu \lesssim 10^{-3}$ {\kayser}).  These cross sections are
then either resampled to an output spectrum at a constant
sampling-rate $\Delta \nu$ (generally, coarser than the internal
sampling rate), or interpolated to an output spectrum at a constant
resolving power ${\rm R} = \nu/\Delta \nu$.  For the study of
low-resolution observations, the typical resolving powers used are
${\rm R} \approx 10\,000$--$15\,000$.

\subsubsection{Radiative Transfer}
\label{sec:radtransfer}

For transmission spectroscopy, {\pyratbay} solves the
radiative-transfer equation for the stellar irradiation traveling
across the planetary atmosphere, considering only the rays parallel to
the star--observer line of sight.  This approximation assumes
spherical symmetry for the planetary atmosphere, and neglects emission
and scattering by the planetary atmosphere into the ray.  The observed
intensity is then given by $I(\nu)=I_0(\nu)e^{-\tau_\nu}$, where
$I_0(\nu)$ is the stellar intensity.  Taking advantage of the
spherical symmetry, the fraction of flux absorbed by the planet can be
computed by integrating over the planetary impact parameter
$b$ \citep[for a more detailed derivation see,
e.g.,][]{Brown2001apjTransmissionSpectra}:
\begin{equation}
\Delta F(\nu) = 2\pi I_0/d^2
    \int_0^{R_{\rm top}} \left(1-e^{-\tau_\nu(b)}\right) b{\rm d}b,
\end{equation}
where $R_{\rm top}$ is the radius at the top of the atmospheric model
and $d$ is the distance to the observed system.  Then, the observed
transit depth is given as:
\begin{equation}
\label{eq:transit_depth}
\frac{\Delta F(\nu)}{F(\nu)} \biggr\rvert_{\rm transit}
\equiv \frac{R_{\rm p}^2}{R_{\rm s}^2}
= \left(R_{\rm top}^2 - 2\int_0^{R_{\rm top}} e^{-\tau_\nu(b)} b{\rm d}b\right)
/ R_{\rm s}^2,
\end{equation}
where $R_{\rm s}$ is the stellar radius, and $R_{\rm p}$ is the
effective radius of the planet in transmission.

For emission spectroscopy, {\pyratbay} solves the radiative-transfer
equation for the emergent intensity under the plane-parallel
approximation, assuming local thermodynamic equilibrium, and
neglecting scattering.  An integration over the observed planetary
hemisphere gives the emergent planetary flux \citep[for a more
detailed derivation see, e.g., Section 2
of][]{SeagerDeming2010araaExoplanetAtmospheres}:
\begin{equation}
F_{\rm p}(\nu) = 2\pi \int_0^1 \int_0^\infty B(\tau_\nu, \nu)
    e^{-\tau_\nu/\mu} {\rm d}\tau_\nu {\rm d}\mu,
\end{equation}
where $B(\tau_\nu, \nu)$ is the blackbody function, $\mu$ is the
cosine of the angle relative to the normal vector.  The integral over
optical depth is done via a trapezoidal rule (we found no significant
variation compared to more accurate integration algorithms such as the
Simpson rule).  The integral over the observed hemisphere is done by
summing the integrand evaluated at a discrete set of $\mu$ values
(either a user-defined set or a Gaussian quadrature).  Then, the
observed eclipse depth is given as:
\begin{equation}
\label{eq:eclipse_depth}
\frac{\Delta F(\nu)}{F(\nu)} \biggr\rvert_{\rm eclipse} =
\frac{F_{\rm p}(\nu)}{F_{\rm s}(\nu)} \left(\frac{R_{\rm p}}{R_{\rm s}}\right)^2.
\end{equation}

\subsubsection{Transmission Cloud Fraction}
\label{sec:patchy}

To account for a non-uniform cloud coverage during transit, the code
implements the patchy cloud-fraction parameterization
of \citet{LineParmentier2016apjPartialClouds}.  This model computes
the transmission spectrum as the linear combination of a cloud-free
and a cloudy spectrum, weighted by a patchy cloud-fraction parameter
$f_{\rm patchy} \in [0,1]$:
\begin{equation}
\frac{\Delta F(\nu)}{F(\nu)}\biggr\rvert_{\rm transit} =
    (1-f_{\rm patchy}) \frac{\Delta F}{F}\biggr\rvert_{\rm clear}
    + f_{\rm patchy} \frac{\Delta F}{F}\biggr\rvert_{\rm cloudy}
\end{equation}

Following \citet{MacDonaldMadhusudhan2017mnrasRetrievalHD209b}, the
clear component neglects the absorption cross section from the cloud
model (Sec.\ \ref{sec:cloud}) and the Lecavelier Rayleigh haze model
(Sec.\ \ref{sec:rayleigh}), if any of these are included in the
atmospheric model.

\subsubsection{Stellar Spectrum}
\label{sec:star}

A stellar spectrum ($F_{\rm s}$) is required to compute planet-to-star
flux ratios for secondary-eclipse observations or other astrophysical
calculations, such as signal-to-noise ratio estimations.  {\pyratbay}
allows the user to define a stellar flux spectrum either from a Kurucz
stellar model\footnote{\href{http://kurucz.harvard.edu/grids.html}
{http://kurucz.harvard.edu/grids.html}} \citep{CastelliKurucz2003aiausATLAS9},
a blackbody, or a custom stellar model provided by the user.

\subsection{The Retrieval}
\label{sec:mcmc}

\subsubsection{Bayesian Sampler}

To sample posterior distributions, {\pyratbay} uses the open-source
{\MCC} Bayesian statistics
package\footnote{\href{https://mc3.readthedocs.io/}
{https://mc3.readthedocs.io/}} \citep{CubillosEtal2017apjRednoise}.
{\MCC} is a multi-core code (using the \textsc{multiprocessing}
Standard Library package) that implements the Differential-evolution
Markov-chain Monte Carlo posterior
sampler \citep[DEMC,][]{terBraak2006DifferentialEvolution,
terBraak2008SnookerDEMC}.  DEMC algorithms automatically adjust the
MCMC proposal trials to optimize the posterior sampling.  The {\MCC}
package enables a simple framework to fix or set free specific model
parameters, set uniform or Gaussian parameter priors, monitor for
the \citet{GelmanRubin1992stascGRstatistics} convergence criteria, and
generate marginal and pairwise posterior distribution plots.

\subsubsection{Atmospheric retrieval}

The {\pyratbay} atmospheric retrieval explores the posterior
distribution of the temperature, altitude, cloud coverage, and
chemical composition models, constrained by given transit- or
eclipse-depth observations (Eqs.~(\ref{eq:transit_depth}) and
(\ref{eq:eclipse_depth}), respectively).  During a retrieval run, the
user can choose to retrieve or keep fixed any of the parameters of
these models (Secs.\ \ref{sec:atmosphere} and \ref{sec:opacities}).
Both the planetary radius (at the reference pressure $p_0$) and
planetary mass can be set as free parameters.  The radius directly
affects the altitude profile and the eclipse depth, whereas the mass
affects the altitude-profile model.

{\pyratbay} considers three states for each of the species in the
atmospheric model.  First, species can be left as `fixed', in which
case their abundances remain fixed at their initial input values.
Second, species can be set as `free', in which case the code will
retrieve the log of their volume mixing ratios, which are then assumed
to be constant-with-altitude.  Third, species can be set as `bulk', in
which their abundances are automatically adjusted to maintain the sum
of all volume mixing ratios equal to one at each atmospheric layer:
\begin{equation}
\label{eq:bulk}
\sum^{\rm bulk} X_i = 1 - (\sum^{\rm free} X_i + \sum^{\rm fixed} X_i).
\end{equation}

The relative abundance ratios between the bulk species are kept fixed
at their initial values.  Therefore, for a primary-atmosphere planet
for example, the user should flag {\molhyd} and He as bulk species.

\section{Benchmark}
\label{sec:benchmark}

To validate the {\pyratbay} framework, we compared its outputs to that
of other open-source packages.  Since an atmospheric characterization
study involves modeling multiple physical processes, differences in
the outputs from each step propagate throughout the calculation.
Thus, it is important to validate each individual component.  In the
following sections we progressively benchmark the main components of
the {\pyratbay} framework: the absorption cross-section sampling, the
radiative-transfer spectrum calculation, and the atmospheric-retrieval
scheme.  As a final benchmark test, we analyzed the transmission
spectrum of {\twoohnineb}, one of the most thoroughly observed
exoplanets from space, hence providing a good-quality target for
comparison.  In each scenario, we adopt inputs and assumptions as
similar as possible between the comparing codes and analyses.

\begin{figure*}
\centering
\includegraphics[width=\linewidth, clip, trim=0 9 0 0]
{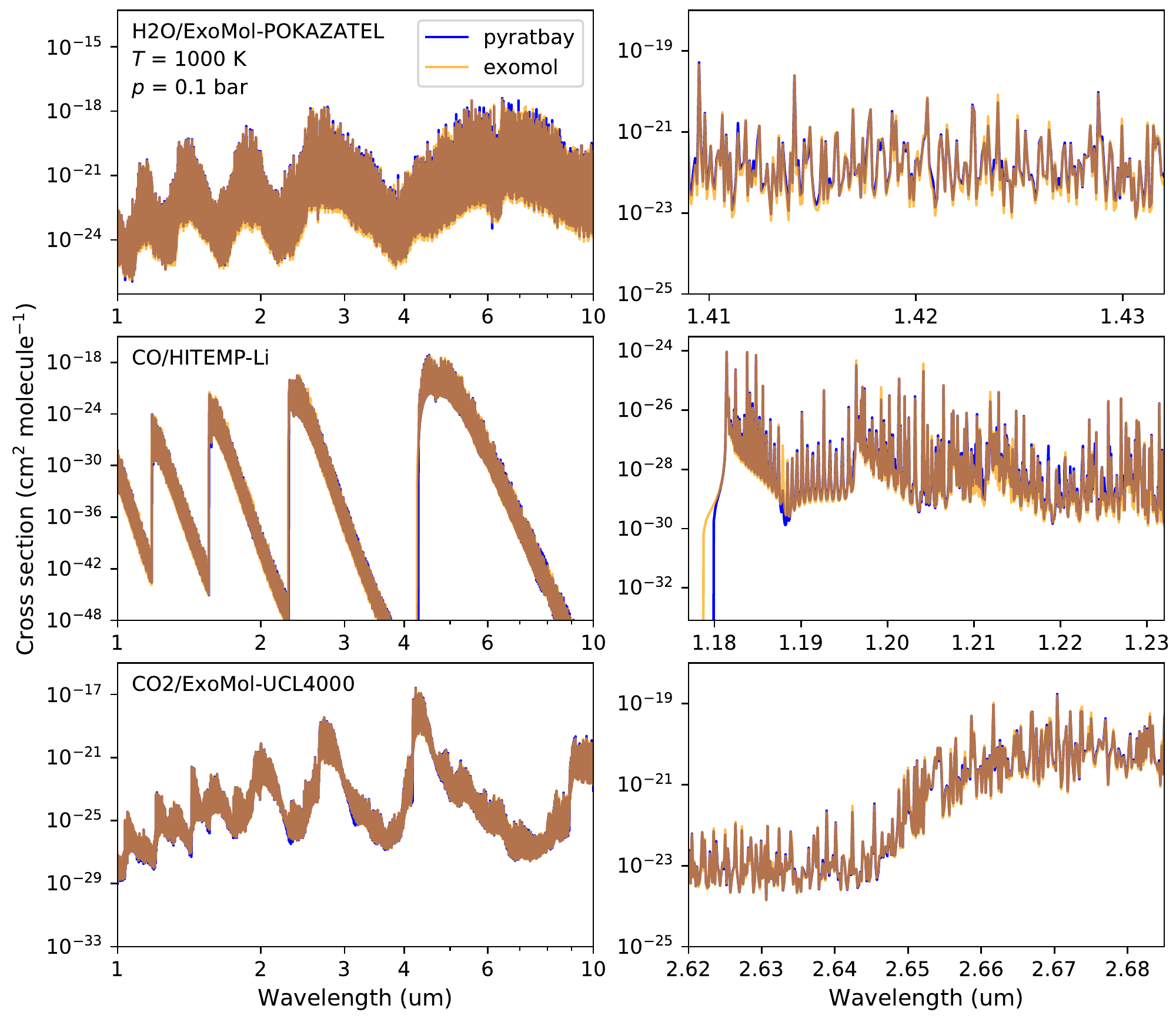}
\caption{
    Line sampling comparison between {\pyratbay} (blue) and
    ExoMol \citep[semi-transparent
    orange,][]{ChubbEtal2021aaExoMolOP}.  The top, middle, and bottom
    panels show {\water}, CO, and {\carbdiox} line-by-line cross
    sections (respectively), sampled at a resolving power of ${\rm R}
    = 10\,000$, evaluated at a temperature of $T=1000$~K and pressure
    of $p=0.1$~bar.  The left panels show the cross sections over the
    1--10~{\micron} region, whereas the right panels show a selected
    narrow region for each molecule.}
\label{fig:validation_opacity}
\end{figure*}

\subsection{Cross Section Spectrum Benchmark}
\label{sec:val_opacity}

We benchmark the line-by-line cross-section sampling of {\pyratbay} by
comparing its cross sections to the ExoMolOP
database\footnote{\href{http://www.exomol.com/data/data-types/opacity/}
{http://www.exomol.com/data/data-types/opacity/}}
of \citet[][]{ChubbEtal2021aaExoMolOP}.  We selected three
spectroscopically relevant species: the {\water} POKAZATEL line list
of \citet{PolyanskyEtal2018mnrasPOKAZATELexomolH2O}, provided in
ExoMol format; the CO line list of \citet{LiEtal2015apjsCOlineList},
provided in HITRAN format; and the {\carbdiox} UCL4000 line list
of \citet{YurchenkoEtal2020mnrasCO2ucl4000}, provided in ExoMol
format.  The only difference in the input line lists is that we
pre-processed the {\water} database with the {\repack} package,
keeping only the strongest $\sim$9 million line transitions in the
tested range (1--10 {\microns}).  Cross sections sampled from the
repacked line lists show relative differences smaller than 1\% by dex
from the original line
lists \citep[submitted][]{Cubillos2021apjRepackExomol}).  We followed
the methodology of \citet{ChubbEtal2021aaExoMolOP} as closely as
possible, considering the same isotopes for each of these molecules,
sampling the spectrum at the same resolving power ${\rm R}=15\,000$
and wavelength locations, and assuming a solar atmospheric composition
consisting primarily of $X_{\rm H2} = 0.86$ and $X_{\rm He} =
0.14$ \citep{AsplundEtal2009araSolarComposition}.
Figure \ref{fig:validation_opacity} shows an example of the resulting
cross sections evaluated at 1000~K and 0.1~bar.  In general, our
results match well those of \citet{ChubbEtal2021aaExoMolOP} at all
temperatures and pressures.  However, it is worth mentioning that the
outputs can vary significantly depending on how the line profiles are
modeled (e.g., line-wings cutoff), for which there is no clear
prescription.  We briefly discuss below how we computed the line
profiles.

\begin{figure*}
\centering
\includegraphics[width=\linewidth, clip]
  {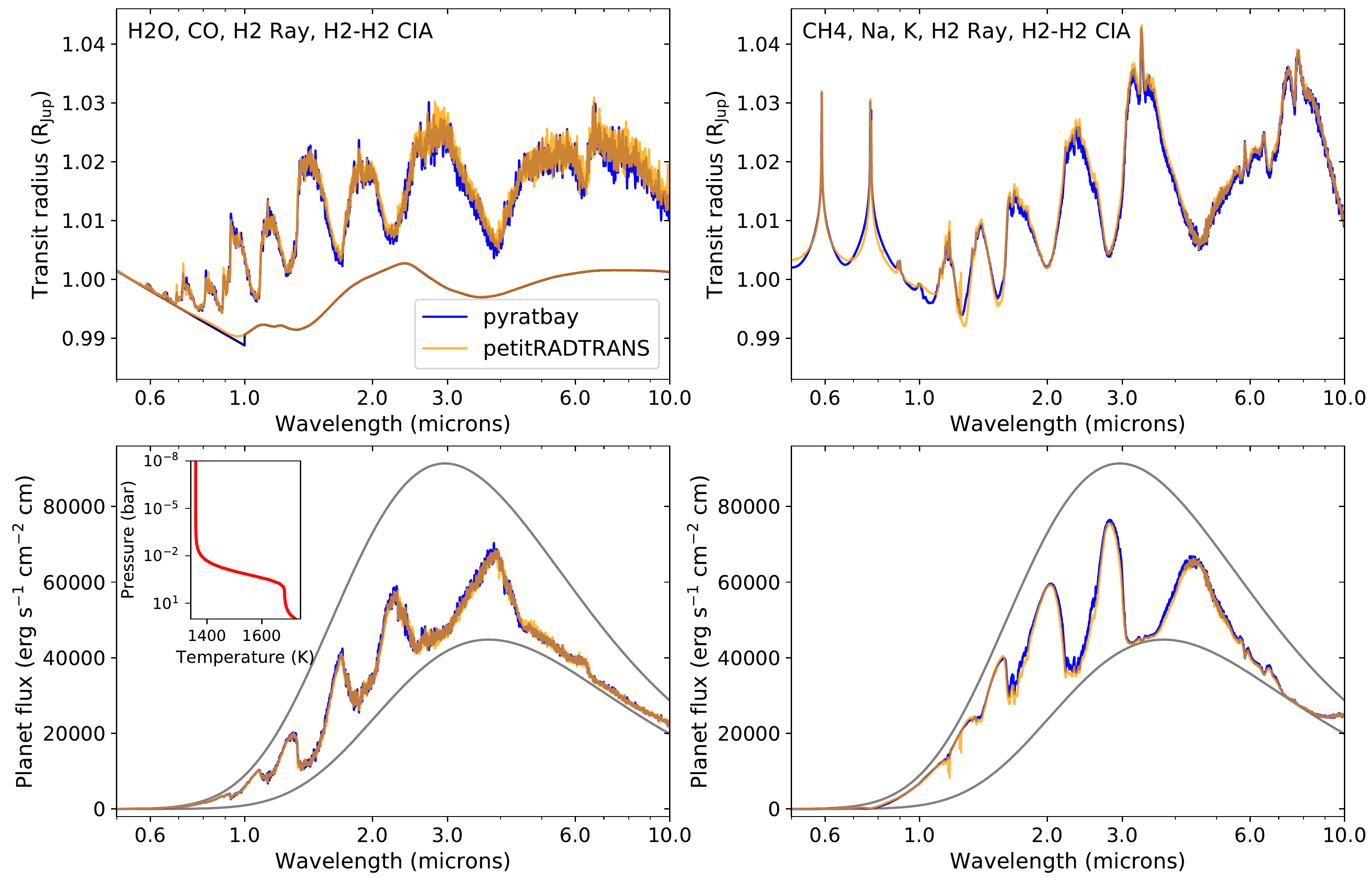}
\caption{Transmission and emission spectra comparison between
    {\pyratbay} and {\petit}. The solid blue and semi-transparent
    orange curves show the transmission (top panels) and emission
    spectra (bottom panels) computed by {\pyratbay} and {\petit},
    respectively.  The models in the left panels consider {\water} and
    CO cross sections, whereas the models in the right panels consider
    {\methane}, Na, and K cross sections.  All models also consider
    {\molhyd} Rayleigh and {\molhyd}--{\molhyd} CIA cross sections.
    The smooth curves at the bottom of the top-left panel show spectra
    considering only Rayleigh and CIA cross section.  The {\pyratbay}
    models have been smoothed from their original sampling rate (${\rm
    R}=10\,000$) to match that of {\petit} (${\rm R}=1000$).  The
    transmission models adopted an isothermal temperature profile at
    $T=1300$~K.  The emission models adopted the temperature profile
    shown in the inset in the bottom left panel.  The gray curves
    above and below the spectra show Planck emission models at the
    highest temperature (bottom layer) and lowest temperature (top
    layer) of the atmosphere, respectively.  All scenarios show a good
    agreement between {\pyratbay} and {\petit}.}
\label{fig:validation_fm}
\end{figure*}

The unknown extent of the line
wings \citep{LevyEtal1992bookCollisionalLines} has one of the largest
impacts on sampled cross sections \citep[see,
e.g.,][]{GrimmHeng2015apjHELIOSK}.  Varying the line-wings cutoff can
modify the sampled {cross section} by several orders of magnitude.
{\citet{ChubbEtal2021aaExoMolOP}} adopted a line-wing cutoff of 500
half-width at half maximum (HWHM) from the center of each line.  For
our sampling routine we adopted both a HWHM cutoff as well as a
maximum fixed cutoff of 25~cm$^{-1}$.  We found that cutoff values of
300 HWHM (for {\water}), 500 HWHM (CO), and 100 HWHM ({\carbdiox})
were sufficient to replicate the sampling
of \citet{ChubbEtal2021aaExoMolOP}.  Larger HWHM cutoff values did not
change the results significantly, since the line profiles overlap due
to the high density of lines and the fixed 25~cm$^{-1}$ cutoff caps
the effect of much larger HWHM cutoff values.  Note that the values
adopted here are ad-hoc values that are typically used in the
literature, they are not derived from theoretical first principles.
For this reason, we deliberately designed the {\pyratbay} package to
enable the users to compute the cross sections with custom cutoff
values.

One main difference between our methodology
and \citet{ChubbEtal2021aaExoMolOP} is collisional broadening
calculation.  Nonetheless, we do not see any major discrepancy between
our cross sections and those of \citet{ChubbEtal2021aaExoMolOP}
despite these different approaches.  \citet{ChubbEtal2021aaExoMolOP}
uses the HITRAN $\gamma$ and $n$ broadening parameters \citep[Appendix
of][]{RothmanEtal1998jqsrtHITRAN1996} to compute the collisional HWHM,
applied for {\molhyd} and He broadening (as opposed to {\it self} and
{\it air} broadening).  {\pyratbay}, instead, computes the collisional
broadening based on the collision diameter of the atmospheric
species \citep{GoodyYung1989bookAtmosphericRadiation}.

\subsection{Spectrum Forward-modeling Benchmark}
\label{sec:val_forward}

We benchmark the radiative-transfer module of {\pyratbay} by comparing
the output transmission and emission spectra to that of the
open-source atmospheric-modeling framework
{\petit} \citep[][commit \texttt{f43559de}]{MolliereEtal2019aaPetitRADTRANS}.
Both {\pyratbay} and {\petit} adopt similar physical assumptions
(e.g., LTE, plane-parallel atmosphere, hydrostatic equilibrium),
however, the treatment of the cross section differs significantly.  In
particular, {\petit} implements the $k$-distribution method under the
correlated-$k$ approximation \citep{FuLiou1992jatsCorrelatedKdist},
whereas {\pyratbay} uses line sampling.

To ensure that both codes use the same input atmospheric model, we
generated the atmospheric profile with the {\petit} code, which we
then fed into {\pyratbay}. This limits the scope of the comparison to
the radiative-transfer calculations.  The atmosphere has 28 layers,
extending from 100 to $\ttt{-5}$ bar.  The composition is
constant-with-altitude, dominated by {\molhyd} and He, with volume
mixing ratios of 0.85 and 0.149, respectively, and traces of {\water}
(4$\tttt{-4}$), {\methane} ($\ttt{-4}$), CO (3$\tttt{-4}$), Na
(3$\tttt{-6}$), and K ($\ttt{-6}$).  For transmission we used an
isothermal profile at 1300~K. For emission we used
the \citet{Guillot2010aaRadiativeEquilibrium} parameterization to
generate a non-inverted profile where the temperature drops from
$\sim$1700~K at the bottom of the atmosphere to $\sim$1350~K at the
top of the atmosphere.  The radius profile corresponds to a
hydrostatic-equilibrium model of a $0.6 {\mjup}$ planet with a radius
of $1 {\rjup}$ at 0.1 bar.

We used the same cross section databases for both {\pyratbay} and
{\petit}: {\molhyd}--{\molhyd} CIA (HITRAN), {\water} (POKAZATEL),
{\methane} (YT10to10), and CO (HITRAN/Li).  However, to speed up the
calculation we pre-processed the {\water} and {\methane} line-list
data with the {\repack} code to preserve the 9 and 90 million stronger
transitions, respectively, rather than the original 4 and 10 billion
transitions in the 0.3--10 {\micron} range.

Figure \ref{fig:validation_fm} shows the resulting spectra for
transmission (top panels) and emission (bottom panels), considering
cross sections only for {\molhyd} (top left panel), {\water} and CO
(left panels) and for {\methane}, Na, and K (right panels).  We obtain
consistent results between {\pyratbay} and {\petit} in all scenarios.
The difference in the Na and K lines is expected, since the two codes
implement different alkali line-profile models (top right panel).
{\petit} assumed a Voigt profile, whereas {\pyratbay} assumed the
power-law and exponential-decay model
of \citet{BurrowsEtal2000apjBDspectra}.  Other minor differences
between the resulting spectra are well understood.  In particular, at
high pressures the {\methane} cross section of {\pyratbay} is stronger
than that of the {\petit} model, since {\petit} applied an intensity
cutoff for this molecule (top right panel).  At low pressures the
{\pyratbay} cross section is weaker than that of the {\petit} models,
since {\pyratbay} applied a line-wing cutoff proportional to the
lines' HWHM in addition to the fixed line-wing cutoff applied by both
codes.

\subsection{Retrieval Benchmark}
\label{sec:val_retrieval}

We benchmark the atmospheric retrieval framework of {\pyratbay} by
retrieving a set of synthetic spectra generated with the open-source
package {\taurex} \citep[][version
3.0.3-beta]{AlRefaieEtal2019arxivTaurex3}, and comparing the posterior
distributions to the input models.  The benchmark sample consists of
12 planets for transmission and 12 planets for emission, selected from
the {\ariel} target sample
of \citet{EdwardsEtal2019ajARIELtargetList}.  These planets span a
broad range of physical properties in equilibrium temperature
(450--2500~K), radius (1--20~{\rearth}), and mass
(0.85~{\mearth}--9.0~{\mjup}).

For the transmission models we adopted isothermal temperature profile
models near the targets' equilibrium temperatures.  For the emission
models we adopted \citet{Guillot2010aaRadiativeEquilibrium}
temperature profiles, generating both inverted and non-inverted
profiles.  For each target, we used the open-source {\rate}
package \citep{CubillosEtal2019apjRate} to estimate the atmospheric
composition under thermochemical equilibrium.  We sampled a variety of
elemental compositions ranging from 0.5 to 30 times solar, and C/O
ratios ranging form 0.05 to 0.9.  We used these estimates as
guidelines to set the composition of the synthetic models as
constant-with-altitude volume mixing ratios.  For the spectral
calculations we considered CIA and Rayleigh cross sections from
{\molhyd}, and line-sampling cross sections from {\water}, {\methane},
CO, and {\carbdiox}.  For the transmission models we also included an
opaque gray cloud deck model.  For both {\taurex} and {\pyratbay} we
used the cross sections provided by
{\taurex}\footnote{\href{https://taurex3-public.readthedocs.io/en/latest/user/taurex/quickstart.html}
{https://taurex3-public.readthedocs.io/en/latest/user/taurex/quickstart.html}}.
These cross sections are sampled at a constant resolving power of
${\rm R} = 15\,000$, extending from 100 to \ttt{-5} bar.

For each target, we simulated {\ariel}-like
observations \citep{TinettiEtal2018exaARIEL}, by integrating the
synthetic spectra (0.45--8.0 {\microns}) over the instrument
bandpasses released for the {\ariel} retrieval
challenge\footnote{\href{https://ariel-datachallenge.azurewebsites.net/retrieval}
{https://ariel-datachallenge.azurewebsites.net/retrieval}}.  We
adopted a simple two-component noise model added in quadrature, where
the first term is a constant noise floor level, and the second term is
proportional to the square root of the stellar flux collected by each
bandpass.  We modeled the observed stellar fluxes as blackbody spectra
according to the stellar effective temperatures, and scaled them
according to their K-band apparent magnitudes.

The {\pyratbay} retrieval adopts the same model parameterization as
the input {\taurex} spectra.  For transmission, the retrieval
parameters are the isothermal temperature, the planetary radius at 100
bar, the volume mixing ratios of {\water}, {\methane}, CO, and
{\carbdiox}, and the pressure of the gray cloud deck.  We assume
uniform priors for all parameters.  For emission, the retrieval
parameters are the $\kappa'$, $\gamma$, and $T_{\rm irr}$ parameters
of the \citet{Guillot2010aaRadiativeEquilibrium} model (we kept
$T_{\rm int}$ fixed at 100~K, as assumed by the {\taurex} models), the
radius at 100 bar, and the volume mixing ratios of {\water},
{\methane}, CO, and {\carbdiox}.  We assume uniform priors for all
parameters, except for the planetary radius.  In the literature, the
radius is generally assumed fixed for secondary-eclipse retrievals,
even though this is not a known value a priori.  We let this parameter
free, and assume a Normal prior distribution according to the transit
radius and 1$\sigma$ uncertainty of each
planet. Table~\ref{table:retrieval_benchmark} summarizes the retrieval
parameterization for these benchmark runs.

{\renewcommand{\arraystretch}{1.1}
\begin{table}
\centering
\caption{Retrieval Benchmark Parameterization}
\label{table:retrieval_benchmark}
\begin{tabular*}{0.85\linewidth} {@{\extracolsep{\fill}} lll}
\hline
Article      & Transmission & Emission \\
Parameter     & Priors$^*$ & Priors$^*$ \\
\hline
$T_0$ (K)              & $\mathcal U(200, 3000)$ & $\cdots$  \\
$\log_{10}(\kappa')$    & $\cdots$ & $\mathcal U(-8.0, 2.0)$  \\
$\log_{10}(\gamma)$     & $\cdots$ & $\mathcal U(-5, 5)$  \\
$T_{\rm irr}$ (K)        & $\cdots$ & $\mathcal U(0, 5000)$  \\
$R_{\rm planet}$ ($\rjup$) & $\mathcal U(0.5,30.0)$ & $\mathcal N(R_{\rm p},\sigma_R)$ \\
$\log_{10}(X_{\rm H2O})$  & $\mathcal U(-12, -1)$ & $\mathcal U(-12, -1)$ \\
$\log_{10}(X_{\rm CH4})$  & $\mathcal U(-12, -1)$ & $\mathcal U(-12, -1)$ \\
$\log_{10}(X_{\rm CO})$   & $\mathcal U(-12, -1)$ & $\mathcal U(-12, -1)$ \\
$\log_{10}(X_{\rm CO2})$  & $\mathcal U(-12, -1)$ & $\mathcal U(-12, -1)$ \\
$\log_{10}(p_{\rm cloud}/{\rm bar})$ & $\mathcal U(-4, 2)$ & $\cdots$ \\
\hline
\end{tabular*}
\begin{tablenotes}
\item $^*$ $\mathcal U(a, b)$ stands for a uniform distribution between $a$ and $b$.  $\mathcal N(\mu, \sigma)$ stands for a normal distribution with mean $\mu$ and standard deviation $\sigma$.
\end{tablenotes}
\end{table}
}

\begin{figure*}
\centering
\includegraphics[width=\linewidth, clip]
{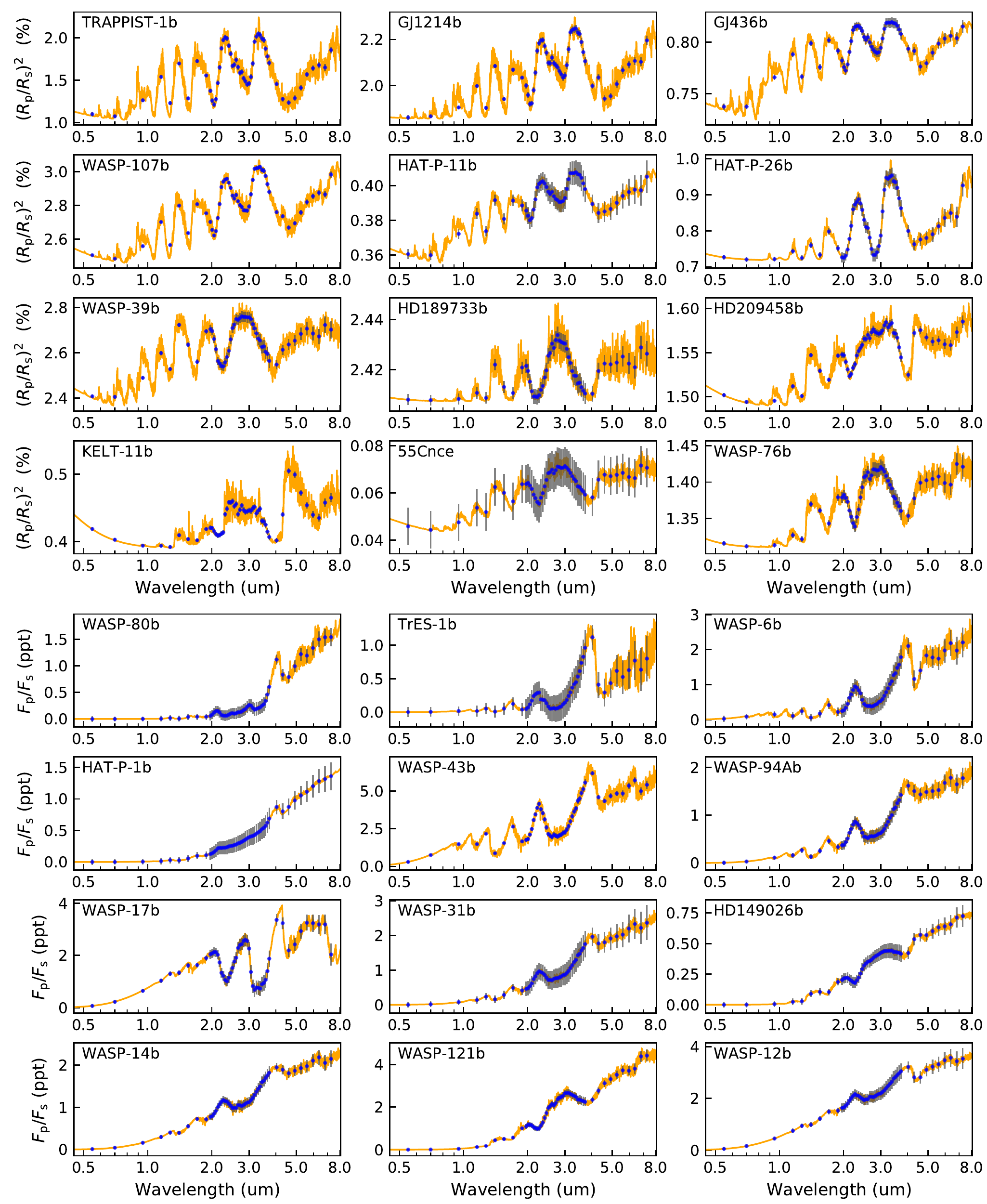}
\caption{
    Synthetic {\ariel} sample of transmission (top panels) and
    emission (bottom panels) spectra.  The blue markers with gray
    error bars show the {\taurex} synthetic models integrated over the
    {\ariel} observing bands and their 1$\sigma$ uncertainties,
    respectively.  The orange curves show the {\pyratbay} retrieved
    best-fitting spectra.}
\label{fig:validation_ret_spectra}
\end{figure*}

\begin{figure*}
\centering
\includegraphics[width=\linewidth, clip]
{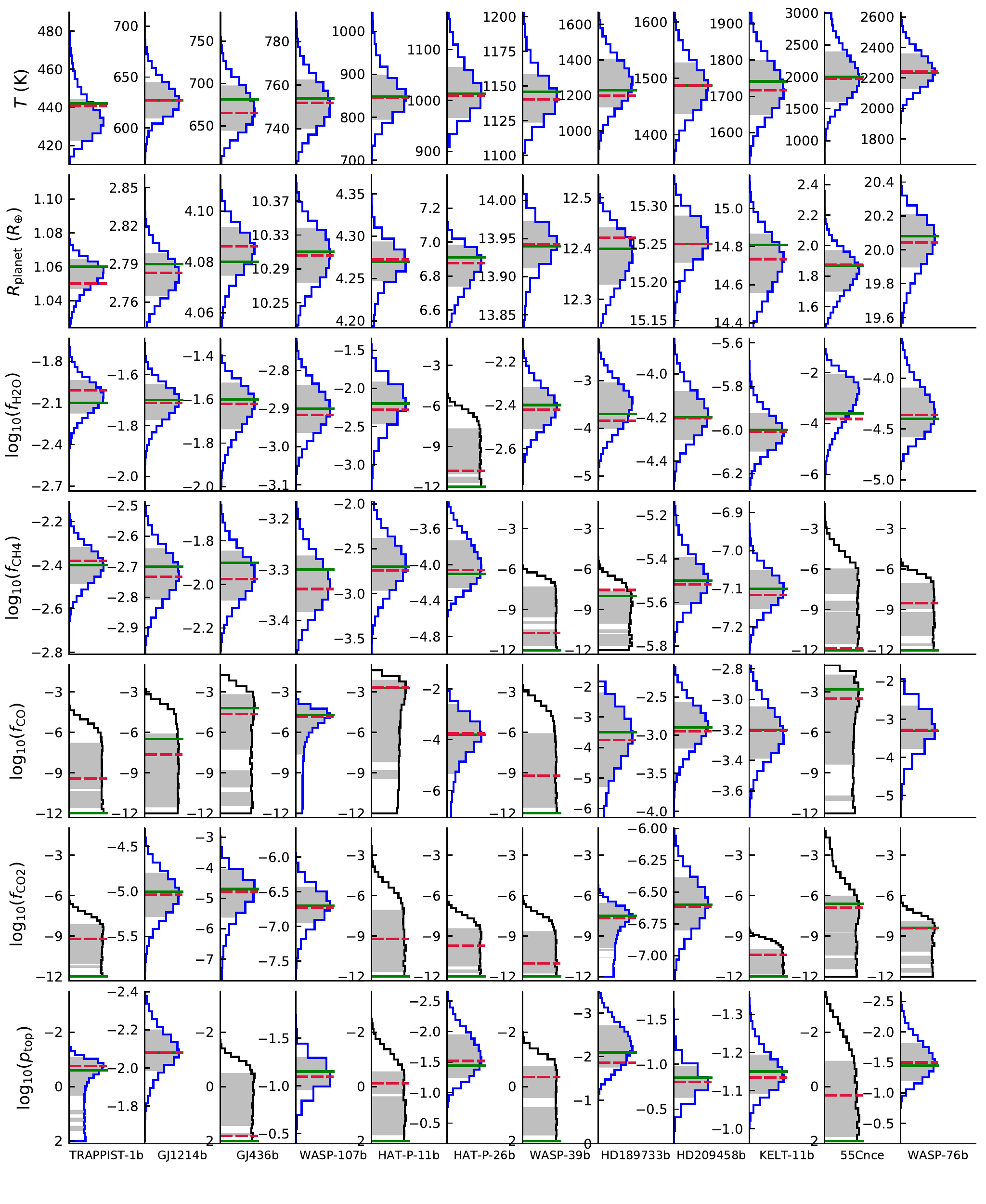}
\caption{
    Retrieved marginal posterior distributions of the
    transmission-spectrum sample.  Each column and row correspond to a
    target and retrieval parameter, respectively (see labels at the
    bottom and left of the figure).  The gray shaded areas denote the
    68\% highest-posterior-density region \citep[HPD, see Appendix
    in][]{CubillosEtal2017apjRednoise} of each parameter (analogous to
    the 1$\sigma$ uncertainty for a Gaussian posterior). The solid
    green and dashed red lines denote the input and retrieved
    best-fitting values, respectively.  The contour of the posterior
    distribution is colored black for the parameters with an upper
    limit constrain.  For all constrained parameters, the input values
    are contained within the 68\% HPD region, whereas for all upper
    limit constrains, the input values are contained within the 99\%
    HPD region.}
\label{fig:validation_ret_transmission}
\end{figure*}

\begin{figure*}
\centering
\includegraphics[width=\linewidth, clip]
{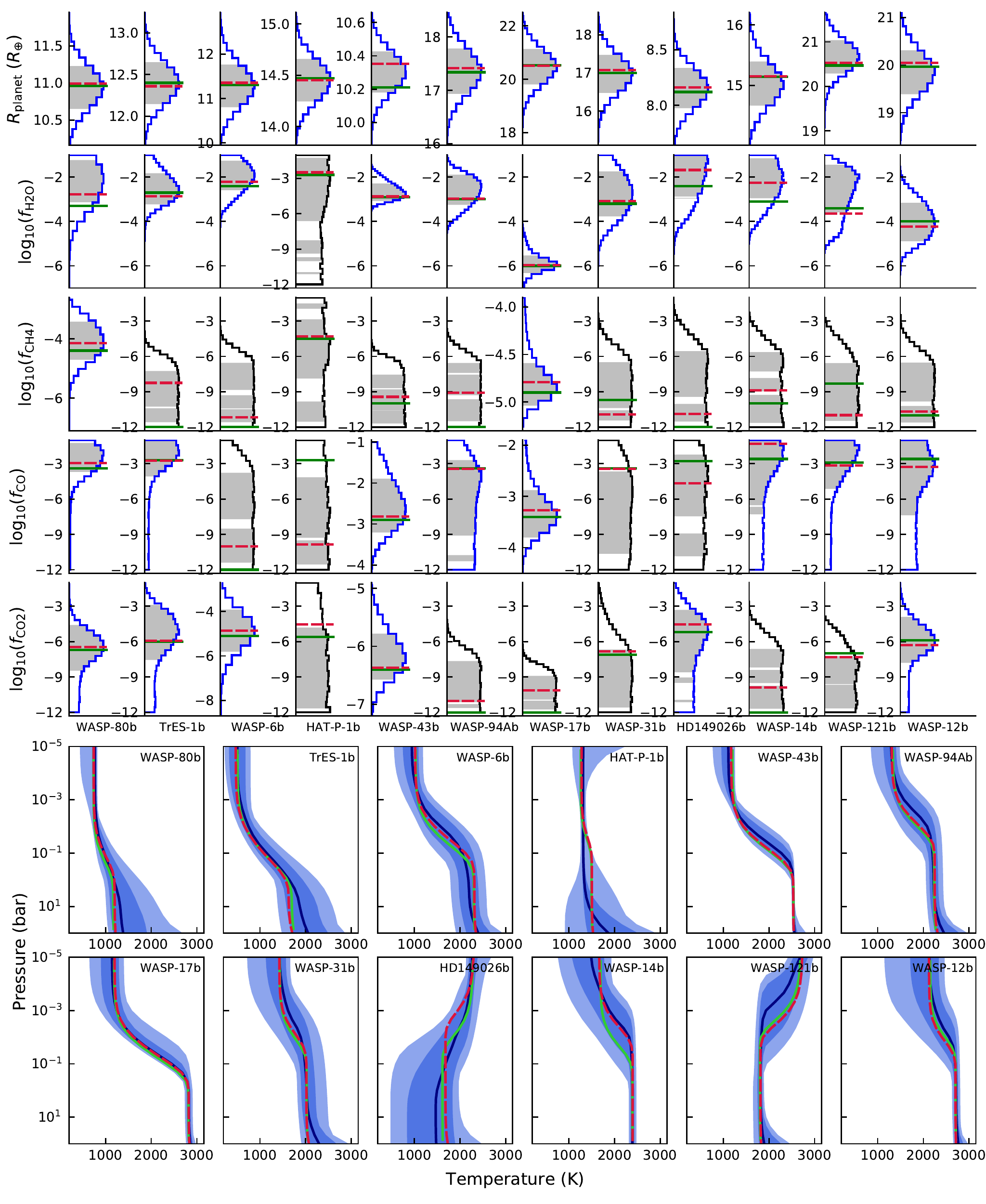}
\caption{
    Retrieved posterior distributions of the emission-spectrum sample.
    {\bf Top:} each column and row correspond to a target and
    retrieval parameter (except for the temperature-profile
    parameters), respectively (see labels at the bottom and left of
    the figure) {\bf Bottom:} Temperature-profile posterior
    distribution.  The dark and light blue shaded area denotes the
    span of the 68th and 95th percentiles of the temperature-profile
    posteriors, respectively. The green, red, and blue curves denote
    the input, best-fit, and median temperature profiles,
    respectively.}
\label{fig:validation_ret_emission}
\end{figure*}

Figures \ref{fig:validation_ret_spectra}, \ref{fig:validation_ret_transmission},
and \ref{fig:validation_ret_emission} show the retrieved spectra,
transmission posterior distributions, and emission posterior
distributions, respectively.  All retrieved best-fitting spectra and
posterior distributions are consistent with the input values from the
{\taurex} models (within 1$\sigma$).  The precision of the posterior
distributions are consistent with expectations based on the
signal-to-noise ratio of the input spectra.  Note that in several
cases it is not possible to constrain all parameters.  Often the most
spectroscopically active and abundant species (e.g., {\water}) cover
the spectral features of other species (e.g., CO or {\carbdiox}).  It
is thus expected to find weakly constrained posteriors or upper limits
for certain molecules.  Overall, the abundance constraints are
stronger for the transit simulations than for the eclipse simulations,
which we can explain in two ways.  First, the transit simulations have
larger signal-to-noise ratios than the eclipse simulations, since they
generally have brighter stars and show stronger spectral modulations
(on the order of percents rather than parts per thousand).  Second,
the eclipse retrievals use a more complex temperature profile than the
transit retrievals.  Since the emission features arise from the
variation of the thermal structure (the location and magnitude of the
thermal gradient), the parameters for the abundances and thermal
structure correlate to generate a degenerate family of solutions.
Figure \ref{fig:validation_ret_emission} shows evidence of this
behavior.  In general, larger temperature gradients lead to stronger
features, which are easier to constrain.  Note, for example, that the
nearly isothermal atmosphere of the HAT-P-1b simulation places the
weakest abundance constraints, whereas the WASP-17b simulation places
the strongest abundance constraints.

\subsection{HD~209458\,b Benchmark}
\label{sec:hd209458b_benckmark}

As a final benchmark test, we applied the {\pyratbay} atmospheric
retrieval to real transmission observations and compare our results to
that of previous studies.  We chose {\twoohnineb}, an inflated hot
Jupiter orbiting one of the brightest and nearest stars known to host
an exoplanet \citep{CharbonneauEtal2000apjHD209458bTransit,
HenryEtal2000apjHD209458bTransit, MazehEtal2000apjlHD209458b}.  The
extraordinarily favorable observing conditions of this planet have
made {\twoohnineb} the poster child for most exoplanet observing
techniques, including atmospheric retrievals.  Here we considered the
space-based transmission spectrum of {\twoohnineb} reported
by \citet{SingEtal2016natHotJupiterTransmission}, obtained from {\HST}
(0.29--1.0 {\micron} with STIS, 1.1--1.7 {\micron} with WFC3) and
{\Spitzer} observations (3.6 and 4.5 {\micron} with IRAC).  We
compared our results with those of the
{\poseidon} \citep{MacDonaldMadhusudhan2017mnrasRetrievalHD209b} and
{\aura} code \citep{PinhasEtal2019mnrasHotJupiterSampleRetrieval},
henceforth
named \citetalias{MacDonaldMadhusudhan2017mnrasRetrievalHD209b}
and \citetalias{PinhasEtal2019mnrasHotJupiterSampleRetrieval},
respectively.  We chose these analyses since we can re-analyze the
data in a similar fashion
to \citetalias{MacDonaldMadhusudhan2017mnrasRetrievalHD209b}
and \citetalias{PinhasEtal2019mnrasHotJupiterSampleRetrieval}, and
therefore, make a more direct comparison of the retrieval results.  We
configured our model setup and parameterization to follow that
of \citetalias{MacDonaldMadhusudhan2017mnrasRetrievalHD209b}
and \citetalias{PinhasEtal2019mnrasHotJupiterSampleRetrieval},
deviating only when we could not replicate their approach (which we
will point out in the text).  Thus, we adopted the same system
parameters \citep{TorresEtal2008apjTransitParameters}, transmission
observations, model parameterization, and priors used
by \citetalias{MacDonaldMadhusudhan2017mnrasRetrievalHD209b}
and \citetalias{PinhasEtal2019mnrasHotJupiterSampleRetrieval}.

Our atmospheric model consisted of 41 pressure layers extending from
$\ttt{-6}$~bar to 100~bar, using the temperature profile
parameterization of \citet{MadhusudhanSeager2009apjRetrieval}, a
primary-atmosphere composition with a solar {\molhyd}/He abundance
ratio, and an altitude profile in hydrostatic equilibrium.  We adopted
constant-with-altitude abundance profiles for all species.

The radiative-transfer model considered line-by-line cross sections
from {\water} \citep{PolyanskyEtal2018mnrasPOKAZATELexomolH2O},
{\methane} \citep{YurchenkoTennyson2014mnrasExomolCH4},
HCN \citep{HarrisEtal2006mnrasHCNlineList,
HarrisEtal2008mnrasExomolHCN},
{\ammonia} \citep{Yurchenko2015jqsrtBYTe15exomolNH3,
ColesEtal2019mnrasNH3coyuteExomol},
CO \citep{LiEtal2015apjsCOlineList}, and
{\carbdiox} \citep{RothmanEtal2010jqsrtHITEMP}.  Prior to the MCMC
runs, we applied the {\repack} algorithm \citep{Cubillos2017apjRepack}
to the ExoMol line lists to extract only the dominant transitions.
Then, we sampled the line-by-line cross sections into $R=15\,000$
cross-section tables evaluated at the pressures of the atmospheric
model and over a temperature array evenly spaced from 100~K to 3000~K
with a step of 100~K.

The model also included collision-induced absorption for
{\molhyd}--{\molhyd} pairs \citep{BorysowEtal2001jqsrtH2H2highT,
Borysow2002jqsrtH2H2lowT} and {\molhyd}--He
pairs \citep{BorysowEtal1988apjH2HeRT, BorysowEtal1989apjH2HeRVRT,
BorysowFrommhold1989apjH2HeOvertones}; Rayleigh-scattering cross
section for {\molhyd} \citep{DalgarnoWilliams1962apjRayleighH2} and
for an unknown haze
particulate \citep{LecavelierEtal2008aaRayleighHD189733b}; Na and K
resonance line cross sections; and a gray cloud deck parameterized by
the cloud top pressure (\pcloud).  Lastly, we included the patchy
cloud/haze fraction parameter $f_{\rm patchy}$.

For \citetalias{MacDonaldMadhusudhan2017mnrasRetrievalHD209b}, we
re-analyzed their `fixed-fraction' retrieval (Section 4.1
of \citetalias{MacDonaldMadhusudhan2017mnrasRetrievalHD209b}).  This
analysis only considers the {\HST} observations, neglects the
contribution from CO and {\carbdiox}, and fixes the patchy cloud/haze
fraction parameter at $f_{\rm patchy}=0.47$.  The main difference
between our and their setup is that we let $X_{\rm K}$ to be a free
parameter,
whereas \citetalias{MacDonaldMadhusudhan2017mnrasRetrievalHD209b} kept
the ratio $X_{\rm K}/X_{\rm Na}$ fixed according to the solar
volume-mixing ratios.

Another difference is that for the hydrostatic equilibrium solution we
fixed a reference pressure point ($p_{\rm ref}$) and let its
corresponding altitude free ($R_{\rm planet}$);
whereas \citetalias{MacDonaldMadhusudhan2017mnrasRetrievalHD209b}
and \citetalias{PinhasEtal2019mnrasHotJupiterSampleRetrieval} fixed a
reference altitude point and retrieved the logarithm of its
corresponding
pressure.  \citet{WelbanksMadhusudhan2019ajRetrievalDegeneracies}
showed that the choice of $p_{\rm ref}$ or $R_{\rm planet}$ as free
parameter does not impact the retrieval results.  To ease the
comparison, for the reference pressure we chose the best-fitting
$p_{\rm ref}$
of \citetalias{MacDonaldMadhusudhan2017mnrasRetrievalHD209b}
and \citetalias{PinhasEtal2019mnrasHotJupiterSampleRetrieval} (thus,
ideally, our retrieval radius should match their reference altitude).
Table \ref{table:benchmark_HD209458b} summarizes our retrieval
parameterization for
the \citetalias{MacDonaldMadhusudhan2017mnrasRetrievalHD209b}
and \citetalias{PinhasEtal2019mnrasHotJupiterSampleRetrieval}
re-analyses.

{\renewcommand{\arraystretch}{1.1}
\begin{table}
\begin{minipage}{\linewidth}
\centering
\caption{HD 209458b Retrieval Parameters Summary}
\label{table:benchmark_HD209458b}
\begin{tabular*}{\linewidth} {@{\extracolsep{\fill}} lll}
\hline
Reference  & \citetalias{MacDonaldMadhusudhan2017mnrasRetrievalHD209b}
           & \citetalias{PinhasEtal2019mnrasHotJupiterSampleRetrieval} \\
Parameter     & Priors$^*$ & Priors$^*$ \\
\hline
$\log_{10}(p_1/{\rm bar})$ & $\mathcal U(-6, 2)$   & $\mathcal U(-6, 2)$ \\
$\log_{10}(p_2/{\rm bar})$ & $\mathcal U(-6, 2)$   & $\mathcal U(-6, 2)$ \\
$\log_{10}(p_3/{\rm bar})$ & $\mathcal U(-2, 2)$ and $p_3>p_1$
                          & $\mathcal U(-2, 2)$ and $p_3>p_1$\\
$a_1$ (K$^{-0.5}$)         & $\mathcal U(0.02, 1.0)$ & $\mathcal U(0.02, 1.0)$ \\
$a_2$ (K$^{-0.5}$)         & $\mathcal U(0.02, 1.0)$ & $\mathcal U(0.02, 1.0)$ \\
$T_0$ (K)                 & $\mathcal U(800, 1600)$ & $\mathcal U(800, 1650)$ \\
$R_{\rm planet}$ ($R_{\rm Jup}$) & $\mathcal U(0.5,2.0)$ & $\mathcal U(0.5,2.0)$ \\
$\log_{10}(X_{\rm Na})$        & $\mathcal U(-10, -2)$ & $\mathcal U(-12, -2)$ \\
$\log_{10}(X_{\rm K})$         & $\mathcal U(-10, -2)$ & $\mathcal U(-12, -2)$ \\
$\log_{10}(X_{\rm H2O})$       & $\mathcal U(-10, -2)$ & $\mathcal U(-12, -2)$ \\
$\log_{10}(X_{\rm CH4})$       & $\mathcal U(-10, -2)$ & $\mathcal U(-12, -2)$ \\
$\log_{10}(X_{\rm NH3})$       & $\mathcal U(-10, -2)$ & $\mathcal U(-12, -2)$ \\
$\log_{10}(X_{\rm HCN})$       & $\mathcal U(-10, -2)$ & $\mathcal U(-12, -2)$ \\
$\log_{10}(X_{\rm CO})$        & Fixed $(-12)$ & $\mathcal U(-12, -2)$ \\
$\log_{10}(X_{\rm CO2})$       & Fixed $(-12)$ & $\mathcal U(-12, -2)$ \\
$\log_{10}(P_{\rm top}/{\rm bar})$ & $\mathcal U(-6, 2)$  & $\mathcal U(-6, 2)$ \\
$\log_{10}(f_{\rm ray}$)       & $\mathcal U(-4, 8)$  & $\mathcal U(-4, 8)$ \\
$\alpha_{\rm ray}$            & $\mathcal U(-20, 2)$ & $\mathcal U(-20, 2)$ \\
$f_{\rm patchy}$               & Fixed $(0.47)$   & $\mathcal U(0, 1)$  \\
\hline
\end{tabular*}
\end{minipage}
\begin{tablenotes}
\item $^*$ $\mathcal U(a, b)$ stands for a uniform distribution between $a$ and $b$.
\end{tablenotes}
\end{table}
}

\begin{figure*}
\centering
\includegraphics[width=0.95\linewidth, clip]{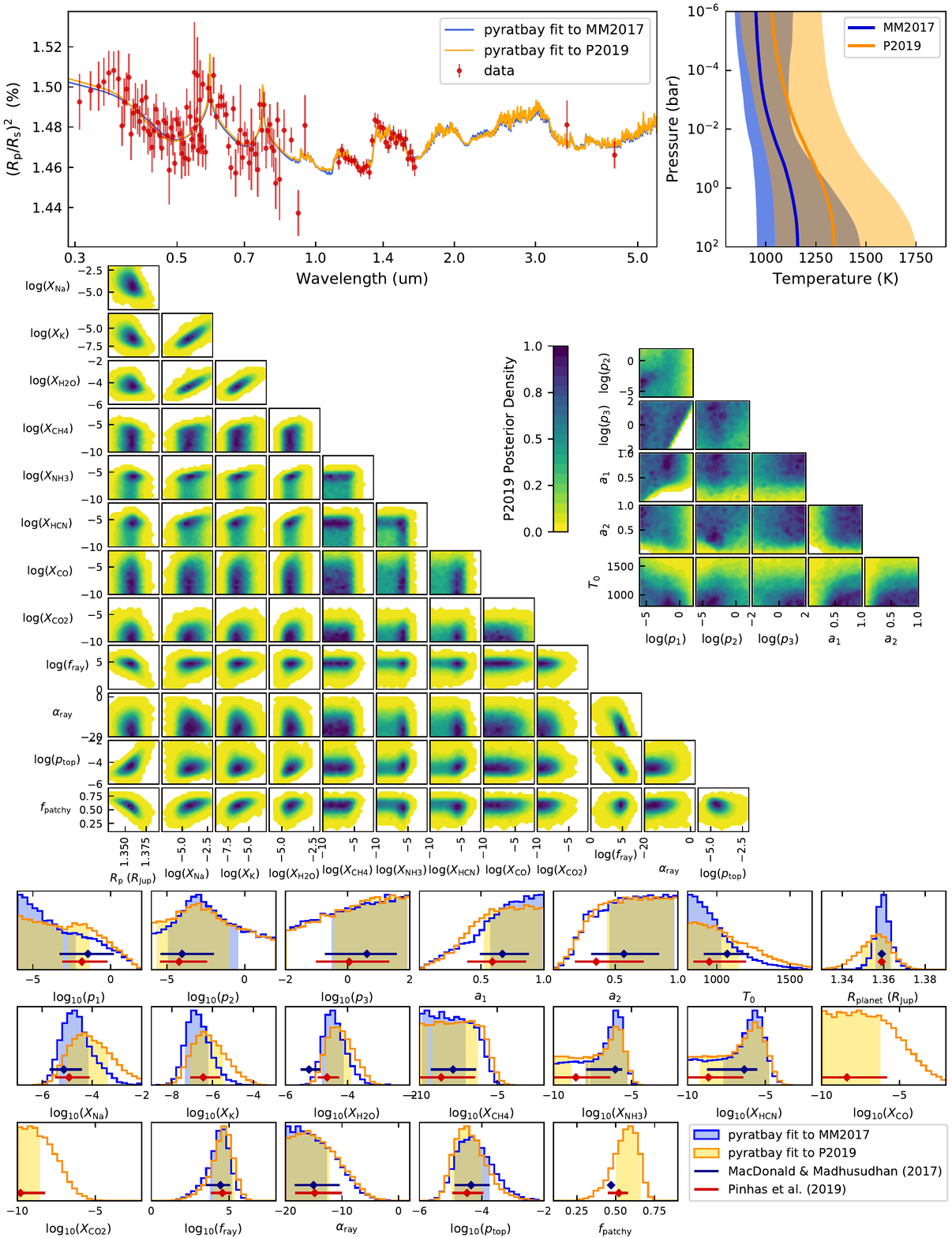}
\caption{
    {\pyratbay} retrieval of the transmission spectra of
    {\twoohnineb}.  In all panels, the blue- and orange-themed curves
    correspond to our re-analysis of
    the \citetalias{MacDonaldMadhusudhan2017mnrasRetrievalHD209b}
    and \citetalias{PinhasEtal2019mnrasHotJupiterSampleRetrieval}
    analyses, respectively.  {\bf Top left:} {\twoohnineb} observed
    spectra (red markers with error bars) and best-fitting model to
    the \citetalias{MacDonaldMadhusudhan2017mnrasRetrievalHD209b}
    and \citetalias{PinhasEtal2019mnrasHotJupiterSampleRetrieval}
    datasets.  {\bf Top right:} posterior temperature profiles. The
    solid curves and shaded areas mark the median and the 68\% central
    credible intervals of the posterior distributions, respectively.
    {\bf Middle:} pair-wise posterior distributions for
    the \citetalias{PinhasEtal2019mnrasHotJupiterSampleRetrieval}
    {\pyratbay} re-analysis (our re-analysis
    of \citetalias{MacDonaldMadhusudhan2017mnrasRetrievalHD209b}
    followed similar patterns).  {\bf Bottom:} marginal posterior
    distributions.  The shaded areas denote the 68\%
    highest-posterior-density credible intervals.  The blue and red
    markers with error bars denote median and 68\% central credible
    intervals obtained
    by \citetalias{MacDonaldMadhusudhan2017mnrasRetrievalHD209b}
    and \citetalias{PinhasEtal2019mnrasHotJupiterSampleRetrieval},
    respectively.}
\label{fig:benchmark_HD209458b}
\end{figure*}

Figure \ref{fig:benchmark_HD209458b} shows our retrieval results
(spectra, temperature profiles, pair-wise posteriors, and marginal
posteriors) and compares our marginal posteriors to those
of \citetalias{MacDonaldMadhusudhan2017mnrasRetrievalHD209b}
and \citetalias{PinhasEtal2019mnrasHotJupiterSampleRetrieval}.
We found a good agreement between our retrieval results and those
of \citetalias{MacDonaldMadhusudhan2017mnrasRetrievalHD209b}
and \citetalias{PinhasEtal2019mnrasHotJupiterSampleRetrieval}.  Our
best-fitting spectra match well the observations ($\chi^2_{\rm red} =
\sim1.5$), and resemble the models retrieved
by \citetalias{MacDonaldMadhusudhan2017mnrasRetrievalHD209b}
and \citetalias{PinhasEtal2019mnrasHotJupiterSampleRetrieval}.  Our
temperature profiles show a similar (non-inverted) behavior to that
of \citetalias{MacDonaldMadhusudhan2017mnrasRetrievalHD209b}
and \citetalias{PinhasEtal2019mnrasHotJupiterSampleRetrieval}.
Although our profiles are consistently $\sim$100--200~K colder than
theirs at the pressures probed by the observations
($p\lesssim0.1$~bar), our temperature-profile credible intervals
overlap with those
of \citetalias{MacDonaldMadhusudhan2017mnrasRetrievalHD209b}
and \citetalias{PinhasEtal2019mnrasHotJupiterSampleRetrieval}.  For
most parameters, our pair-wise posterior distributions show the same
trends and correlations as those
of \citetalias{MacDonaldMadhusudhan2017mnrasRetrievalHD209b}
and \citetalias{PinhasEtal2019mnrasHotJupiterSampleRetrieval}; for
example, compare our pair-wise posteriors for
the \citetalias{PinhasEtal2019mnrasHotJupiterSampleRetrieval}
re-analysis from Fig.~\ref{fig:benchmark_HD209458b} (our re-analysis
of \citetalias{MacDonaldMadhusudhan2017mnrasRetrievalHD209b} followed
similar patterns) to Fig.~9
of \citetalias{MacDonaldMadhusudhan2017mnrasRetrievalHD209b}.  The
major discrepancy lies in the $p_2$-{\vs}-$p_1$ panel, the posterior
distribution
of \citetalias{MacDonaldMadhusudhan2017mnrasRetrievalHD209b} shows an
abrupt edge at $p_2 = p_1$, suggesting that their retrieval imposed a
$p_2 < p_1$ condition (i.e., requiring a non-inverted profile).  We
allowed for any combination of $p_1$ and $p_2$ values within their
prior ranges.  Our marginal posterior distributions match well those
of \citetalias{MacDonaldMadhusudhan2017mnrasRetrievalHD209b}
and \citetalias{PinhasEtal2019mnrasHotJupiterSampleRetrieval}
(significant overlap of the credible intervals), except for some
temperature-profile parameters, which is not surprising considering
the potential discrepancy noted above.  As in the previous studies
by \citetalias{MacDonaldMadhusudhan2017mnrasRetrievalHD209b}
and \citetalias{PinhasEtal2019mnrasHotJupiterSampleRetrieval}, our
retrievals constrain the abundances of {\water}, Na, K, and a
combination of HCN or {\ammonia}.  For CO, {\carbdiox}, and
{\methane}, our retrievals place upper limits to their abundances.
Our retrievals favor a cloud top located at
$\pcloud \approx5\tttt{-5}$~bar with a partially cloudy atmosphere for
the \citetalias{PinhasEtal2019mnrasHotJupiterSampleRetrieval} dataset
($f_{\rm patchy}\sim0.5$).  Lastly, we also recover a super-enhanced
and super-Rayleigh haze absorption ($f_{\rm
ray} \approx \ttt{4}$--$\ttt{5}$, $\alpha_{\rm
ray}\approx-15$).  \citet{OhnoKawashima2020apjSuperRayleigh} propose
that such super-Rayleigh slopes can arise from photochemical hazes
present in an atmosphere exhibiting with strong vertical mixing.

\input{targetsample}

\section{A Systematic Atmospheric Retrieval of Exoplanet Transmission Observations}
\label{sec:analysis}

Here, we begin a systematic analysis of the existing space-based
exoplanet transmission spectra.  The goals of this study are to
produce a large sample of targets analyzed under a standard set of
retrievals and present an independent reanalysis of published results.
The sample includes all targets with published transit spectra with
sufficient wavelength coverage and resolution to resolve atmospheric
spectral features, discarding targets with only broadband photometric
observations.  By applying a standardized retrieval analysis across
the sample, we can compare the results of the different retrievals for
a given target, as well as the results between targets.

\subsection{Target Sample}

In this article, we analyze the targets without observations at
wavelengths shorter than 1 {\micron} (as of early 2020).  This sample
includes 26 datasets of 19 different systems observed with the
{\Hubble} Wide Field Camera 3 (WFC3), using the G141 grism (1.1 to
1.7~{\microns}).  Among them, three datasets (K2-18b, WASP-43b, and
WASP-103b) also include observations from the {\Spitzer} Infrared
Array Camera (IRAC), using the 3.6 and 4.5 {\micron} photometric
filters.  Future articles will present the analysis of the targets
with shorter-wavelength observations (namely, from the {\Hubble} STIS
and WFC3/G104 data).  Table \ref{table:sample} lists the datasets
considered in this sample and the main properties of their systems.
To enable a more direct comparison between our results and those
published, for each dataset we adopted the same system parameters as
those assumed in the original article.

We focus our analysis on characterizing the {\water} absorption
signature, since this molecule is expected to dominate the spectral
range probed by the {\HST} WFC3/G141 grism.  We pay particular
attention to the strong degeneracies between the {\water} abundance,
cloud coverage, and atmospheric mean molecular mass, which can lead to
similar transmission absorption spectra, especially for observations
over a narrow wavelength range such as this one \citep[see,
e.g.,][]{Griffith2014rsptaDegenerateSolutions,
MadhusudhanEtal2014apjlH2OabundancesIn3HotJupiters,
LineParmentier2016apjPartialClouds, FuEtal2017apjWFC3statistics}.  To
determine in a statistically robust manner whether the data can
distinguish between such scenarios, for each dataset we run a set of
retrievals including or excluding clouds, and with or without a
parameter that varies the atmospheric mean molecular mass.  We compare
the results using the Bayesian Information
Criterion \citep[BIC,][]{Liddle2007mnrasBIC}.  To assess the
significance of the model comparison, we adopt
the \citet{Raftery1995BIC} rule of thumb, ranking the evidence as weak
(for $\Delta$BIC values within 0--2), positive (2--6), strong (6--10),
and very strong (values greater than 10), where $\Delta$BIC is the
difference in BIC between a given model and the lowest-BIC model for
the dataset.

\subsection{Retrieval Setup}
\label{sec:setup}

To yield a uniformly analyzed sample, we apply the {\pyratbay}
retrieval under the same set of assumptions for each dataset, where we
set free parameters for the atmospheric temperature, composition,
cloud coverage, and altitude profile models.  For specific cases we
perform additional retrieval tests, for example, search for additional
molecules when there is {\Spitzer} data or when there is prior
evidence for them.

We retrieve the atmospheric abundances as constant-with-altitude
volume-mixing-ratio profiles, setting a lower boundary at $\ttt{-12}$,
at which their spectral features become negligible.  For the
Jupiter-type planets ($M_{\rm p}\gtrsim0.3 \mjup$), we limit the sum
of the mixing ratios of the metals to be less than 0.2 (i.e., the MCMC
rejects samples where $X_{\rm H2} + X_{\rm He} < 0.8$, see
Eq. (\ref{eq:bulk})), which roughly corresponds to a maximum
metallicity of $\sim$250$\times$ solar (thus, ensuring
{\molhyd}/He-dominated atmospheres).  For the sub-Saturn mass planets
($M_{\rm p}\lesssim0.3 \mjup$) we relax the limit the sum of the metal
mixing ratios to be less than 0.9.

As in previous studies \citep[e.g.,][]{LineEtal2013apjRetrievalI,
TsiarasEtal2018ajPopulationStudy,
WelbanksEtal2019apjMasssMetallicityTrends}, these metallicity upper
boundaries are rough estimates based on the bulk properties of the
planets.  More precise estimates could be obtained on a case-by-case
basis from interior-structure and evolution models, constrained by the
planet's mass, radius, and age \citep[see,
e.g.,][]{ThorngrenEtal2016apjMassMetallicity,
KreidbergEtal2018apjlWASP107bWFC3}, but such analysis lies beyond the
scope of this work.

For HAT-P-38b and K2-18b we set the temperature upper limit at a
stricter $T_{\rm eq}^{\rm max} + 3\sigma_T$, since the posterior
distributions were returning temperatures far in excess above their
maximum estimated equilibrium temperatures ($T_{\rm eq}^{\rm max}$,
for zero albedo and no day--night energy redistribution).  These upper
limits are 1500~K and 390~K for HAT-P-38b and K2-18b, respectively.
We consider uniform priors in log scale for the mixing ratios.  For
all datasets we retrieve the {\water} abundance.  For the datasets
that include {\Spitzer} observations, we also retrieve the {\methane}
abundance (expected to have features at 3.6 {\micron}) and the CO and
{\carbdiox} abundances (expected to have features at 4.5 {\microns}).
Lastly, for WASP-63b we also retrieve the HCN abundance, given the
prior evidence for its
detection \citep{KilpatrickEtal2018apjWASP63bWFC3}.

We retrieve the temperature profile adopting the isothermal model.
This is an appropriate choice for these datasets given the weak
sensitivity of transmission spectra to temperature gradients and the
narrow pressure range probed by these
observations \citep{BarstowEtal2013mnrasEchoAtmospheresCharacterization,
RocchettoEtal2016apjJWSTbiases}.  We let the temperature vary between
100 and 3000~K (the range allowed by the cross sections), adopting a
uniform prior.

We retrieve the altitude profile adopting Eq.~(\ref{eq:hydrostatic})
to compute hydrostatic equilibrium.  To solve this equation we
retrieve the planetary radius $R_{\rm planet}$ at a reference pressure
level of $p\sb{0}=0.1$~bar.  We let the planetary radius vary within
0.1 and 4 {\rjup} (though none of the retrievals reached these
boundaries), adopting a uniform prior.  Additionally, we retrieve the
abundance of N$_2$ (without considering its absorption cross section),
which works as a proxy to vary the mean molecular mass $\mu$
independently of the other free-abundance species that affect the
atmospheric cross section.  In practice, we monitor the metal mass
fraction relative to the sun:
\begin{equation}
\label{eq:mass_fraction}
[Z/X] = \log_{10}(Z/X) - \log_{10}(Z/X)_\odot,
\end{equation}
where $X$ and $Z$ are the hydrogen and metal mass fractions,
respectively.  We adopt the solar mass fraction values
from \citet{AsplundEtal2009araSolarComposition}: $X_\odot=0.7381$ and
$Z_\odot=0.0134$.  Recalling that $X + Y + Z = 1$ \citep[see,
e.g.,][]{CarrollOstlie1996bookModernAstrophysics}, with $Y$ the helium
mass fraction, we compute the metal mass fraction from $Z = 1 - X -
Y$.  To compute $X$ and $Y$ we need to account for the volume mixing
ratios $X_i$ of all atmospheric species containing hydrogen and helium
(respectively), weighted by their stoichiometric coefficients, i.e.:
$X = (2X_{\rm H2} + 2X_{\rm H2O} + X_{\rm HCN} + 4X_{\rm CH4})\,m_{\rm
H}/\mu$ and $Y=X_{\rm He}\,m_{\rm He}/\mu$.  We prefer to show the
metal mass fraction (Eq.~(\ref{eq:mass_fraction})) instead of the mean
molecular mass because it offers a metric much more sensitive to the
composition of the atmosphere, since it can trace the metal content
across several orders of magnitude.  Instead, the mean molecular mass
converges to $\mu\approx2.3$ for all solar/sub-solar metallicities.

Finally, we consider the gray cloud deck model.  We retrieve the
pressure at the top of the cloud deck ({\pcloud}), which can vary
within 100 and $\ttt{-6}$ bar, adopting a uniform prior in log scale.
Table \ref{table:WFC3_retrieval} summarizes the retrieval
parameterization for the WFC3 sample.

{\renewcommand{\arraystretch}{1.1}
\begin{table}
\centering
\caption{Retrieval Parameters Summary}
\label{table:WFC3_retrieval}
\begin{tabular*}{0.7\linewidth} {@{\extracolsep{\fill}} ll}
\hline
Parameter     & Priors$^*$ \\
\hline
$T$ (K)                         & $\mathcal U (100, 3000)$ \\
$R_{\rm planet}$ ($R_{\rm Jup}$)    & $\mathcal U (0.1, 4.0)$ \\
$\log_{10}(X_{\rm H2O})$           & $\mathcal U (-12, 0)$\\
$\log_{10}(X_{\rm CO})^\dagger$         & $\mathcal U (-12, 0)$ \\
$\log_{10}(X_{\rm CO2})^\dagger$       & $\mathcal U (-12, 0)$ \\
$\log_{10}(X_{\rm CH4})^\dagger$       & $\mathcal U (-12, 0)$ \\
$\log_{10}(X_{\rm HCN})^\ddagger$   & $\mathcal U (-12, 0)$ \\
$\log_{10}(X_{\rm N2})^\S$   & $\mathcal U (-12, 0)$ \\
$\log_{10}(\pcloud/{\rm bar})$ & $\mathcal U (-6, 2)$ \\
                                & $ X_{\rm H2} + X_{\rm He} > 0.8^\P$ \\
\hline
\end{tabular*}
\begin{tablenotes}
\item $^*$ $\mathcal U(a, b)$ stands for a uniform distribution between $a$ and $b$. 
  \item $^\dagger$ Only considered for datasets with {\Spitzer} observations.
  \item $^\ddagger$ Only considered for WASP-63b datasets.
  \item $^\S$ Used as proxy to enhance the atmospheric metal mass fraction.
  \item $^\P$ Only imposed for Saturn/Jupiter-mass planets
        ($M_{\rm p}\gtrsim0.2\mjup$).
\end{tablenotes}
\end{table}
}

For the molecular line-by-line cross sections we consider the ExoMol
databases for
{\water} \citep{PolyanskyEtal2018mnrasPOKAZATELexomolH2O},
{\methane} \citep{YurchenkoTennyson2014mnrasExomolCH4}, and
HCN \citep{HarrisEtal2006mnrasHCNlineList,
HarrisEtal2008mnrasExomolHCN}; and the HITEMP databases for
CO \citep{LiEtal2015apjsCOlineList} and
{\carbdiox} \citep{RothmanEtal2010jqsrtHITEMP}.  Before computing the
cross sections, we apply the {\repack} compression to the large ExoMol
line lists to reduce the number of transitions to compute.  We then
sample the line-by-line cross sections at a constant resolving power
of ${\rm R} = 15\,000$ between 1.0 and 5.5 {\micron}.  We sample the
cross-section spectra into a grid of 41 pressure layers between 100
and $\ttt{-8}$ bar and 30 temperature samples between 100 and 3000~K.
During the retrieval, the MCMC sampler interpolates in temperature
from this grid.  Additionally, the atmospheric model includes Rayleigh
scattering cross section for {\molhyd} and CIA cross section for
{\molhyd}--{\molhyd} and
{\molhyd}--He \citep{BorysowEtal1988apjH2HeRT,
BorysowEtal1989apjH2HeRVRT, BorysowEtal2001jqsrtH2H2highT,
Borysow2002jqsrtH2H2lowT, BorysowFrommhold1989apjH2HeOvertones}.

We sample the posterior parameter space with the Snooker DEMC
algorithm \citep{terBraak2008SnookerDEMC} implemented via the {\MCC}
package.  We obtain between 1 and 6 million samples, distributed into
24 parallel chains (discarding the first 10\,000 iterations from each
chain).  To check for convergence we apply the Gelman-Rubin
statistics \citep{GelmanRubin1992stascGRstatistics}, checking that the
potential scale reduction factor approaches $\sim$1.01 or lower for
each free parameter.

\begin{figure*}
\centering
\includegraphics[width=\linewidth, clip, trim= 0 91 0 0]
{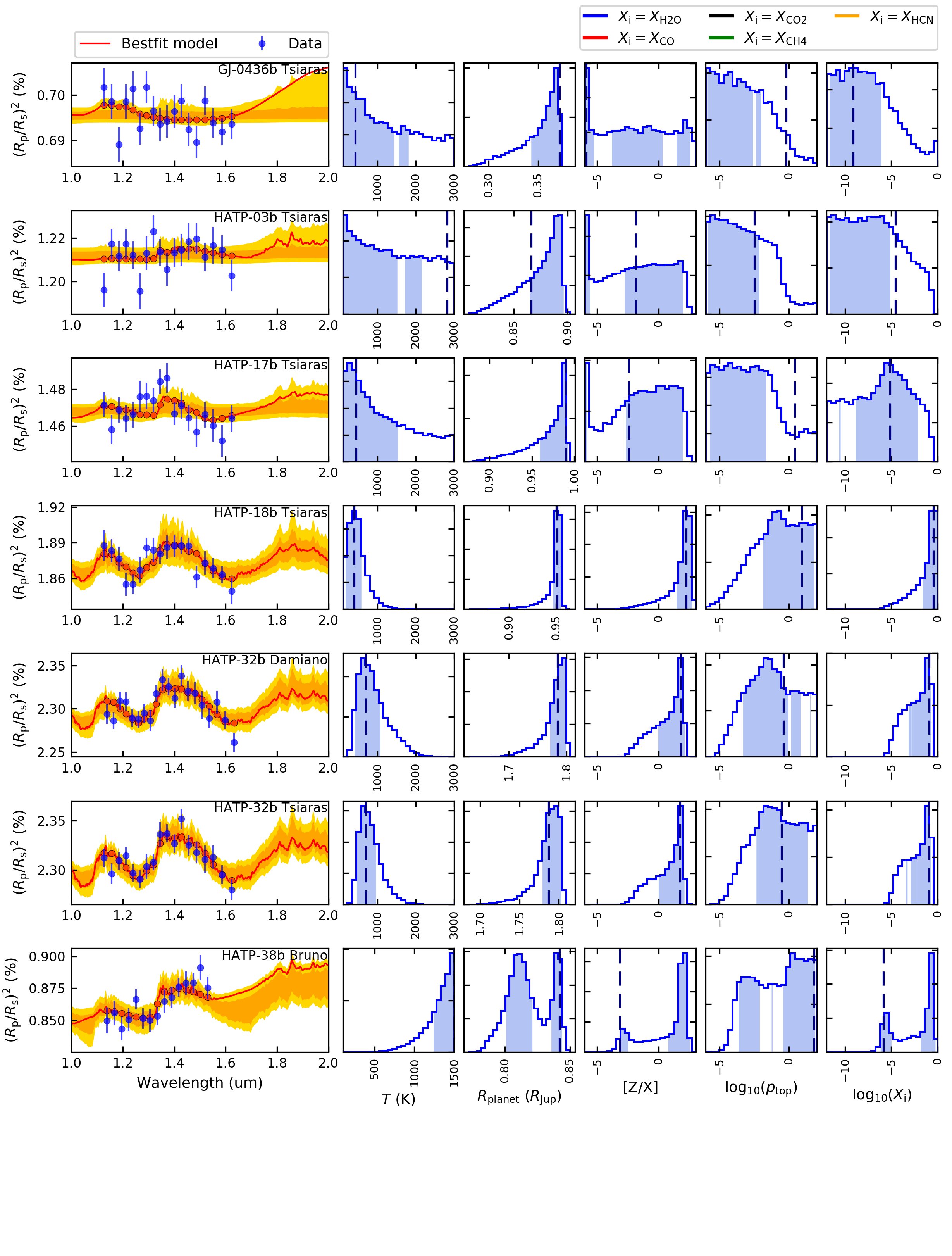}
\caption{
    {\pyratbay} retrieval of the WFC3 transmission sample.  {\bf Left
    panels:} observed transmission spectra with 1$\sigma$
    uncertainties (blue dots with error bars) and retrieved
    best-fitting models (red curves).  The orange and yellow contours
    denote the span of the 68th and 95th percentile of the MCMC
    posterior distribution, respectively.  The label at the top right
    corner of each panel indicates the name of the dataset.  {\bf
    Following panels:} retrieved marginal posterior histograms of the
    corresponding datasets shown in the left panels.  The shaded areas
    denote the 68\% HPD credible regions of the distributions.  The
    panels on the far right might show the posterior for multiple
    species, color coded according to the labels at the top of the
    figure.  The dashed vertical lines denote the best-fitting values
    for each parameter.}
\label{fig:spectra1}
\end{figure*}

\begin{figure*}
\centering
\includegraphics[width=\linewidth, clip, trim=0 1 0 0]
{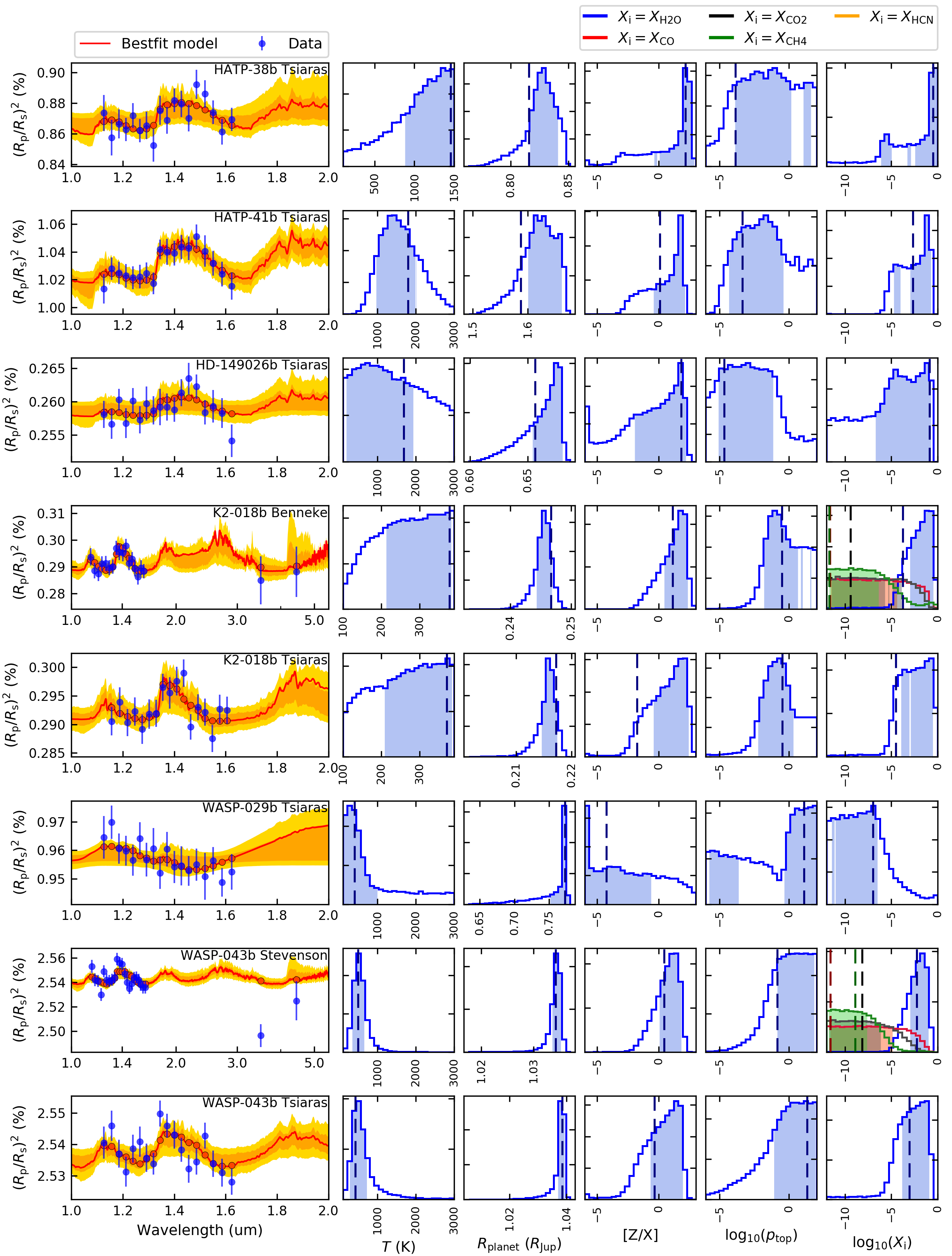}
\caption{
    {\pyratbay} retrieval of the WFC3 transmission sample
    (continuation).  See caption in Figure \ref{fig:spectra1}.}
\label{fig:spectra2}
\end{figure*}

\begin{figure*}
\centering
\includegraphics[width=\linewidth, clip, trim=0 1 0 0]
{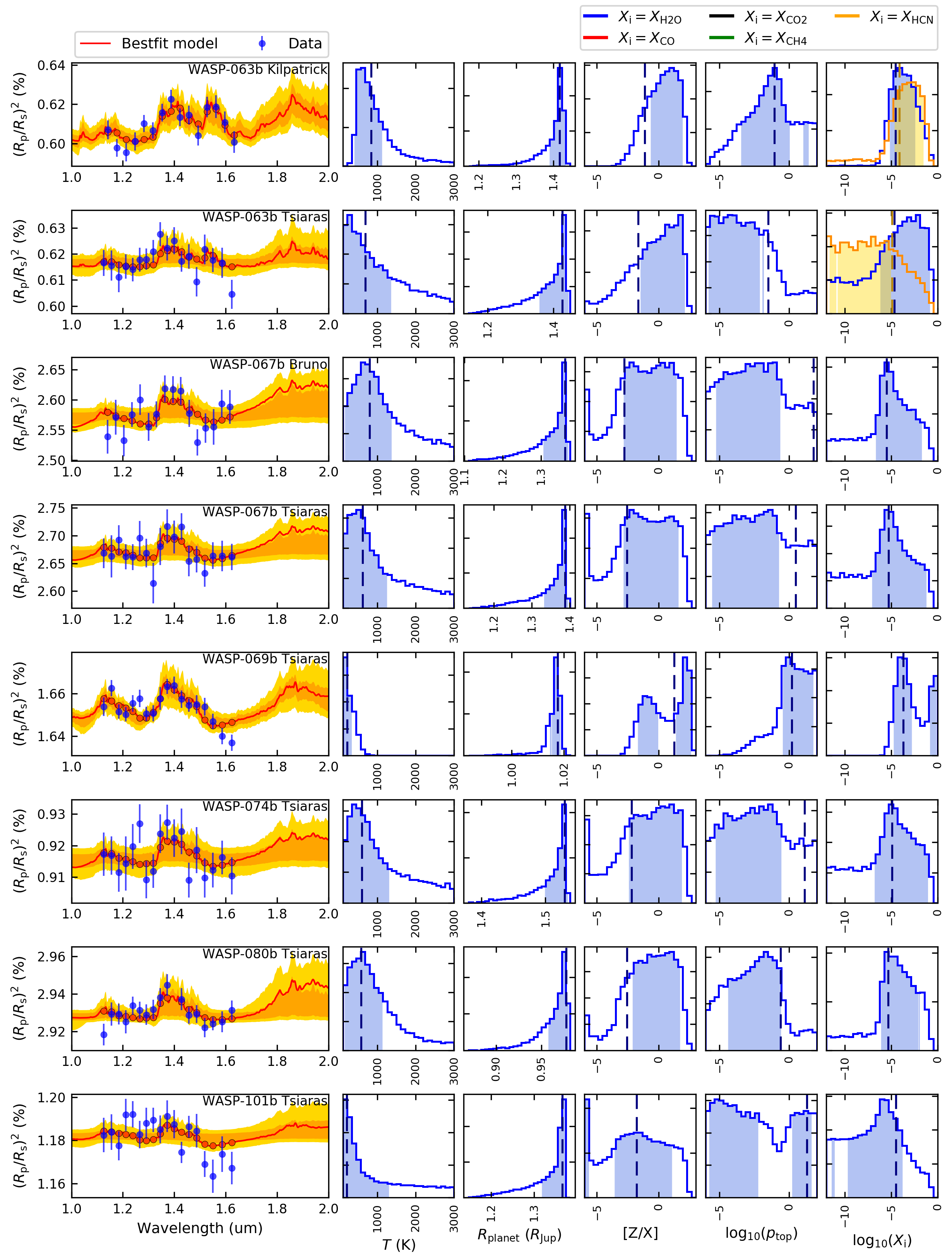}
\caption{
    {\pyratbay} retrieval of the WFC3 transmission sample
    (continuation).  See caption in Figure \ref{fig:spectra1}.}
\label{fig:spectra3}
\end{figure*}

\begin{figure*}
\centering
\includegraphics[width=\linewidth, clip, trim=0 455 0 0]
{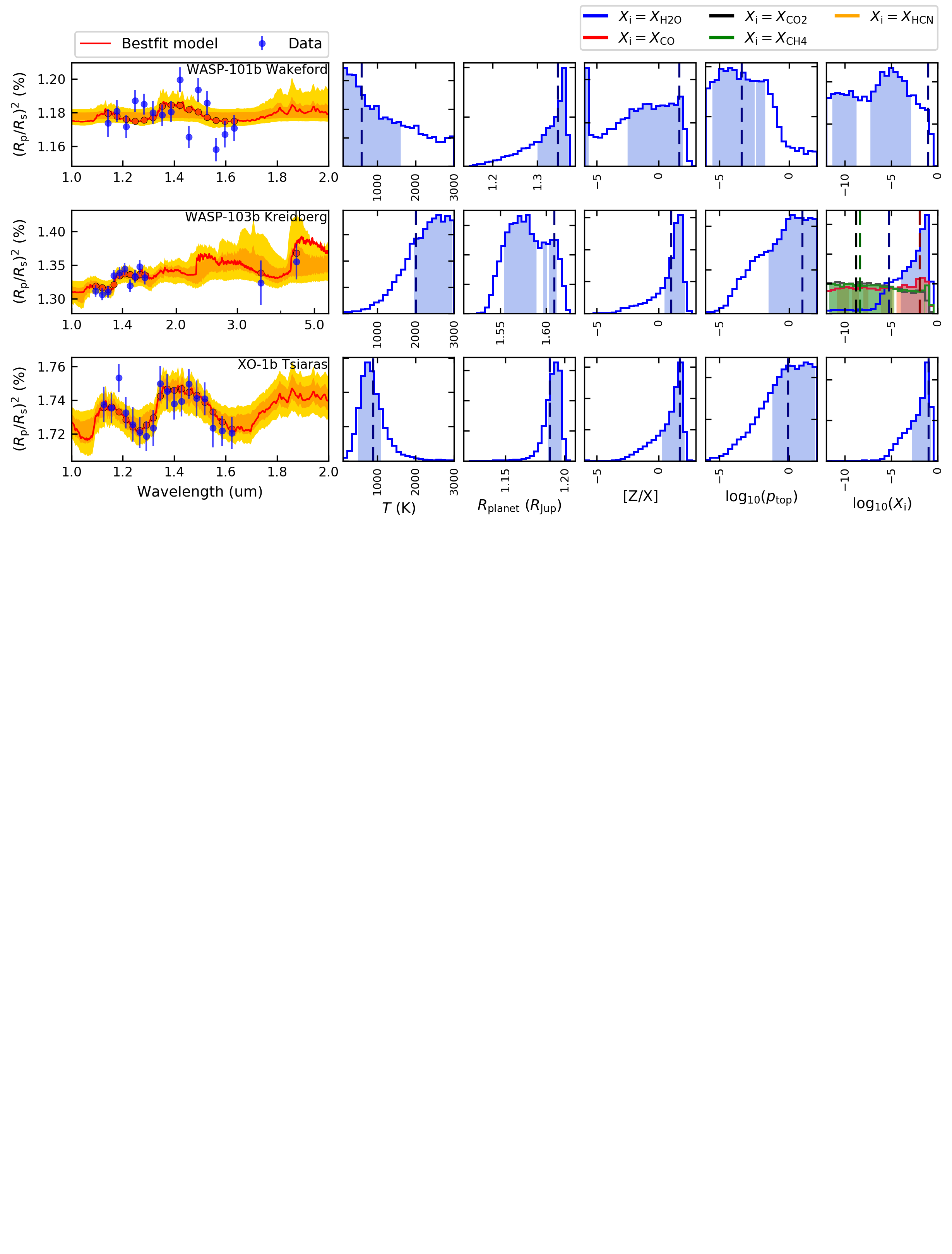}
\caption{
    {\pyratbay} retrieval of the WFC3 transmission sample
    (continuation).  See caption in Figure \ref{fig:spectra1}.}
\label{fig:spectra4}
\end{figure*}

\begin{figure}
\centering
\includegraphics[width=\linewidth, clip]{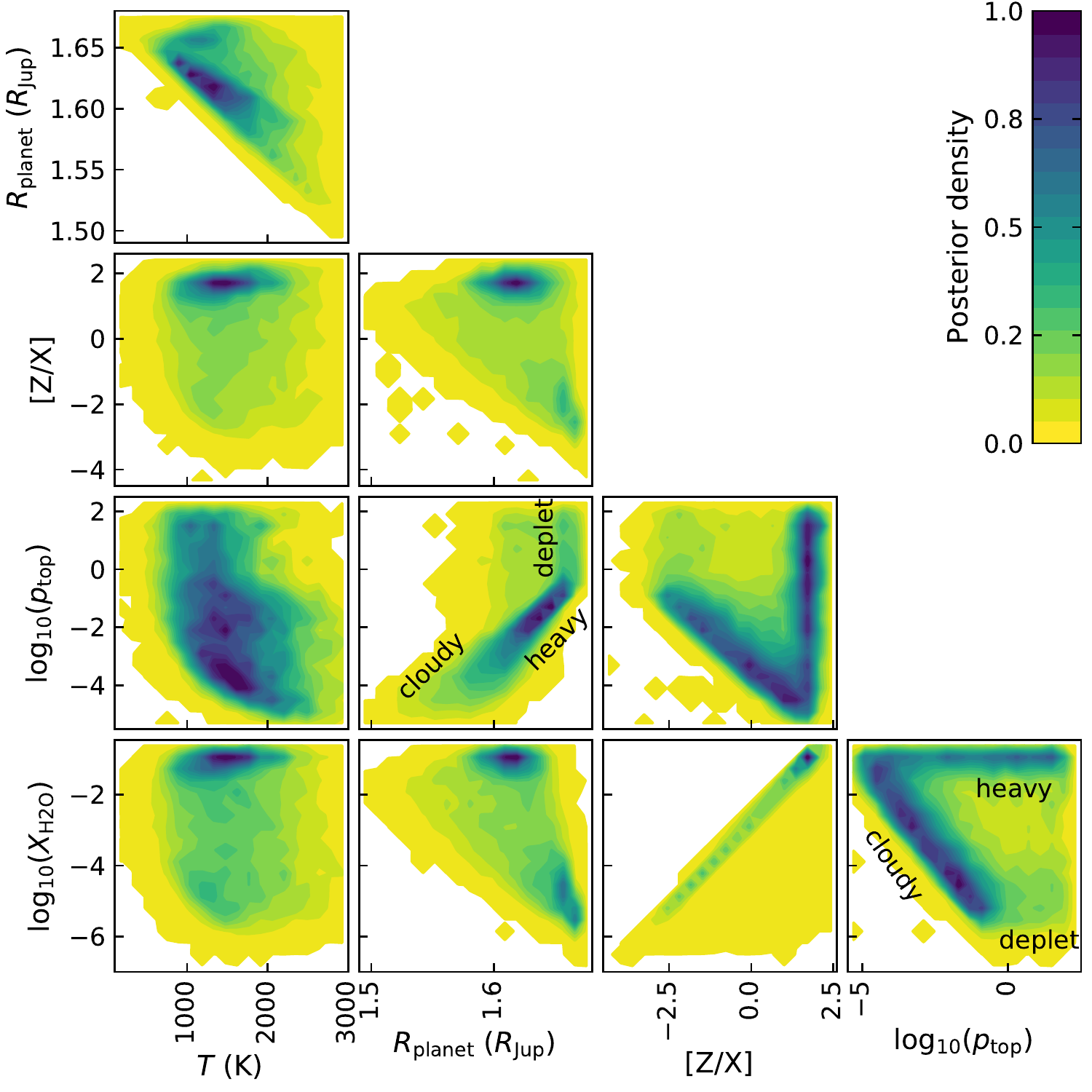}
\caption{
    {\pyratbay} pair-wise posterior distributions of the
    HAT-P-41b/Tsiaras WFC3 transmission spectrum.  This figure
    exemplifies a typical correlation between parameters found in the
    overall sample analysis.  We selected two panels to label the
    different solution modes where the atmospheric models are
    preferentially cloudy (`cloudy'), high in metal mass fractions
    (`heavy'), and depleted in {\water} abundance (`deplet').  The
    posterior distribution transitions smoothly from one region to
    another.}
\label{fig:WFC3_correlation_example}
\end{figure}

\subsection{Results}
\label{sec:results}

The datasets in this sample show a wide variety of spectral features,
ranging from featureless and noisy spectra to well defined absorption
bands that are consistent with the {\water} absorption signature.
Figures \ref{fig:spectra1}--\ref{fig:spectra4} show the retrieved
spectra and marginal posterior distributions for our sample.  In these
runs we retrieve the cloud top pressure, the metal mass fraction [Z/X]
(via the N$_2$ free abundance), and the {\methane}, CO, {\carbdiox},
and HCN abundances when applicable.
Tables \ref{table:model_comparison1}--\ref{table:model_comparison4} in
the appendix show the model comparison statistics for each dataset.
Based on the BIC model comparison, we found that 15 out of the 26
datasets are better explained by a {\water} absorption feature rather
than a flat transmission spectrum; however, none of these datasets
yield a $\Delta$BIC significant enough to distinguish between a cloudy
or a high metal-mass-fraction scenario (i.e., $|{\rm BIC}_{\rm
M1}-{\rm BIC}_{\rm M2}|<2$).  The retrieved model spectra reproduce
well the observations for most datasets, finding $\redchisq$ values
close to one (with the exception of WASP-43b, WASP-69b, and
WASP-101b).  However, the posterior distributions reveal strong
correlations between many of the free parameters.  These correlations
lead to broad, multi-modal, or unconstrained marginal posteriors.

Figure \ref{fig:WFC3_correlation_example} shows a typical pair-wise
posterior distribution, obtained from the HAT-P-41b/Tsiaras
{\pyratbay} retrieval.  We found that the temperature correlates
strongly with the planets' reference altitude at 0.1 bar.  The
temperature also often shows non-linear correlation with the cloud top
pressure.  The posterior multimodality is more clearly seen in the
{\pcloud}, [Z/X], and $X_{\rm H2O}$ correlation panels (see labeled
regions in Fig.~\ref{fig:WFC3_correlation_example}).  The two main
solution modes are a cloudy mode, characterized by a strong
{\pcloud}--$X_{\rm H2O}$ correlation, and a heavy atmosphere,
characterized by a high metal mass fraction (driven mainly by the high
$X_{\rm H2O}$), where clouds become irrelevant.  We also often see a
{\water}-depleted solution.  Note that the {\water}-depleted, cloudy,
and high metal-mass-fraction modes are not mutually excluding
solutions, the posterior sampling transitions seamlessly from one mode
to another, for example, connecting the regions with solutions that
are both cloudy and high in metal mass fraction.  These correlations
propagate to the $T$ and $R_{\rm planet}$ parameters as well.

Finding such strongly correlated posteriors is expected, given the
combination of limited data available and the degenerate nature of the
atmospheric parameters.  The retrieval model essentially fits for two
main characteristics in the WFC3/G141 data, the absolute transit depth
level (i.e., the altitude where the transmission photosphere sits) and
the relative transit depth between the {\water} absorption band
(centered at 1.4 {\microns}) and the baseline depth at the surrounding
wavelengths.  All free parameters modify these absolute and relative
transit depths in slightly different ways.  Increasing the
temperature, increasing the {\water} abundance, or decreasing the
metal mass fraction increases the depth of the {\water} band over the
baseline, but also changes the absolute transit depth to a minor
extent. Increasing the altitude of the cloud deck raises the baseline
transit depth, leading to a shallower relative depth of the {\water}
band.  Finally, varying the reference radius of the planet mainly
modifies the absolute transit depth, but it also affects the relative
transit depth of the {\water} band by changing the atmospheric scale
height.

The limited spectral resolution and signal-to-noise ratio of these
observations do not provide the precision required to distinguish the
different shapes of the spectral features under different
scenarios \citep[see, e.g.,][]{LineParmentier2016apjPartialClouds}.
Therefore, the retrieval sampling finds a degenerate family of
solutions that reproduce equally well the observations, in statistical
terms.
Usually, a broader wavelength coverage helps to break the degeneracies
between the free
parameters \citep[e.g.,][]{PinhasEtal2019mnrasHotJupiterSampleRetrieval}.
However, in this case the two photometric {\Spitzer}/IRAC bands do not
significantly increase the constraining power since they probe the
unresolved contribution from CO, {\carbdiox}, and {\methane}, thus
adding three more variables to the retrieval.  In addition, combining
non-simultaneous observations introduces a possible transit-depth
offset between the observations, further adding a degree of degeneracy
to the composition retrieval
constraints \citep{BarstowEtal2015mnrasTransmissionJWST}.

\begin{figure}
\centering
\includegraphics[width=\linewidth, clip]{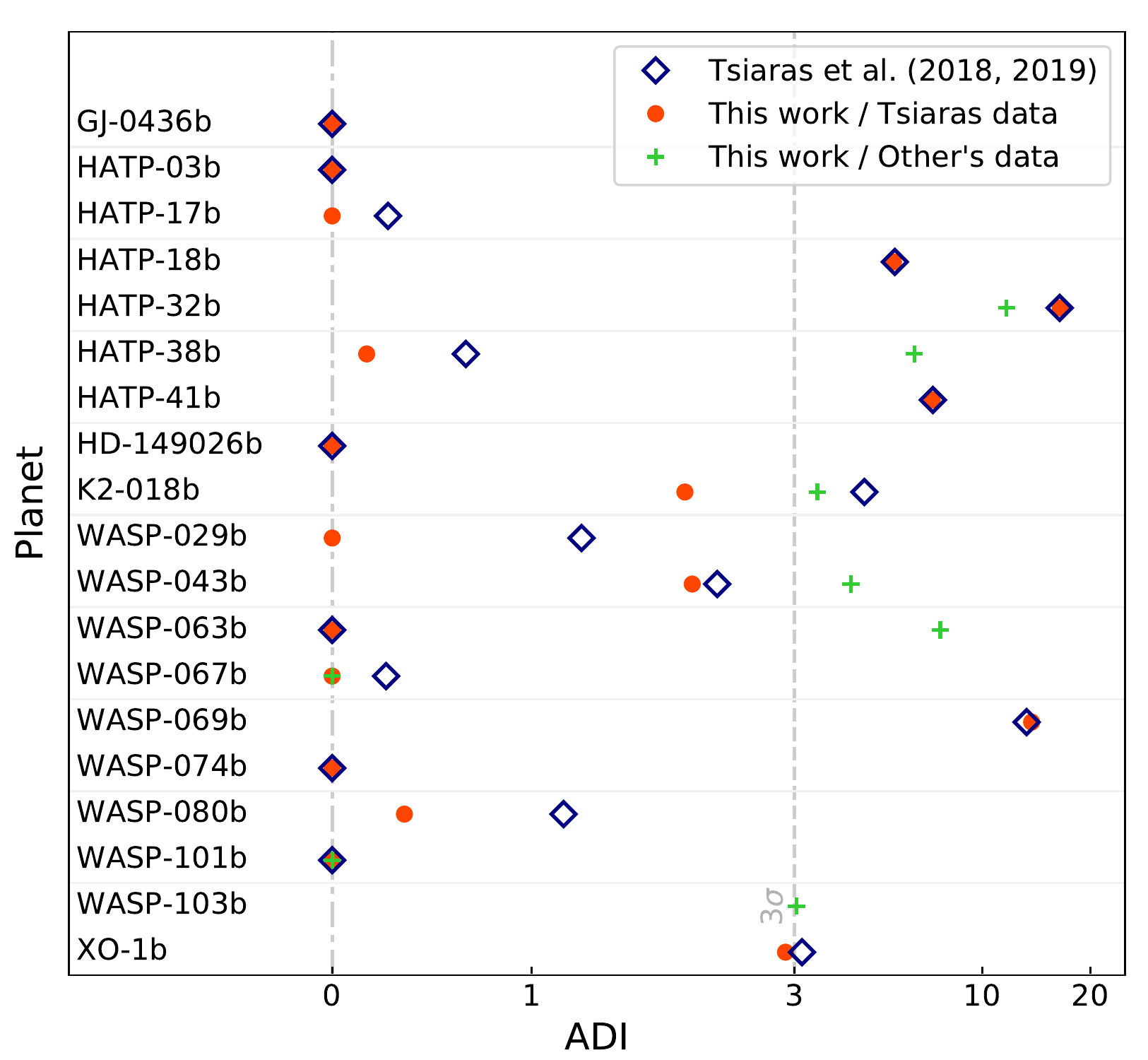}
\caption{
    Atmospheric Detectability Index for the datasets in our sample.
    The red circles and green crosses show our estimation of the ADI
    values for the datasets
    from \citet{TsiarasEtal2018ajPopulationStudy,
    TsiarasEtal2019natasK218b} and other groups, respectively.  The
    blue diamonds show the ADI values estimated
    by \citet{TsiarasEtal2018ajPopulationStudy,
    TsiarasEtal2019natasK218b}.  For our analyses, we have selected
    the lowest-BIC model for each dataset (see
    Tables \ref{table:model_comparison1}--\ref{table:model_comparison4}).
    Note that the X axis is linear for $0 < {\rm ADI} < 1$ and
    logarithmic for $1 < {\rm ADI}$.  The vertical dashed line marks
    the ADI level where the flat (featureless) fit is rejected at a
    3$\sigma$ significance.}
\label{fig:adi}
\end{figure}

\subsubsection{{\water} Detection}
\label{sec:adi_planets}

To further assess the significance of the spectral features seen in
our models, we compared these detections to a featureless model fit
using the Atmospheric Detectability Index ${\rm ADI} = \max\,\{0, \log
B\}$ \citep{TsiarasEtal2018ajPopulationStudy}, where $\log B$ is the
logarithmic Bayes factor \citep[see, e.g.,][]{Raftery1995BIC}:
\begin{equation}
\log B = \frac{1}{2} ({\rm BIC}_i - {\rm BIC\sb{flat}}),
\end{equation}
where BIC\sb{flat} is the Bayesian information criterion of a
flat-curve fit and BIC$_i$ is the Bayesian information criterion of
the best-fit model (for the optimal model setup according to
Tables \ref{table:model_comparison1}--\ref{table:model_comparison4},
i.e., the lowest-BIC model).

Figure \ref{fig:adi} shows our ADI values for the datasets in the
sample.  We find that 10 out of the 19 targets show strong evidence in
favor of molecular features over a flat spectrum \citep[${\rm
ADI} \gtrsim 3$, which corresponds to the $3\sigma$ criterion
of][]{TsiarasEtal2018ajPopulationStudy}.  In general, the targets that
have multiple light-curve analyses show similar ADI values.  However,
there are exceptional cases like HAT-P-38b and WASP-63b, where each
dataset produces widely different ADI values.  These results suggest
that although the WFC3/G141 data alone are often not sufficient to
distinguish between different scenarios, these observations can still
tell us whether there is evidence for molecular features in a
statistically robust manner.

\begin{figure}
\centering
\includegraphics[width=\linewidth, clip]{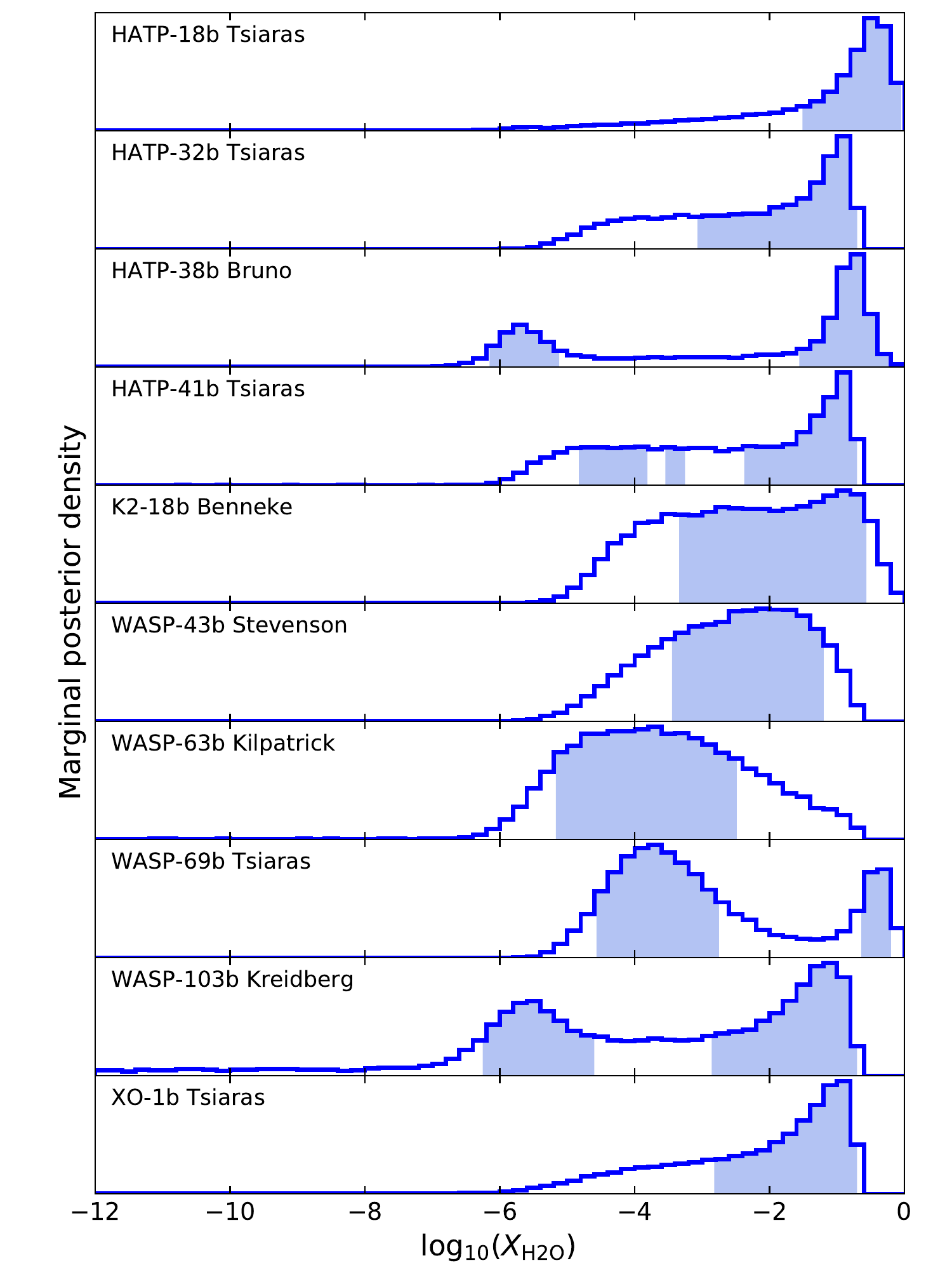}
\caption{
    {\water} volume mixing ratio marginal posterior for the targets
    with ${\rm ADI} \gtrsim 3$ (Fig.~\ref{fig:adi}).  The shaded areas
    denote the 68\% highest-posterior-density credible intervals.}
\label{fig:WFC3_H2O}
\end{figure}

For the rest of this section we will focus on the datasets with
significant spectral feature detections (i.e., ${\rm ADI} \gtrsim 3$),
since the datasets with ${\rm ADI}<3$ produced largely unconstrained
posteriors that provide little insight.
Table \ref{table:wfc3_retrieval} shows our retrieval results for the
10 targets with ${\rm ADI} \gtrsim 3$ (the highest ADI dataset and
lowest-BIC cloudy model for each target).  Figure \ref{fig:WFC3_H2O}
shows the marginal posterior of $X_{\rm H2O}$, the only molecule
consistently detected.  Due to the degeneracy between the model
parameters the $X_{\rm H2O}$ posteriors are asymmetric and the
credible intervals often span several orders of magnitude.  Since the
amplitude of the 1.4-{\micron} {\water} feature is often smaller than
that of a clear solar-composition atmosphere, most posteriors in
Fig.~\ref{fig:WFC3_H2O} show either very high $X_{\rm H2O}$ values
(small scale-height atmospheres), broad solar-to-supersolar $X_{\rm
H2O}$ values correlated with the {\pcloud} parameter (cloudy
atmospheres), sub-solar $X_{\rm H2O}$ values ($\water$-depleted
atmospheres), or a combination of them.  In particular, HAT-P-38b,
WASP-69b, and WASP-103b show bimodal $X_{\rm H2O}$ posterior
distributions.  The only other molecule observed in these datasets is
HCN for the WASP-63b analysis
of \citet{KilpatrickEtal2018apjWASP63bWFC3}.  For K2-18b, WASP-43b,
and WASP-103b, which have {\Spitzer} observations, we obtained flat or
upper-limit posteriors for $X_{\rm CO}$, $X_{\rm CO2}$, and $X_{\rm
CH4}$ (we will discuss these constraints in more detail in
Section \ref{sec:literature}).

\input{table_results}

\subsubsection{Comparison to Literature}
\label{sec:literature}

The ADI metric allows us to make a global comparison of our retrieval
results with those of \citet{TsiarasEtal2018ajPopulationStudy,
TsiarasEtal2019natasK218b}.  Note that ADI is a model-dependent
metric; even though two analyses may be fitting the same data, we do
not necessarily expect to find the same ADI results if the underlying
models differ.  Nevertheless, the ADI is a simple metric that can tell
whether different analyses find evidence for spectral features in a
same dataset.  Figure \ref{fig:adi} shows that the ADI values found in
this work agree relatively well with those
of \citeauthor{TsiarasEtal2018ajPopulationStudy} for most of the
samples, in particular for the most significant detections where ${\rm
ADI} \gtrsim 3$.

For individual retrieval analyses we see a good agreement between our
results and those from the literature, as long as both analyses adopt
a similar set of assumptions (i.e., same models to parameterize the
atmosphere, same set of parameters, and similar parameter priors and
boundaries).  We can make a direct comparison with the HAT-P-32b,
HAT-P-38b, and WASP-67b analyses \citep{DamianoEtal2017ajHAT32bWFC3,
BrunoEtal2018ajWASP67bHAT38bWFC3}, which have published pairwise
posterior distributions and adopt the same atmospheric
parameterization as we do.  In these cases, both retrieval analyses
find consistent best-fitting solutions, posterior distributions, and
correlations between the parameters.

Notably, HAT-P-38b shows seemingly anomalously high values for the
retrieved temperature (see Fig.\ \ref{fig:spectra1} or
Table \ref{table:wfc3_retrieval}).  This could be attributed to
systematics in the reduced data, reported
by \citet{BrunoEtal2018ajWASP67bHAT38bWFC3}, particularly for the
longer-wavelength data points.

{\citet{BennekeEtal2019apjK218b}} retrieved the WFC3 and {\Spitzer}
spectrum of K2-18b by fitting an isothermal temperature, a gray cloud
deck, and the volume mixing ratios of {\water}, {\methane}, CO,
{\carbdiox}, {\ammonia}, HCN, and N$_2$.  Qualitatively, our retrieval
results match those of \citet{BennekeEtal2019apjK218b}, finding a
strong correlation between $X_{\rm H2O}$ and the cloud top pressure
(see their Fig.~4).  We obtained posterior boundaries fairly similar
to those of \citet{BennekeEtal2019apjK218b}; at a 1$\sigma$
significance we constrained $X_{\rm H2O}$ between $1\tttt{-3}$ and
$0.3$ (they, between 3$\tttt{-4}$ and 0.1) and $\pcloud$ between
2$\tttt{-2}$ and 4 bar (they, between 8$\tttt{-3}$ and 0.1 bar).  For
the remaining molecular abundances we obtained 2$\sigma$ upper limits
of 0.04, 0.01, 8$\tttt{-4}$, and 0.09 for $X_{\rm CO}$, $X_{\rm CO2}$,
$X_{\rm CH4}$, and $X_{\rm N2}$, respectively,
whereas \citet{BennekeEtal2019apjK218b} found upper limits of 0.07,
0.02, 2$\tttt{-3}$, and 0.1, respectively.  The {\water} and cloud
detection make K2-18b an interesting candidate for further
characterization, since the planet receives a stellar insulation
similar to that of the Earth, enabling the possibility to characterize
water clouds.

{\citet{KreidbergEtal2014apjWASP43bWFC3}} analyzed the WFC3
transmission spectrum of WASP-43b.  Their retrieval parameterization
included an isothermal temperature, a gray cloud deck, a reference
pressure level ($\log p_0$), and the volume mixing ratios of {\water},
{\methane}, CO, and {\carbdiox}.  This modeling setup is equivalent to
ours, except that they fit for a reference pressure level ($\log p_0$)
instead of a reference radius ($R_{\rm p}$).  However, as mentioned in
Section \ref{sec:hd209458b_benckmark}, \citet{WelbanksMadhusudhan2019ajRetrievalDegeneracies}
showed that fixing $R_{\rm p}$ and fitting for $\log p_0$ is
equivalent to fixing $\log p_0$ and fitting for $R_{\rm p}$.
Qualitatively, our retrieval posteriors show the same correlations
found by \citet{KreidbergEtal2014apjWASP43bWFC3}, with $X_{\rm H2O}$
correlating mainly with the reference $R_{\rm p}$--$p_0$ point.  The
cloud top pressure is generally constrained below the pressures probed
by the observations.  Quantitatively, our 1$\sigma$ credible intervals
overlap with those of \citet{KreidbergEtal2014apjWASP43bWFC3}, though
our ranges are slightly offset from theirs.  We constrained $X_{\rm
H2O}$ between $4\tttt{-4}$ and $0.06$, $T$ between 360 and 660~K, and
$\pcloud>0.2$ bar (at 1$\sigma$);
whereas \citet{KreidbergEtal2014apjWASP43bWFC3} constrained $X_{\rm
H2O}$ between 3$\tttt{-5}$ and 5$\tttt{-3}$, T between 510 and 780~K,
and $\pcloud>0.1$ bar.

{\citet{KilpatrickEtal2018apjWASP63bWFC3}} present a further example
of consistent retrieval results between different frameworks. They
studied the {\HST} WFC3/G141 transmission spectrum of WASP-63b,
comparing our retrieval framework with
CHIMERA \citep{LineEtal2016ajHD209458bRetrieval},
{\taurex} \citep{Waldmann2016apjDreamingAtmospheres}, and
POSEIDON \citep{MacDonaldMadhusudhan2017mnrasRetrievalHD209b}.  Even
though the different frameworks used different statistical samplers,
cross section databases, and atmospheric parameterizations, all four
retrieval analyses produced consistent results, finding a strong
detection of the {\water} feature. The amplitude of the {\water}
features is however muted when compared to that of a clear
solar-composition atmospheric model.  A weaker detection of an
absorption feature is also observed at 1.55 {\micron}.  This feature
is consistent with a super-solar abundance of HCN that would require
the presence of strong disequilibrium-chemistry processes like
quenching and photochemistry \citep[see,
e.g.,][]{Moses2014rsptaChemicalKinetics}.

For WASP-103b it is not possible to make a direct comparison
with \citet{KreidbergEtal2018apjWASP103bPhaseCurve}, since they did
not retrieve the transmission spectrum, but rather analyzed the
phase-curve emission spectrum of the planet.  Under the assumption of
thermochemical and radiative-convective equilibrium, they found that
WASP-103b does not show {\water} spectral features in emission,
attributed to {\water} dissociation or additional absorption from
other species like H$^-$.  The emission spectra are consistent with
blackbody emission ranging from 1900~K (at the night side) to 2900~K
(at the day side).  From our transmission spectrum retrieval, we found
a temperature posterior that is consistent with the range of
temperatures obtained
by \citet{KreidbergEtal2018apjWASP103bPhaseCurve}, although very
broad, $T\sim$2100--3000~K.  Our retrieved {\water} composition is
affected by strong parameter correlations as well as by the choice of
retrieval parameters: when we included all abundance free parameters
in the retrieval ($X_{\rm H2O}$, $X_{\rm CO}$, $X_{\rm CO2}$, $X_{\rm
CH4}$, and $X_{\rm N2}$) we obtained a broad $X_{\rm H2O}$ posterior
peaking at super-solar values (see Fig.~\ref{fig:spectra4}); however,
when we retrieved the abundance of only {\water} the $X_{\rm H2O}$
posterior showed a bimodal $X_{\rm H2O}$ distribution, peaking at
sub-solar and super-solar compositions (see Fig.~\ref{fig:WFC3_H2O}
and Table \ref{table:wfc3_retrieval}).  Our lower {\water}-abundance
solution mode ($X_{\rm H2O} \approx 3\tttt{-6}$) is more consistent
with the emission models
of \citet{KreidbergEtal2018apjWASP103bPhaseCurve}.  Although, this
should not be considered a direct comparison since transmission and
emission spectroscopy follow a different geometry and, thus, probe
different regions in the atmosphere.

{\citet{FisherHeng2018mnrasWFC3SampleRetrieval}} performed a
comprehensive analysis of 38 WFC3 exoplanet transmission spectra, with
several targets in common with those in our sample.  Their atmospheric
models fit the same components as ours, however, their formulation of
the radiative transfer equation differs from ours.  They employed a
semi-analytic model evaluating physical properties at a single
pressure level.  They also, tested a suit of clear, gray, and non-gray
cloudy models.  For the molecular cross sections, they included the
contribution from {\water}, HCN, and
{\ammonia}. \citet{FisherHeng2018mnrasWFC3SampleRetrieval} found that
for the majority of targets the Bayesian evidence favors isothermal
temperature profiles and the presence of gray clouds.  {\water} is the
only molecule consistently detected.  They found no evidence of trends
between the retrieved {\water} abundances and other physical
parameters, like the planetary mass or temperature.  Considering the
targets in common, our retrieved $X_{\rm H2O}$ values are generally
consistent (within the 68\% credible intervals) with those
of \citet{FisherHeng2018mnrasWFC3SampleRetrieval}.  We also observed
no clear trend between physical parameters.

Overall, the main conclusion from these analyses is that {\water} is
the only molecule consistently detected with WFC3 transmission data.
However, with only WFC3 observations, the $X_{\rm H2O}$ posterior
distributions are highly correlated with other atmospheric parameters,
like the cloud top pressure or the pressure--radius reference point.
These degeneracies lead to broad, poorly constrained {\water}
abundances.

\section{Summary}
\label{sec:summary}

In this article, we presented the {\pyratbay} framework for exoplanet
atmospheric characterization.  The modular design of {\pyratbay}
allows the users to carry out specific or general atmospheric modeling
tasks, that is, compute 1D atmospheric models, sample line-by-line
molecular cross sections, compute transmission or emission spectra,
and perform Bayesian atmospheric retrievals.  {\pyratbay} is an
open-source (GNU GPL v2 license), well documented, and thoroughly
tested Python package (compatible with Python versions 3.6 and above).
The code is readily available for installation from PyPI
(\mintinline{shell}{pip install pyratbay}) or conda
(\mintinline{shell}{conda install -c conda-forge pyratbay}).  The
source code is hosted at \href{http://pcubillos.github.io/pyratbay}
{http://pcubillos.github.io/pyratbay} and the documentation is located
at \href{https://pyratbay.readthedocs.io}
{https://pyratbay.readthedocs.io}.  We benchmarked the individual
components of the {\pyratbay} framework (line sampling, spectral
synthesis, and atmospheric retrieval) by successfully reproducing the
line-by-line sampling cross sections
of \citet{ChubbEtal2021aaExoMolOP}, by reproducing {\petit}
transmission and emission spectra, and by retrieving synthetic
{\ariel}-like transmission and emission spectra generated with
{\taurex}.  We also re-analyzed with {\pyratbay} the
0.3--5.5~{\micron} transmission spectrum of {\twoohnineb} studied
by \citet{MacDonaldMadhusudhan2017mnrasRetrievalHD209b}
and \citet{PinhasEtal2019mnrasHotJupiterSampleRetrieval}, finding a
good agreement with them for most retrieved parameters.

We also initiated a population study of the existing exoplanet
transmission spectra obtained from space-based observatories, where we
applied a standardized retrieval approach.  As a starting point, in
this article we considered the targets without observations at
$\lambda<1$ {\micron}.  This sample consists of 26 datasets of 19
different systems, most of them observed exclusively with the {\HST}
WFC3/G141 detector (1.1--1.7 {\microns}).  Thus, we focused our
analysis on the characterization of the {\water} absorption band
centered at $1.4$ {\microns}.  A Bayesian model comparison indicates
that 9 out of the 19 systems show strong evidence for {\water}
molecular features in the data.  However, it is not possible to obtain
precise {\water} abundance constraints.  The limited wavelength
coverage and signal-to-noise ratio of the data do not allow us to
distinguish cloudy from high metal-mass-fraction scenarios
($\Delta$BIC<2, see
Tables \ref{table:model_comparison1}--\ref{table:model_comparison4}),
leading to degenerate posterior distributions.  Our results are well
in agreement with previous studies of these datasets, particularly
when both retrieval analyses adopt similar assumptions.

Follow up articles will present the atmospheric retrieval analysis of
planets that have been observed in transmission at shorter
wavelengths.  Ongoing development of the {\pyratbay} framework
includes self-consistent radiative- and thermochemical-equilibrium
atmospheric modeling, posterior sampling under thermochemical
equilibrium, sophisticated kinetic and thermal-stability cloud models,
and physically-consistent multi-dimensional retrievals for phase-curve
observations.

In conclusion, current atmospheric retrievals rely on a number of
modeling choices, either to simplify the already computationally
intensive calculations (e.g., assuming 1D atmospheric
parameterizations) or to simplify the parameter space exploration
(e.g., assuming thermochemical-equilibrium abundances).  These choices
can have a significant, or even dominating, influence on the retrieval
results.  Carrying out independent retrieval analyses of a given
observation help us test the impact of the model assumptions, making
our interpretation of the observation more robust.

Despite the limited characterizing power of the WFC3/G141 grism alone,
these observations are highly valuable to determine the targets with
the most favorable conditions for further characterization.  We expect
that better targets with stronger features and additional spectral
coverage will enable us to perform more in-depth population
studies \citep[e.g.,][]{SingEtal2016natHotJupiterTransmission,
CrossfieldKreidberg2017ajNeptuneTrends}.  In
particular, \citet{PinhasEtal2019mnrasHotJupiterSampleRetrieval}
showed that by combining infrared WFC3 data with optical {\HST}/STIS
data helps to break the degeneracy between the reference
pressure--radius point and the {\water} abundance.  For the
transmission spectrum of HD~209458\,b, the retrieval of the combined
STIS and WFC3 data lead to an uncorrelated $X_{\rm H2O}$ constraint,
three times tighter than when retrieving the WFC3 data alone.
Observations from future telescopes (e.g., {\JWST} or {\ariel}), with
better sensitivity, wavelength coverage, and spectral resolution will
be crucial to construct a more robust picture of the physical
processes dominating exoplanet atmospheres.  For
example, \citet{GreeneEtal2016apjJWSTtransitCharacterization}
estimated that {\JWST} will be able to discriminate between cloudy and
high metal-mass-fraction atmospheres.  {\JWST} observations between
1--2.5 {\microns} will often suffice to constrain the volume mixing
ratios of dominant molecular species, C/O ratios, and metallicity of
clear planetary atmospheres.  A broader wavelength coverage (1--11
{\microns}) provides significantly better constraints and has the
potential to constrain the properties of cloudy atmospheres.

\section*{Data Availability}

To allow the readers to fully reproduce the analyses presented in this
article, we provide a research compendium available in Zenodo
at \href{https://doi.org/10.5281/zenodo.4749530}
{https://doi.org/10.5281/zenodo.4749530}.

\section*{Acknowledgments}

We thank the anonymous referee for his/her time and valuable comments.
We thank contributors to the Python Programming Language and the free
and open-source community (see Software Section below).  J. B. was
partially supported by NASA through the NASA ROSES-2016/Exoplanets
Research Program, grant NNX17AC03G.  We drafted this article using the
MNRAS latex template
(\href{https://www.ctan.org/tex-archive/macros/latex/contrib/mnras}
{https://www.ctan.org/tex-archive/macros/latex/contrib/mnras}, version
3.1), with further style modifications that are available
at \href{https://github.com/pcubillos/ApJtemplate}
{https://github.com/pcubillos/ApJtemplate}.  This work has made use of
the VALD database, operated at Uppsala University, the Institute of
Astronomy RAS in Moscow, and the University of Vienna.  Part of this
work is based on observations made with the NASA/ESA Hubble Space
Telescope, obtained from the data archive at the Space Telescope
Science Institute.  STScI is operated by the Association of
Universities for Research in Astronomy, Inc. under NASA contract NAS
5-26555.  This work is based in part on observations made with the
{\SST}, which is operated by the Jet Propulsion Laboratory, California
Institute of Technology under a contract with NASA.  This research has
made use of NASA's Astrophysics Data System Bibliographic Services.

The following software and packages were used in this work:
{\pyratbay} (\href{https://pyratbay.readthedocs.io/en/latest/}
                  {https://pyratbay.readthedocs.io}),
{\MCC} \citep{CubillosEtal2017apjRednoise},
{\TEA} \citep{BlecicEtal2016apsjTEA},
{\repack} \citep{Cubillos2017apjRepack},
{\rate} \citep{CubillosEtal2019apjRate},
{\taurex} \citep{AlRefaieEtal2019arxivTaurex3},
{\petit} \citep{MolliereEtal2019aaPetitRADTRANS},
\textsc{Numpy} \citep{vanderWaltEtal2011numpy},
\textsc{SciPy} \citep{JonesEtal2001scipy},
\textsc{sympy} \citep{MeurerEtal2017pjcsSYMPY},
\textsc{Astropy} \citep{AstropyCollaboration2013aaAstropy,
    AstropyCollaboration2018ajAstropy},
\textsc{Matplotlib} \citep{Hunter2007ieeeMatplotlib},
\textsc{IPython} \citep{PerezGranger2007cseIPython},
and \textsc{bibmanager}\footnote{
\href{http://pcubillos.github.io/bibmanager}
     {http://pcubillos.github.io/bibmanager}}
\citep{Cubillos2019zndoBibmanager}.

\bibliographystyle{mnras}
\bibliography{pyratbayI}

\appendix
\section{Model Comparison Statistics}

Tables \ref{table:model_comparison1}--\ref{table:model_comparison4} show
the model-comparison statistics for the WFC3 sample.

{\renewcommand{\arraystretch}{1.0}
\setlength{\tabcolsep}{3pt}
\begin{table}
\centering
\caption{
Model comparison statistics. $\Delta$BIC is the difference in BIC with
respect to the lowest-BIC (optimal) model for the given
dataset. $N\sb{\rm free}$ is the number of free retrieval parameters.}
\label{table:model_comparison1}
\begin{tabular}{@{\extracolsep{\fill}}c@{\hskip 6pt}lccc}
\multicolumn{5}{l}{ GJ-436b Tsiaras }\\ 
\hline
ID & Free parameters & $N\sb{\rm free}$ & $\Delta$BIC & \redchisq \\
\hline 
M1 & $T$, $R\sb{\rm p}$, {\water}, {\pcloud}            & 4 & $ 6.8$ & $1.33\pm0.38$ \\ 
M2 & $T$, $R\sb{\rm p}$, {\water}, [M/H]                & 4 & $ 6.8$ & $1.33\pm0.38$ \\ 
M3 & $T$, $R\sb{\rm p}$, {\water}, [M/H], {\pcloud}     & 5 & $ 9.7$ & $1.43\pm0.39$ \\ 
M0 & Flat fit                                           & 1 & $ 0.0$ & $1.21\pm0.34$ \\ 
\hline
\\ 
\multicolumn{5}{l}{ HATP-3b Tsiaras }\\
\hline
ID & Free parameters & $N\sb{\rm free}$ & $\Delta$BIC & \redchisq \\
\hline
M1 & $T$, $R\sb{\rm p}$, {\water}, {\pcloud}            & 4 & $ 7.6$ & $1.07\pm0.38$ \\
M2 & $T$, $R\sb{\rm p}$, {\water}, [M/H]                & 4 & $ 8.2$ & $1.11\pm0.38$ \\
M3 & $T$, $R\sb{\rm p}$, {\water}, [M/H], {\pcloud}     & 5 & $10.5$ & $1.15\pm0.39$ \\
M0 & Flat fit                                           & 1 & $ 0.0$ & $0.94\pm0.34$ \\
\hline
\\
\multicolumn{5}{l}{ HATP-17b Tsiaras }\\
\hline
ID & Free parameters & $N\sb{\rm free}$ & $\Delta$BIC & \redchisq \\
\hline
M1 & $T$, $R\sb{\rm p}$, {\water}, {\pcloud}            & 4 & $ 5.1$ & $1.34\pm0.38$ \\
M2 & $T$, $R\sb{\rm p}$, {\water}, [M/H]                & 4 & $ 5.1$ & $1.34\pm0.38$ \\
M3 & $T$, $R\sb{\rm p}$, {\water}, [M/H], {\pcloud}     & 5 & $ 8.0$ & $1.45\pm0.39$ \\
M0 & Flat fit                                           & 1 & $ 0.0$ & $1.32\pm0.34$ \\
\hline
\\
\multicolumn{5}{l}{ HATP-18b tsiaras }\\
\hline
ID & Free parameters & $N\sb{\rm free}$ & $\Delta$BIC & \redchisq \\
\hline
M1 & $T$, $R\sb{\rm p}$, {\water}, {\pcloud}            & 4 & $ 0.0$ & $1.05\pm0.38$ \\
M2 & $T$, $R\sb{\rm p}$, {\water}, [M/H]                & 4 & $ 0.0$ & $1.05\pm0.38$ \\
M3 & $T$, $R\sb{\rm p}$, {\water}, [M/H], {\pcloud}     & 5 & $ 2.9$ & $1.13\pm0.39$ \\
M0 & Flat fit                                           & 1 & $11.3$ & $2.04\pm0.34$ \\
\hline
\\
\multicolumn{5}{l}{ HATP-32b Damiano }\\
\hline
ID & Free parameters & $N\sb{\rm free}$ & $\Delta$BIC & \redchisq \\
\hline
M1 & $T$, $R\sb{\rm p}$, {\water}, {\pcloud}            & 4 & $ 0.0$ & $1.38\pm0.35$ \\
M2 & $T$, $R\sb{\rm p}$, {\water}, [M/H]                & 4 & $ 0.0$ & $1.38\pm0.35$ \\
M3 & $T$, $R\sb{\rm p}$, {\water}, [M/H], {\pcloud}     & 5 & $ 3.0$ & $1.48\pm0.37$ \\
M0 & Flat fit                                           & 1 & $23.4$ & $2.87\pm0.32$ \\
\hline
\\
\multicolumn{5}{l}{ HATP-32b Tsiaras }\\
\hline
ID & Free parameters & $N\sb{\rm free}$ & $\Delta$BIC & \redchisq \\
\hline
M1 & $T$, $R\sb{\rm p}$, {\water}, {\pcloud}            & 4 & $ 0.0$ & $0.92\pm0.38$ \\
M2 & $T$, $R\sb{\rm p}$, {\water}, [M/H]                & 4 & $ 0.0$ & $0.92\pm0.38$ \\
M3 & $T$, $R\sb{\rm p}$, {\water}, [M/H], {\pcloud}     & 5 & $ 2.9$ & $0.99\pm0.39$ \\
M0 & Flat fit                                           & 1 & $33.0$ & $3.21\pm0.34$ \\
\hline
\\
\multicolumn{5}{l}{ HATP-38b bruno }\\
\hline
ID & Free parameters & $N\sb{\rm free}$ & $\Delta$BIC & \redchisq \\
\hline
M1 & $T$, $R\sb{\rm p}$, {\water}, {\pcloud}            & 4 & $ 0.0$ & $1.33\pm0.43$ \\
M2 & $T$, $R\sb{\rm p}$, {\water}, [M/H]                & 4 & $ 0.0$ & $1.33\pm0.43$ \\
M3 & $T$, $R\sb{\rm p}$, {\water}, [M/H], {\pcloud}     & 5 & $ 2.8$ & $1.47\pm0.45$ \\
M0 & Flat fit                                           & 1 & $13.0$ & $2.55\pm0.38$ \\
\hline
\end{tabular}
\end{table}
\normalsize
}

{\renewcommand{\arraystretch}{1.0}
\setlength{\tabcolsep}{3pt}
\begin{table}
\centering
\caption{Model comparison statistics (continuation).}
\label{table:model_comparison2}
\begin{tabular}{@{\extracolsep{\fill}}c@{\hskip 6pt}lccc}
\multicolumn{5}{l}{ HATP-38b tsiaras }\\
\hline
ID & Free parameters & $N\sb{\rm free}$ & $\Delta$BIC & \redchisq \\
\hline
M1 & $T$, $R\sb{\rm p}$, {\water}, {\pcloud}            & 4 & $ 0.0$ & $0.80\pm0.38$ \\
M2 & $T$, $R\sb{\rm p}$, {\water}, [M/H]                & 4 & $ 0.1$ & $0.81\pm0.38$ \\
M3 & $T$, $R\sb{\rm p}$, {\water}, [M/H], {\pcloud}     & 5 & $ 3.0$ & $0.87\pm0.39$ \\
M0 & Flat fit                                           & 1 & $ 0.3$ & $1.19\pm0.34$ \\
\hline
\\
\multicolumn{5}{l}{ HATP-41b Tsiaras }\\
\hline
ID & Free parameters & $N\sb{\rm free}$ & $\Delta$BIC & \redchisq \\
\hline
M1 & $T$, $R\sb{\rm p}$, {\water}, {\pcloud}            & 4 & $ 0.0$ & $0.32\pm0.38$ \\
M2 & $T$, $R\sb{\rm p}$, {\water}, [M/H]                & 4 & $ 1.6$ & $0.43\pm0.38$ \\
M3 & $T$, $R\sb{\rm p}$, {\water}, [M/H], {\pcloud}     & 5 & $ 2.9$ & $0.34\pm0.39$ \\
M0 & Flat fit                                           & 1 & $14.5$ & $1.62\pm0.34$ \\
\hline
\\
\multicolumn{5}{l}{ HD 149026 b Tsiaras }\\
\hline
ID & Free parameters & $N\sb{\rm free}$ & $\Delta$BIC & \redchisq \\
\hline
M1 & $T$, $R\sb{\rm p}$, {\water}, {\pcloud}            & 4 & $ 4.1$ & $0.85\pm0.38$ \\
M2 & $T$, $R\sb{\rm p}$, {\water}, [M/H]                & 4 & $ 4.4$ & $0.87\pm0.38$ \\
M3 & $T$, $R\sb{\rm p}$, {\water}, [M/H], {\pcloud}     & 5 & $ 6.9$ & $0.91\pm0.39$ \\
M0 & Flat fit                                           & 1 & $ 0.0$ & $0.97\pm0.34$ \\
\hline
\\ 
\multicolumn{5}{l}{ K2-18b Benneke }\\ 
\hline
ID & Free parameters & $N\sb{\rm free}$ & $\Delta$BIC & \redchisq \\ 
\hline
M1 & $T$, $R\sb{\rm p}$, {\water}, {\pcloud}            & 4 & $ 0.0$ & $0.57\pm0.37$ \\
M2 & $T$, $R\sb{\rm p}$, {\water}, [M/H]                & 4 & $ 0.0$ & $0.57\pm0.37$ \\
M3 & $T$, $R\sb{\rm p}$, {\water}, [M/H], {\pcloud}     & 5 & $ 2.8$ & $0.60\pm0.38$ \\
M4 & $T$, $R\sb{\rm p}$, {\water}, CO, {\carbdiox}, {\methane}, {\pcloud} & 7 & $ 8.9$ & $0.72\pm0.41$ \\
M5 & $T$, $R\sb{\rm p}$, {\water}, CO, {\carbdiox}, {\methane}, [M/H] & 7 & $ 9.0$ & $0.73\pm0.41$ \\
M6 & $T$, $R\sb{\rm p}$, {\water}, CO, {\carbdiox}, {\methane}, [M/H], {\pcloud} & 8 & $11.7$ & $0.77\pm0.43$ \\
M0 & Flat fit                                           & 1 & $ 7.0$ & $1.35\pm0.33$ \\
\hline
\\ 
\multicolumn{5}{l}{ K2-18b tsiaras }\\
\hline
ID & Free parameters & $N\sb{\rm free}$ & $\Delta$BIC & \redchisq \\
\hline
M1 & $T$, $R\sb{\rm p}$, {\water}, {\pcloud}            & 4 & $ 0.0$ & $0.98\pm0.39$ \\
M2 & $T$, $R\sb{\rm p}$, {\water}, [M/H]                & 4 & $ 0.4$ & $1.01\pm0.39$ \\
M3 & $T$, $R\sb{\rm p}$, {\water}, [M/H], {\pcloud}     & 5 & $ 2.8$ & $1.06\pm0.41$ \\
M0 & Flat fit                                           & 1 & $ 4.0$ & $1.57\pm0.35$ \\
\hline
\\ 
\multicolumn{5}{l}{ WASP-29b Tsiaras }\\ 
\hline
ID & Free parameters & $N\sb{\rm free}$ & $\Delta$BIC & \redchisq \\
\hline
M1 & $T$, $R\sb{\rm p}$, {\water}, {\pcloud}            & 4 & $ 2.4$ & $0.58\pm0.38$ \\
M2 & $T$, $R\sb{\rm p}$, {\water}, [M/H]                & 4 & $ 2.4$ & $0.58\pm0.38$ \\
M3 & $T$, $R\sb{\rm p}$, {\water}, [M/H], {\pcloud}     & 5 & $ 5.3$ & $0.62\pm0.39$ \\
M0 & Flat fit                                           & 1 & $ 0.0$ & $0.85\pm0.34$ \\
\hline
\end{tabular}
\end{table}
\normalsize
}

{\renewcommand{\arraystretch}{1.0}
\setlength{\tabcolsep}{3pt}
\begin{table}
\centering
\caption{Model comparison statistics (continuation).}
\label{table:model_comparison3}
\begin{tabular}{@{\extracolsep{\fill}}c@{\hskip 6pt}lccc}
 
\multicolumn{5}{l}{ WASP-43b Stevenson }\\ 
\hline
ID & Free parameters & $N\sb{\rm free}$ & $\Delta$BIC & \redchisq \\ 
\hline 
M1 & $T$, $R\sb{\rm p}$, {\water}, {\pcloud}            & 4 & $ 0.0$ & $2.64\pm0.32$ \\
M2 & $T$, $R\sb{\rm p}$, {\water}, [M/H]                & 4 & $ 0.2$ & $2.65\pm0.32$ \\
M3 & $T$, $R\sb{\rm p}$, {\water}, [M/H], {\pcloud}     & 5 & $ 3.2$ & $2.78\pm0.32$ \\
M4 & $T$, $R\sb{\rm p}$, {\water}, CO, {\carbdiox}, {\methane}, {\pcloud} & 7 & $ 9.6$ & $3.11\pm0.34$ \\
M5 & $T$, $R\sb{\rm p}$, {\water}, CO, {\carbdiox}, {\methane}, [M/H]     & 7 & $ 9.7$ & $3.12\pm0.34$ \\
M6 & $T$, $R\sb{\rm p}$, {\water}, CO, {\carbdiox}, {\methane}, [M/H], {\pcloud} & 8 & $12.8$ & $3.31\pm0.35$ \\
M0 & Flat fit                                           & 1 & $ 8.6$ & $3.09\pm0.29$ \\
\hline
\\
\multicolumn{5}{l}{ WASP-43b Tsiaras }\\ 
\hline
ID & Free parameters & $N\sb{\rm free}$ & $\Delta$BIC & \redchisq \\
\hline
M1 & $T$, $R\sb{\rm p}$, {\water}, {\pcloud}            & 4 & $ 0.0$ & $1.53\pm0.38$ \\
M2 & $T$, $R\sb{\rm p}$, {\water}, [M/H]                & 4 & $ 0.0$ & $1.53\pm0.38$ \\
M3 & $T$, $R\sb{\rm p}$, {\water}, [M/H], {\pcloud}     & 5 & $ 2.9$ & $1.64\pm0.39$ \\
M0 & Flat fit                                           & 1 & $ 3.6$ & $1.98\pm0.34$ \\
\hline
\\
\multicolumn{5}{l}{ WASP-63b Kilpatrick }\\ 
\hline
ID & Free parameters & $N\sb{\rm free}$ & $\Delta$BIC & \redchisq \\ 
\hline 
M1 & $T$, $R\sb{\rm p}$, {\water}, {\pcloud}            & 4 & $ 5.5$ & $2.04\pm0.43$ \\
M2 & $T$, $R\sb{\rm p}$, {\water}, [M/H]                & 4 & $ 6.0$ & $2.08\pm0.43$ \\
M3 & $T$, $R\sb{\rm p}$, {\water}, [M/H], {\pcloud}     & 5 & $ 8.2$ & $2.24\pm0.45$ \\
M4 & $T$, $R\sb{\rm p}$, {\water}, HCN, {\pcloud}       & 5 & $ 0.0$ & $1.42\pm0.45$ \\
M5 & $T$, $R\sb{\rm p}$, {\water}, HCN, [M/H]           & 5 & $ 1.9$ & $1.62\pm0.45$ \\
M6 & $T$, $R\sb{\rm p}$, {\water}, HCN, [M/H], {\pcloud} & 6 & $ 2.7$ & $1.58\pm0.47$ \\
M0 & Flat fit                                           & 1 & $15.3$ & $2.88\pm0.38$ \\

\hline
\\ 
\multicolumn{5}{l}{ WASP-63b Tsiaras }\\ 
\hline
ID & Free parameters & $N\sb{\rm free}$ & $\Delta$BIC & \redchisq \\ 
\hline 
M1 & $T$, $R\sb{\rm p}$, {\water}, {\pcloud}            & 4 & $ 3.6$ & $0.99\pm0.38$ \\
M2 & $T$, $R\sb{\rm p}$, {\water}, [M/H]                & 4 & $ 4.0$ & $1.02\pm0.38$ \\
M3 & $T$, $R\sb{\rm p}$, {\water}, [M/H], {\pcloud}     & 5 & $ 6.5$ & $1.07\pm0.39$ \\
M1 & $T$, $R\sb{\rm p}$, {\water}, HCN, [M/H], {\pcloud} & 6 & $ 8.8$ & $1.11\pm0.41$ \\
M0 & Flat fit                                           & 1 & $ 0.0$ & $1.11\pm0.34$ \\
\hline
\\ 
\multicolumn{5}{l}{ WASP-67b Bruno }\\ 
\hline
ID & Free parameters & $N\sb{\rm free}$ & $\Delta$BIC & \redchisq \\
\hline
M1 & $T$, $R\sb{\rm p}$, {\water}, {\pcloud}            & 4 & $ 1.3$ & $1.32\pm0.41$ \\
M2 & $T$, $R\sb{\rm p}$, {\water}, [M/H]                & 4 & $ 1.3$ & $1.32\pm0.41$ \\
M3 & $T$, $R\sb{\rm p}$, {\water}, [M/H], {\pcloud}     & 5 & $ 4.1$ & $1.44\pm0.43$ \\
M0 & Flat fit                                           & 1 & $ 0.0$ & $1.52\pm0.37$ \\
\hline
\\
\multicolumn{5}{l}{ WASP-67b Tsiaras }\\
\hline
ID & Free parameters & $N\sb{\rm free}$ & $\Delta$BIC & \redchisq \\
\hline
M1 & $T$, $R\sb{\rm p}$, {\water}, {\pcloud}            & 4 & $ 2.5$ & $0.57\pm0.38$ \\
M2 & $T$, $R\sb{\rm p}$, {\water}, [M/H]                & 4 & $ 2.5$ & $0.57\pm0.38$ \\
M3 & $T$, $R\sb{\rm p}$, {\water}, [M/H], {\pcloud}     & 5 & $ 5.4$ & $0.62\pm0.39$ \\
M0 & Flat fit                                           & 1 & $ 0.0$ & $0.84\pm0.34$ \\
\hline
\end{tabular}
\end{table}
\normalsize
}

{\renewcommand{\arraystretch}{1.0}
\setlength{\tabcolsep}{3pt}
\begin{table}
\centering
\caption{Model comparison statistics (continuation).}
\label{table:model_comparison4}
\begin{tabular}{@{\extracolsep{\fill}}c@{\hskip 6pt}lccc}

\multicolumn{5}{l}{ WASP-69b tsiaras }\\
\hline
ID & Free parameters & $N\sb{\rm free}$ & $\Delta$BIC & \redchisq \\
\hline
M1 & $T$, $R\sb{\rm p}$, {\water}, {\pcloud}            & 4 & $ 0.1$ & $1.65\pm0.38$ \\
M2 & $T$, $R\sb{\rm p}$, {\water}, [M/H]                & 4 & $ 0.0$ & $1.65\pm0.38$ \\
M3 & $T$, $R\sb{\rm p}$, {\water}, [M/H], {\pcloud}     & 5 & $ 2.8$ & $1.77\pm0.39$ \\
M0 & Flat fit                                           & 1 & $27.5$ & $3.48\pm0.34$ \\
\hline
\\
\multicolumn{5}{l}{ WASP-74b Tsiaras }\\
\hline
ID & Free parameters & $N\sb{\rm free}$ & $\Delta$BIC & \redchisq \\
\hline
M1 & $T$, $R\sb{\rm p}$, {\water}, {\pcloud}            & 4 & $ 2.9$ & $0.89\pm0.38$ \\
M2 & $T$, $R\sb{\rm p}$, {\water}, [M/H]                & 4 & $ 2.8$ & $0.89\pm0.38$ \\
M3 & $T$, $R\sb{\rm p}$, {\water}, [M/H], {\pcloud}     & 5 & $ 5.7$ & $0.96\pm0.39$ \\
M0 & Flat fit                                           & 1 & $ 0.0$ & $1.07\pm0.34$ \\
\hline
\\
\multicolumn{5}{l}{ WASP-80b Tsiaras }\\
\hline
ID & Free parameters & $N\sb{\rm free}$ & $\Delta$BIC & \redchisq \\
\hline
M1 & $T$, $R\sb{\rm p}$, {\water}, {\pcloud}            & 4 & $ 0.0$ & $1.02\pm0.39$ \\
M2 & $T$, $R\sb{\rm p}$, {\water}, [M/H]                & 4 & $ 0.3$ & $1.04\pm0.39$ \\
M3 & $T$, $R\sb{\rm p}$, {\water}, [M/H], {\pcloud}     & 5 & $ 2.8$ & $1.10\pm0.41$ \\
M0 & Flat fit                                           & 1 & $ 0.7$ & $1.40\pm0.35$ \\
\hline
\\
\multicolumn{5}{l}{ WASP-101b Tsiaras }\\ 
\hline
ID & Free parameters & $N\sb{\rm free}$ & $\Delta$BIC & \redchisq \\
\hline
M1 & $T$, $R\sb{\rm p}$, {\water}, {\pcloud}            & 4 & $ 2.5$ & $1.57\pm0.38$ \\
M2 & $T$, $R\sb{\rm p}$, {\water}, [M/H]                & 4 & $ 2.4$ & $1.57\pm0.38$ \\
M3 & $T$, $R\sb{\rm p}$, {\water}, [M/H], {\pcloud}     & 5 & $ 5.3$ & $1.69\pm0.39$ \\
M0 & Flat fit                                           & 1 & $ 0.0$ & $1.66\pm0.34$ \\
\hline
\\
\multicolumn{5}{l}{ WASP-101b Wakeford }\\
\hline
ID & Free parameters & $N\sb{\rm free}$ & $\Delta$BIC & \redchisq \\
\hline
M1 & $T$, $R\sb{\rm p}$, {\water}, {\pcloud}            & 4 & $ 5.0$ & $2.76\pm0.43$ \\
M2 & $T$, $R\sb{\rm p}$, {\water}, [M/H]                & 4 & $ 4.5$ & $2.72\pm0.43$ \\
M3 & $T$, $R\sb{\rm p}$, {\water}, [M/H], {\pcloud}     & 5 & $ 7.6$ & $3.03\pm0.45$ \\
M0 & Flat fit                                           & 1 & $ 0.0$ & $2.40\pm0.38$ \\
\hline
\\ 
\multicolumn{5}{l}{ WASP-103b Kreidberg }\\ 
\hline
ID & Free parameters & $N\sb{\rm free}$ & $\Delta$BIC & \redchisq \\ 
\hline 
M1 & $T$, $R\sb{\rm p}$, {\water}, {\pcloud}            & 4 & $ 0.0$ & $1.40\pm0.50$ \\
M2 & $T$, $R\sb{\rm p}$, {\water}, [M/H]                & 4 & $ 0.0$ & $1.40\pm0.50$ \\
M3 & $T$, $R\sb{\rm p}$, {\water}, [M/H], {\pcloud}     & 5 & $ 2.5$ & $1.60\pm0.53$ \\
M4 & $T$, $R\sb{\rm p}$, {\water}, CO, {\carbdiox}, {\methane}, {\pcloud} & 7 & $ 6.6$ & $2.07\pm0.63$ \\
M5 & $T$, $R\sb{\rm p}$, {\water}, CO, {\carbdiox}, {\methane}, [M/H]     & 7 & $ 6.7$ & $2.09\pm0.63$ \\
M6 & $T$, $R\sb{\rm p}$, {\water}, CO, {\carbdiox}, {\methane}, [M/H], {\pcloud} & 8 & $ 9.2$ & $2.62\pm0.71$ \\
M0 & Flat fit                                           & 1 & $ 6.1$ & $2.25\pm0.43$ \\
\hline
\\ 
\multicolumn{5}{l}{ XO-1b Tsiaras }\\ 
\hline
ID & Free parameters & $N\sb{\rm free}$ & $\Delta$BIC & \redchisq \\
\hline
M1 & $T$, $R\sb{\rm p}$, {\water}, {\pcloud}            & 4 & $ 0.0$ & $0.76\pm0.38$ \\
M2 & $T$, $R\sb{\rm p}$, {\water}, [M/H]                & 4 & $ 0.0$ & $0.76\pm0.38$ \\
M3 & $T$, $R\sb{\rm p}$, {\water}, [M/H], {\pcloud}     & 5 & $ 2.9$ & $0.82\pm0.39$ \\
M0 & Flat fit                                           & 1 & $ 5.7$ & $1.47\pm0.34$ \\
\hline
\end{tabular}
\end{table}
\normalsize
}

\bsp
\label{lastpage}
\end{document}

%% file: targetsample.tex
{
\begin{table*}
\centering
\caption{Physical Properties and Datasets of our WFC3 Sample.}
\label{table:sample}
\begin{tabular}{@{\extracolsep{\fill}}lllrlccclll}
\hline
Planet      & $R\sb{\rm p}$ & $M\sb{\rm p}$  & ${T\sb{\rm eq}^*}$ &  $R\sb{\rm s}$ & $M\sb{\rm s}$ & $a$ & $T\sb{\rm eff}$ & \multicolumn{1}{c}{System-parameter} & \multicolumn{1}{c}{Transmission-data} & Observing Proposal \\
            & \rjup & \mjup & \multicolumn{1}{c}{K}   & \rsun & \msun & AU   & \multicolumn{1}{c}{K}   & \multicolumn{1}{c}{References} &  \multicolumn{1}{c}{References} & \multicolumn{1}{c}{PI (ID)}   \\
\hline
GJ 436b     & 0.366 & 0.08  &  630 & 0.455 & 0.556 & 0.0308 & 3416 & \citet{LanotteEtal2014aaGJ436bGlobalAnalysis} & \citet{TsiarasEtal2018ajPopulationStudy} & Stevenson (13338) \\
HAT-P-3b    & 0.899 & 0.596 & 1160 & 0.833 & 0.93  & 0.0388 & 5185 & \citet{TorresEtal2008apjTransitParameters}    & \citet{TsiarasEtal2018ajPopulationStudy} & Deming (14260) \\
HAT-P-17b   & 1.01  & 0.534 &  780 & 0.838 & 0.857 & 0.0882 & 5246 & \citet{HowardEtal2012apjHATP17system}    & \citet{TsiarasEtal2018ajPopulationStudy} & Huitson (12956) \\
HAT-P-18b   & 0.947 & 0.196 &  840 & 0.717 & 0.77  & 0.0559 & 4870 & \citet{EspositoEtal2014aaHATP18bGAPS}    & \citet{TsiarasEtal2018ajPopulationStudy} & Deming (14260) \\
HAT-P-32b   & 1.790 & 0.86  & 1780 & 1.22  & 1.16  & 0.034 & 6210 & \citet{HartmanEtal2011apjHATP32andHATP33} & \citet{DamianoEtal2017ajHAT32bWFC3} & Deming (14260) \\
HAT-P-32b   & 1.79  & 0.86  & 1780 & 1.22  & 1.16  & 0.034 & 6210 & \citet{HartmanEtal2011apjHATP32andHATP33} & \citet{TsiarasEtal2018ajPopulationStudy} & Deming (14260) \\
HAT-P-38b   & 0.82  & 0.267 & 1080 & 0.92  & 0.89  & 0.052 & 5330 & \citet{SatoEtal2012pasjHATP38b}              & \citet{BrunoEtal2018ajWASP67bHAT38bWFC3} & Deming (14260) \\
HAT-P-38b   & 0.82  & 0.267 & 1080 & 0.92  & 0.89  & 0.052 & 5330 & \citet{SatoEtal2012pasjHATP38b}              & \citet{TsiarasEtal2018ajPopulationStudy} & Deming (14260)  \\
HAT-P-41b   & 1.685 & 0.8   & 1940 & 1.683 & 1.418 & 0.0426 & 6390 & \citet{HartmanEtal2012ajHATP39b-HATP41b}    & \citet{TsiarasEtal2018ajPopulationStudy} & Deming (14260) \\
HD 149026b  & 0.654 & 0.359 & 1670 & 1.368 & 1.29  & 0.0431 & 6160 & \citet{TorresEtal2008apjTransitParameters}    & \citet{TsiarasEtal2018ajPopulationStudy} & Deming (14260) \\
K2-18b      & 0.242 & 0.027 &  290 & 0.469 & 0.495 & 0.1591 & 3503 & \citet{CloutierEtal2019aaK2-18bHarpsCarmenes} & \citet{BennekeEtal2019apjK218b} & Benneke (13665) \\
K2-18b      & 0.203 & 0.025 &  290 & 0.411 & 0.359 & 0.1429 & 3457 & \citet{BennekeEtal2017apjK2-18bSpitzer}   & \citet{TsiarasEtal2019natasK218b} & Benneke (13665) \\
WASP-29b    & 0.792 & 0.244 &  970 & 0.808 & 0.825 & 0.0457 & 4800 & \citet{HellierEtal2010apjWASP29b}    & \citet{TsiarasEtal2018ajPopulationStudy} & Deming (14260) \\
WASP-43b    & 1.04  & 2.03  & 1460 & 0.67  & 0.72  & 0.015 & 4520 & \citet{GillonEtal2012aaTrappistWASP43b}  & \citet{StevensonEtal2017ajWASP43bSpitzerPhase} & Bean (13467) \\
WASP-43b    & 0.93  & 1.78  & 1440 & 0.67  & 0.71  & 0.0142 & 4400 & \citet{HellierEtal2011aaWASP43bdisc}    & \citet{TsiarasEtal2018ajPopulationStudy} & Bean (13467) \\
WASP-63b    & 1.43  & 0.38  & 1540 & 1.88  & 1.32  & 0.057 & 5550 & \citet{HellierEtal2012mnrasSevenHotJupiters} & \citet{KilpatrickEtal2018apjWASP63bWFC3} & Stevenson (14642) \\
WASP-63b    & 1.43  & 0.38  & 1540 & 1.88  & 1.32  & 0.057 & 5550 & \citet{HellierEtal2012mnrasSevenHotJupiters} & \citet{TsiarasEtal2018ajPopulationStudy} & Stevenson (14642) \\
WASP-67b    & 1.40  & 0.42  & 1040 & 0.87  & 0.87  & 0.052 & 5200 & \citet{HellierEtal2012mnrasSevenHotJupiters} & \citet{BrunoEtal2018ajWASP67bHAT38bWFC3} & Deming (14260) \\
WASP-67b    & 1.40  & 0.42  & 1040 & 0.87  & 0.87  & 0.052 & 5200 & \citet{HellierEtal2012mnrasSevenHotJupiters}    & \citet{TsiarasEtal2018ajPopulationStudy} & Deming (14260) \\
WASP-69b    & 1.057 & 0.26  &  960 & 0.813 & 0.826 & 0.0452 & 4715 & \citet{AndersonEtal2014mnrasWASP84b-WASP69b-WASP70Ab}    & \citet{TsiarasEtal2018ajPopulationStudy} & Deming (14260)  \\
WASP-74b    & 1.56  & 0.95  & 1920 & 1.64  & 1.48  & 0.037  & 5970 & \citet{HellierEtal2015ajWASP74b-WASP83b-WASP89b}    & \citet{TsiarasEtal2018ajPopulationStudy} & Sing (14767) \\
WASP-80b    & 0.999 & 0.538 & 820 & 0.586 & 0.577 & 0.0344 & 4143 & \citet{TriaudEtal2015mnrasWASP80b}    & \citet{TsiarasEtal2018ajPopulationStudy} & Deming (14260) \\
WASP-101b   & 1.41  & 0.50  & 1560 & 1.29  & 1.34  & 0.051 & 6400 & \citet{HellierEtal2014mnrasWASP95b-101b}  & \citet{WakefordEtal2017apjWASP101bWFC3} & Sing (14767) \\
WASP-101b   & 1.41  & 0.50  & 1560 & 1.29  & 1.34  & 0.051 & 6400 & \citet{HellierEtal2014mnrasWASP95b-101b}    & \citet{TsiarasEtal2018ajPopulationStudy} & Sing (14767) \\
WASP-103b   & 1.528 & 1.49  & 2510 & 1.436  & 1.22  & 0.0198 & 6110 & \citet{GillonEtal2014aaWASP103b} & \citet{KreidbergEtal2018apjWASP103bPhaseCurve} & Kreidberg (14050) \\
XO-1b       & 1.206 & 0.918 & 1210 & 0.934 & 1.03  & 0.0493 & 5750 & \citet{TorresEtal2008apjTransitParameters}    & \citet{TsiarasEtal2018ajPopulationStudy} & Deming (12181) \\
\hline
\end{tabular}
\begin{tablenotes}
  \item $^*$ Assuming zero albedo and efficient energy redistribution
        (i.e., reemission over $4\pi$ radians).
\end{tablenotes}
\end{table*}
}

%% file: table_results.tex
{\renewcommand{\arraystretch}{1.3}
\begin{table*}
\centering
\caption{WFC3 Retrieval Results for planets with ${\rm ADI} \gtrsim 3$.}
\label{table:wfc3_retrieval}
\begin{tabular}{@{\extracolsep{\fill}}lrrccr}
\hline
Dataset & $T ({\rm K})$ &  $R_{\rm planet}$ ($R_{\rm Jup}$) & $\log_{10}(X_{\rm H2O})$ &  $\log_{10}(X_{\rm HCN})$ &  $\log_{10}(p_{\rm top})$  \\
\hline
HATP-18b Tsiaras       &   $423_{-261}^{+149}$  &  $0.952_{-0.007}^{+0.004}$  &  $-0.44_{-1.07}^{+0.40}$  &  $\cdots$  &  $-1.07_{-1.13}^{+2.76}$ \\
HATP-32b Tsiaras       &   $643_{-192}^{+321}$  &  $1.797_{-0.019}^{+0.006}$  &  $-0.94_{-2.13}^{+0.24}$  &  $\cdots$  &  $-1.19_{-1.36}^{+2.55}$ \\
HATP-38b Bruno         &   $1460_{-215}^{+40}$  &  $0.841_{-0.041}^{+0.004}$  &  $-5.71_{-0.43}^{+0.59}$ and $-0.77_{-0.78}^{+0.53}$  &  $\cdots$  &  $0.46_{-4.34}^{+1.40}$ \\
HATP-41b Tsiaras       &  $1392_{-418}^{+635}$  &  $1.619_{-0.025}^{+0.043}$  &  $-0.93_{-3.90}^{+0.23}$  &  $\cdots$  &  $-1.74_{-2.56}^{+1.08}$ \\
K2-18b Benneke         &   $352_{-142}^{ +30}$  &  $0.246_{-0.001}^{+0.002}$  &  $-0.87_{-2.47}^{+0.31}$  &  $\cdots$  &  $-0.91_{-0.94}^{+2.00}$ \\
WASP-43b Stevenson     &   $502_{-144}^{+156}$  &  $1.035_{-0.001}^{+0.001}$  &  $-2.16_{-1.29}^{+0.96}$  &  $\cdots$  &  $0.01_{-0.68}^{+1.82}$ \\
WASP-63b Kilpatrick    &   $569_{-169}^{+530}$  &  $1.418_{-0.031}^{+0.014}$  &  $-3.82_{-1.34}^{+1.33}$  &  $-3.21_{-1.36}^{+1.67}$  &  $-1.07_{-2.54}^{+2.39}$ \\
WASP-69b Tsiaras       &   $154_{-45}^{+109}$  &  $1.017_{-0.001}^{+0.001}$  &  $-3.76_{-0.80}^{+1.01}$ and $-0.40_{-0.24}^{+0.20}$  &  $\cdots$  &  $2.28_{-3.86}^{+0.32}$ \\
WASP-103b Kreidberg    &  $2853_{-725}^{+127}$  &  $1.615_{-0.069}^{+0.008}$  &  $-5.60_{-0.65}^{+1.01}$ and $-1.27_{-1.59}^{+0.57}$ &  $\cdots$  &  $0.71_{-1.75}^{+1.29}$ \\
XO-1b Tsiaras          &   $714_{-233}^{+352}$  &  $1.195_{-0.010}^{+0.002}$  &  $-1.02_{-1.79}^{+0.32}$  &  $\cdots$  &  $0.79_{-2.05}^{+1.08}$ \\
\hline
\end{tabular}
\begin{tablenotes}
  \item The reported values correspond to the marginal posterior
distribution's mode and boundaries of the 68\% HPD credible intervals.
\end{tablenotes}
\end{table*}
}